\documentclass{article}

\usepackage{amsmath,mathrsfs}
\usepackage{ upgreek }
\usepackage{ amssymb }
\usepackage{graphicx,subfigure}
\graphicspath{{v-phi/}}
\usepackage{multirow,booktabs}
\usepackage[numbers,sort&compress]{natbib}

\def\abstract#1{\vskip 7mm 
        \begin{center}{\large Abstract}\par \smallskip
                \begin{minipage}[c]{12cm}
                        \small #1
                \end{minipage}
        \end{center}
}
\def\title#1{\begin{center}{\Large\bf #1}\end{center}}
\def\author#1{\vskip 5mm \begin{center}{#1}\end{center}}
\def\address#1{\begin{center}{\it #1}\end{center}}
\makeatletter
\@ifundefined{lesssim}{}{}
\@ifundefined{gtrsim}{}{}
\def\vereq#1#2{\lower3pt\vbox{\baselineskip1.5pt \lineskip1.5pt
\ialign{$\m@th#1\hfill##\hfil$\crcr#2\crcr\sim\crcr}}}
\makeatother

\begin{document}

\title{	Spatial Curvature and Large Scale Lorentz Violation
}
\author{ Jing Li$^a$, Yongxiang Zhou$^a$ and Xun Xue$^{a,b,}\footnote{Corresponding author:xxue@phy.ecnu.edu.cn}$}
\address{ $^a$Department of Physics, East China Normal University, Shanghai 200241, China}
\address{ $^b$Center for Theoretical Physics, Xinjiang University, Urumqi 830046, China}
\abstract{
	The tension between the Hubble constant obtained from the local measurements and from cosmic microwave background (CMB) measurements motivated us to consider the cosmological model beyond $\Lambda$CDM one. We investigate the cosmology in the large scale Lorentz violation model with non-vanishing spatial curvature. The degeneracy among spatial curvature, cosmological constant and cosmological contortion distribution makes the model viable in describing the known observation date. We get some constraints on the spatial curvature by the comparison of the relation between measured distance modulus and red-shift with the predicted one, the evolution of matter density over time and the evolution of effective cosmological constant. The performance of large scale Lorentz violation model with non-vanishing spatial curvature under these constrains is discussed.
	}

\section{Introduction}

The present Hubble constant value can be measured in many ways, e.g., by the $\Lambda$CDM model with CMB measurement fixing its parameters at relatively early universe, or by the distance ladder method at relatively late universe. The discrepancy between the values of these different methods is known as the $H_0$ tension problem. 

In the late 1990s the use of Type Ia supernovae as standard candles led to the discovery that the expansion of the universe is accelerating. Large-scale observations indicates a phase of cosmological acceleration which occurs at late time\cite{Planck:2018vyg}. Hence, several attempts have been proposed to describe the cosmic accelerated scenario. In general, there are two kinds of interpretations for this cosmic phase: i) postulating an exotic form of energy with negative pressure usually called dark energy or ii) modifying the laws of gravity. Numerous models have been proposed based on these two branches, but it is difficult to determine which one is correct due to the degeneracies in the parameter space. Despite this, the $\Lambda$CDM model with six parameters could excellently fit almost all observational data, and has been set as the standard model of cosmology\cite{ade2014planckI}.
By assuming flat space in $\Lambda$CDM model, the final full-mission Planck measurements of the cosmic microwave background (CMB) anisotropies has found a value of the Hubble constant${{H}_{0}}=\left( 67.27\pm 0.60 \right)km/s/Mpc$. This is compatible with many earlier and recent estimates of $H_0$. In contrast, multiple local expansion rate measurements find slightly higher $H_0$ values and slightly larger error bars. And the latest value from the Supernovae and $H_0$ for the Dark Energy Equation of State (S$H_0$ES) project together with GAIA DR2 parallaxes is ${{H}_{0}}=\left( 73.52\pm 1.62 \right)km/s/Mpc$, which has a more than $3\sigma $ tension with the Planck CMB data\cite{Riess:2018byc}. This tension is one of the most intriguing problems in modern cosmology. There have been many attempts to solve the problem, such as introducing new physics beyond the standard $\Lambda$CDM cosmological model. The non-zero spatial curvature may be an choice of solution to the $H_0$ tension problem. Observing that the Friedmann equation ${{\dot{a}}^{2}}+K=\frac{8\pi G}{3}\rho {{a}^{2}}$ $\Lambda$CDM model with spatial curvature can be written as
\begin{equation}
1-\frac{1}{\Omega }=\frac{3K}{8\pi G\rho {{a}^{2}}}
\end{equation}
where $\Omega =\frac{\rho }{{{\rho }_{c}}}=\frac{8\pi G\rho }{3}{{\left( \frac{a}{{\dot{a}}} \right)}^{2}}$ is cosmic density, one can find $\frac{3K}{8\pi G\rho {{a}^{2}}}\propto {{a}^{2}}$ in the radiation dominated era while $\frac{3K}{8\pi G\rho {{a}^{2}}}\propto a$ in the matter-dominated era for  $\rho {{a}^{3}}=const$ then. In a word, the absolute value of $1-\frac{1}{\Omega }$ should grow all the time from the early universe until now if $K\ne 0$ because the cosmic scale factor is growing all the time. From the Planck period until now, $a\left( t \right)$ has increased by dozens of magnitudes, then the value of $1-\frac{1}{\Omega }$ has also increased by dozens of magnitudes. Nowadays, the deviation of $\Omega $ to 1 is observed to be a magnitude of up to 1, then the current value of $1-\frac{1}{\Omega }$ should also be a magnitude of 1. Based on existing conclusions, we can get the cosmological density of the Planck period ${{\Omega }_{p}}=1\pm {{10}^{-N}}$, where N is a constant more than dozens. It suggests that at the end of the Planck period, the critical density and cosmic mass density were the same in many significant numbers, and the two physical quantity were not exactly equal. Otherwise, after the long evolution, there is a universe without the present spatial quasi-flatness. The non-zero spatial curvature encounters the problem of fine-tuning somehow in the inflation scenario. However it is still sensible to figure out whether the present distance observations prefer a non-zero spatial curvature or not. Although the current cosmological observations strongly favor a spatially flat Universe, e.g., the combined Planck 2018 cosmic microwave background (CMB) and baryon acoustic oscillation measurements, which suggest that ${{\Omega }_{K}}=0.001\pm 0.002$\cite{Planck:2018vyg}, these constraints, however, are based on the pre-assumption of a specific cosmological model (e.g., the standard $\Lambda$CDM model). Because of the strong degeneracy between spatial curvature and the equation of state of dark energy\cite{Virey:2008nu}, it is rather difficult to constrain the two quantities simultaneously. In general, dark energy is assumed to be a cosmological constant for the estimation of curvature, or conversely, the Universe is assumed to be flat in a dark energy analysis. Indeed, if the density of dark energy is allowed to vary freely with time, constraints on the geometry of the Universe may not only become less stringent, but may even depend on the properties that dark energy is assumed to have at early times\cite{Wang:2007mza}. But a simple flatness assumption may result in an incorrect reconstruction of the dark energy equation of state, even if the real curvature is very tiny, and a cosmological constant assumption may lead to confusion between $\Lambda$CDM and a dynamical dark energy model\cite{Clarkson:2007bc}. Moreover, the tension between the most recent SneIa data and spatial flat assumption in the $\Lambda$CDM model\cite{Rest:2013mwz} may eventually support a spatial non-flat universe\cite{Kumar:2015saa}. It is pointed out that the Planck 2018 CMB spectra preferring a positive spatial curvature at more than $99\%$ confidence level(CL)\cite{DiValentino:2019qzk}. The combined analysis of Planck CMB anisotropy and luminosity distant data simultaneously excludes a flat universe and a cosmological constant at $99\%$ CL\cite{DiValentino:2020hov}.

The motivation behind the choice of non-flat cosmology is that, as usually believed, an early inflationary phase leads today to almost flat universe, albeit not exactly with a perfectly zero spatial curvature. This is not necessary if the number of e-foldings is not very large\cite{Huang:2004ai}. It is still possible that there is a contribution to the Friedmann equations from the spatial curvature when studying latetime universe, though much smaller than other energy components according to observations.
Recent observations have complicated the problem by suggesting that the deceleration factors of the expansion of the universe change over time, and that the universe went from decelerating to accelerated expansion about 6 billion years ago. 

In the $\Lambda$CDM model, the cosmological constant can be viewed as from vacuum energy density, which is responsible for the accelerated expansion of the universe. Considering the large scale Lorentz violation, the large-scale torsion distribution caused by the Lorentz violation effect combined with the vacuum energy density is responsible for the accelerating expansion\cite{Shen:2018elj}. Therefor we can introduce the effective cosmological constant which is responsible for the accelerating expansion of the universe. Using the evolution of matter density ${{\Omega }_{M}}$ and initial conditions to constrain the bare cosmological constant to a desirable range. The performance of large scale Lorentz violation model with non-vanishing spatial curvature under these constrains will be discussed in this paper .

\section{ Cosmology of the Gravitation Theory with Large Scale Lorentz Violation}

The large scale Lorentz violation gravitation model with zero spatial curvature is discussed in Reference \cite{Shen:2018elj}. The general FRW metric including the open and closed spatial geometry is 
\begin{equation}
d{{s}^{2}}=d{{t}^{2}}-{{a}^{2}}\left( t \right)\left( d{{r}^{2}}/\left( 1-K{{r}^{2}} \right)+{{r}^{2}}d{{\theta }^{2}}+{{r}^{2}}d{{\theta }^{2}}+{{r}^{2}}si{{n}^{2}}\theta d{{\varphi }^{2}} \right).
\end{equation}
The equations of motion for gravitational field equation is
\begin{equation}
{{\tilde{G}}^{a}}_{\ \ b}={{\tilde{R}}^{a}}_{\ \ b}-1/2\tilde{R}{{\delta }^{a}}_{b}=8\pi G/{{c}^{4}}{{\left( {{T}_{M}}+{{T}_{\Lambda }} \right)}^{a}}_{b},
\end{equation}
in which ${{\tilde{G}}^{a}}_{\ \ b}$, ${{\tilde{R}}^{a}}_{\ \ b}$ and $\tilde{R}$ are respectively the Einstein tensor, Ricci curvature tensor and Ricci curvature scalar of the spacetime with Levi-Civita connection. ${{T}_{\Lambda }}$ is energy-momentum tensor of the dark partner contributed by the contortion tensor, which is responsible for the accelerated expansion of the universe, ${{\left[ {{T}_{\Lambda }} \right]}^{a}}_{b}=diag\left( {{\rho }_{\Lambda }},-{{p}_{\Lambda }},-{{p}_{\Lambda }},-{{p}_{\Lambda }} \right)$. The dark partner energy density ${{\rho }_{\Lambda }}$ and pressure ${{p}_{\Lambda }}$ are expressed as
\begin{align}
& {{\rho }_{\Lambda }}=-\frac{{{c}^{4}}}{8\pi G}\left( 3{{\mathcal K}^{2}}+6{\mathcal K}\frac{{\dot{a}}}{a}-{{\Lambda }_{0}} \right) \\ 
& {{p}_{\Lambda }}=-\frac{{{c}^{4}}}{8\pi G}\left( {{\mathcal K}^{2}}+4{\mathcal K}\frac{{\dot{a}}}{a}+2\dot{\mathcal K}-{{\Lambda }_{0}} \right)
\end{align}
where ${\mathcal K}\left( t \right)=K_{11}^{0}=K_{22}^{0}=K_{33}^{0}$ is the non-zero components of the contortion tensor and ${{\Lambda }_{0}}$ is the bare cosmological constant, the geometrical contribution by vacuum energy density. The modified Friedmann Equation in geometrical unit $\frac{{{c}^{4}}}{8\pi G}\text{=}1$ can be written as
\begin{equation}\label{mofridmeq1}
{{\left( \frac{{\dot{a}}}{a} \right)}^{2}}\text{+}2\frac{{\dot{a}}}{a}{\mathcal K}+{{\mathcal K}^{2}}+\frac{K}{{{a}^{2}}}\text{=}\frac{1}{3}\left( \rho \text{+}{{\Lambda }_{0}} \right)
\end{equation}
and
\begin{equation}\label{mofridmeq2}
\ddot{a}=-\frac{a}{2}\left( p+\frac{1}{3}\rho  \right)\text{+}\frac{1}{3}a{{\Lambda }_{0}}-\frac{d}{dt}\left( a{\mathcal K} \right)
\end{equation}
This paper aims to explore the performance of the large scale Lorentz violation model with non-vanishing spatial curvature. It should be noted that the model lacks the evolution of ${\mathcal K}\left( t \right)$, the set of equations \eqref{mofridmeq1} and \eqref{mofridmeq2} is not closed. As discussed in \cite{Shen:2018elj,Han-Yu:2019tmf,Zhai:2019std}, the scales of Lorentz violation region during quantum gravity dominating era experience the stretching out beyond the horizon by inflation and the region with sub-scale may reenter the horizon during normal expansion. In principle it is needed the specific quantum gravity model and inflation model to give the prediction on the evolution of ${\mathcal K}\left( t \right)$. However, the evolution of ${\mathcal K}\left( t \right)$ can be approximated by employing the $\Lambda$CDM model or assuming an equation of state satisfied by the dark partner phenomenologically. One can add an independent equation from the set of Friedmann equations of the $\Lambda$CDM model to close the set of modified Friedmann equations  with two choices as in \cite{Han-Yu:2019tmf,Zhai:2019std}. The third approximation comes from adding the equation of state for the dark partner to the modified Friedmann equations set. The three kinds of approximation can be given as follows as in \cite{Han-Yu:2019tmf,Zhai:2019std}.

The first kind of approximation named Case A is
\begin{equation}\label{appA}
\frac{1}{3}a{{\Lambda }_{0}}-\frac{1}{3}a\Lambda \text{=}\frac{d}{dt}\left( a{\mathcal K} \right),
\end{equation}
and the second kind named Case B is 
\begin{equation}\label{appB}
\left( 3w+2 \right)\frac{{\dot{a}}}{a}{\mathcal K}\text{+}\dot{\mathcal K}\text{+}\frac{3w+1}{2}{{\mathcal K}^{2}}\text{+}\frac{3w+1}{2}\frac{K}{{{a}^{2}}}\text{+}\frac{w+1}{2}\left( \Lambda -{{\Lambda }_{0}} \right)\text{=0},
\end{equation}
while the third one named Case C is
\begin{equation}\label{appC}
\left( 4\text{+}6{{w}_{0}} \right)\frac{{\dot{a}}}{a}{\mathcal K}\text{+}\left( 1+3{{w}_{0}} \right){{\mathcal K}^{2}}+2\dot{\mathcal K}\text{=}\left( {{w}_{0}}+1 \right){{\Lambda }_{0}},
\end{equation}
where $w$ is the equation of state parameter for cosmic media in $p=w\rho $ and ${{w}_{\text{0}}}$ is the equation of state parameter for dark partner part in ${{p}_{\Lambda }}={{w}_{0}}{{\rho }_{\Lambda }}$.

The Friedmann equation of $\Lambda$CDM model can be written as
\begin{equation}\label{fridmlcdmeq1}
\Lambda =3{{\left( \frac{{\dot{a}}}{a} \right)}^{2}}-\rho
\end{equation}
Similarly, the modified Friedmann equation of the large scale Lorentz violation model can be written as
\begin{equation}\label{mdffridmeq1}
{{\Lambda }_{0}}-\text{6}{\mathcal K}\frac{{\dot{a}}}{a}-3{{\mathcal K}^{2}}-3\frac{K}{{{a}^{2}}}\text{=}3{{\left( \frac{{\dot{a}}}{a} \right)}^{2}}-\rho
\end{equation}
Comparing \eqref{fridmlcdmeq1} and \eqref{mdffridmeq1}, it is not only the bare cosmological constant ${{\Lambda }_{0}}$ that contributes to the accelerated expansion of the universe in the large scale Lorentz violation model, but also the contribution from ${\mathcal K}\left( t \right)$, whose total contribution can define the effective cosmological constant
\begin{equation}
{{\Lambda }_{eff}}\equiv {{\Lambda }_{0}}-6H{\mathcal K}-3{{\mathcal K}^{2}}-\frac{3K}{{{a}^{2}}}.
\end{equation}
The initial value of ${\mathcal K}\left( t \right)$, the contortion component value at present time, can be obtained by the Friedmann equation of $\Lambda$CDM model with two possible choices. The first kind of initial value is
\begin{equation}\label{inivlu1}
{\mathcal K}\left( {{t}_{0}} \right)\text{=}{{H}_{0}}\left\{ -1+\sqrt{1-\frac{K}{a_{0}^{2}H_{0}^{2}}-\frac{\Lambda }{3H_{0}^{2}}+\frac{{{\Lambda }_{0}}}{3H_{0}^{2}}} \right\}
\end{equation}
while the second one is  
\begin{equation}\label{inivlu2}
{\mathcal K}\left( {{t}_{0}} \right)\text{=}{{H}_{0}}\left\{ -1-\sqrt{1-\frac{K}{a_{0}^{2}H_{0}^{2}}-\frac{\Lambda }{3H_{0}^{2}}+\frac{{{\Lambda }_{0}}}{3H_{0}^{2}}} \right\}.
\end{equation}
To ensure the value of contortion components take real numbers, there should be a constrain condition  satisfied by the bare cosmological constant, the cosmological constant observed and the spatial curvature, $K/a_{0}^{2}+\Lambda /3-{{\Lambda }_{0}}/3\le H_{0}^{2}$, i.e.
\begin{equation}\label{cstrL0}
{{\Lambda }_{0}}\ge \Lambda \text{+}\frac{\text{3}K}{a_{0}^{2}}-3H_{0}^{2}.
\end{equation}

For a specific large scale Lorentz violation model, the spatial curvature constant $K$, the current cosmic scale factor ${{a}_{0}}$, Hubble constant ${{H}_{0}}$ and the bare cosmological constant ${{\Lambda }_{0}}$ are the free input parameters given by observation or predicted by more fundamental theory model. It can be defined that 
\begin{equation}\label{L0min}
	{{\Lambda }_{min}}=\Lambda \text{+}\frac{\text{3}K}{a_{0}^{2}}-3H_{0}^{2}
\end{equation}
from condition \eqref{cstrL0} so that there is a minimum for the free input bare cosmological constant, $\Lambda _0 \ge \Lambda _{min}$, to give a reasonable evolution of the universe.

\begin{figure}[h]
	\centering
	\subfigure[]
	{\label{A1}
		\includegraphics[width=2.5in]{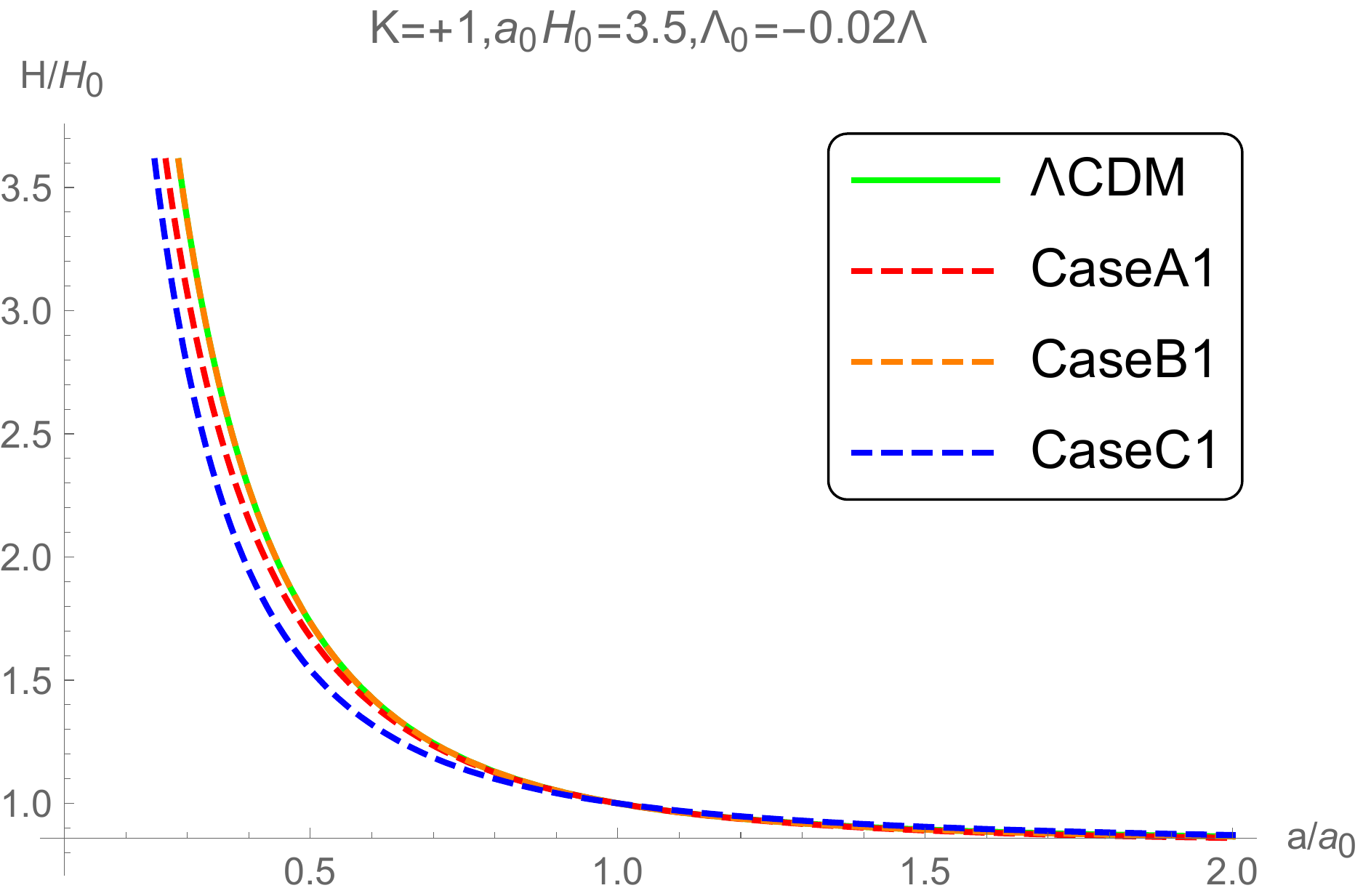}}
	\subfigure[]
	{\label{B1}
		\includegraphics[width=2.5in]{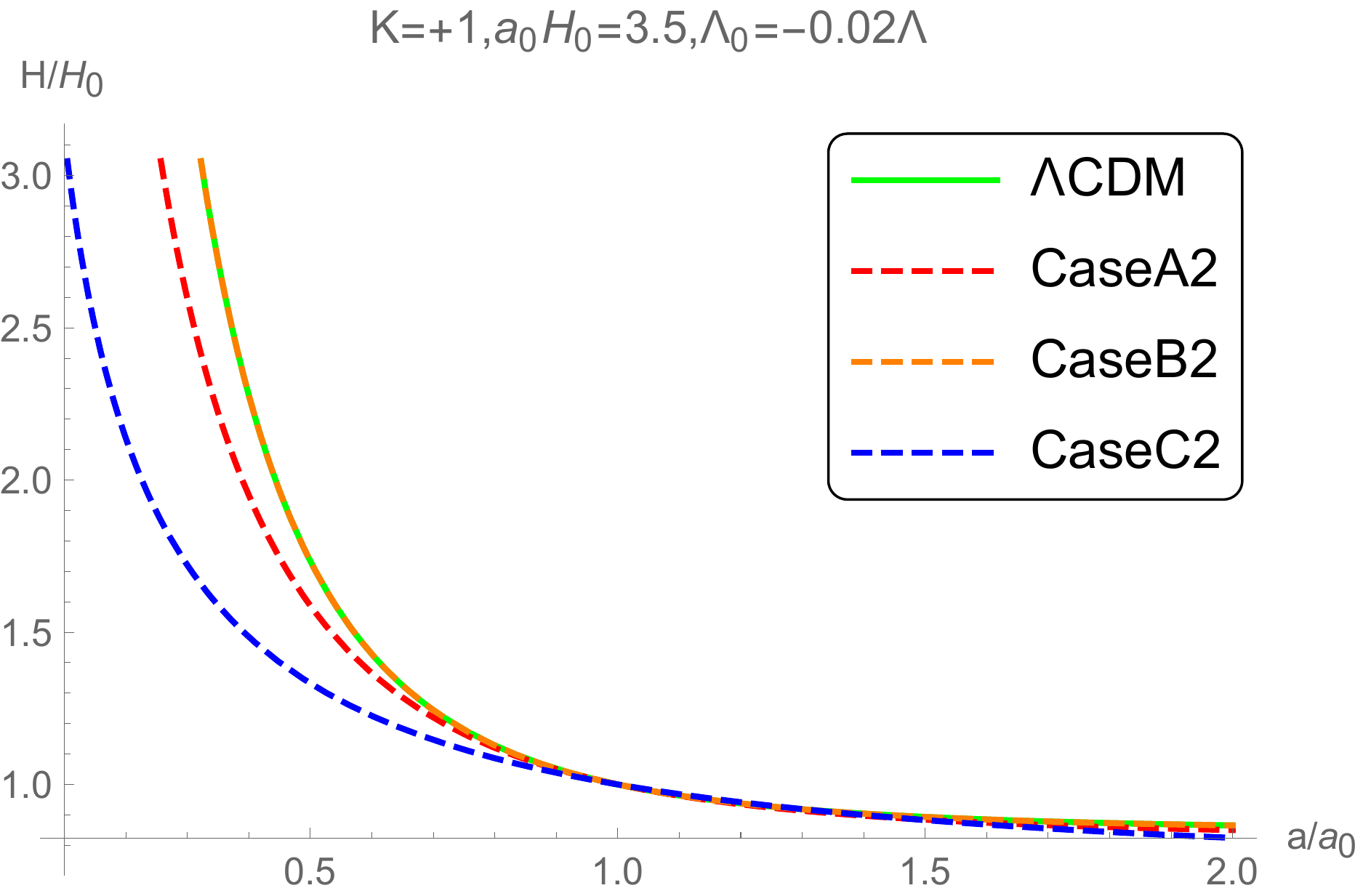}}
	\subfigure[]
	{\label{C1}
		\includegraphics[width=2.5in]{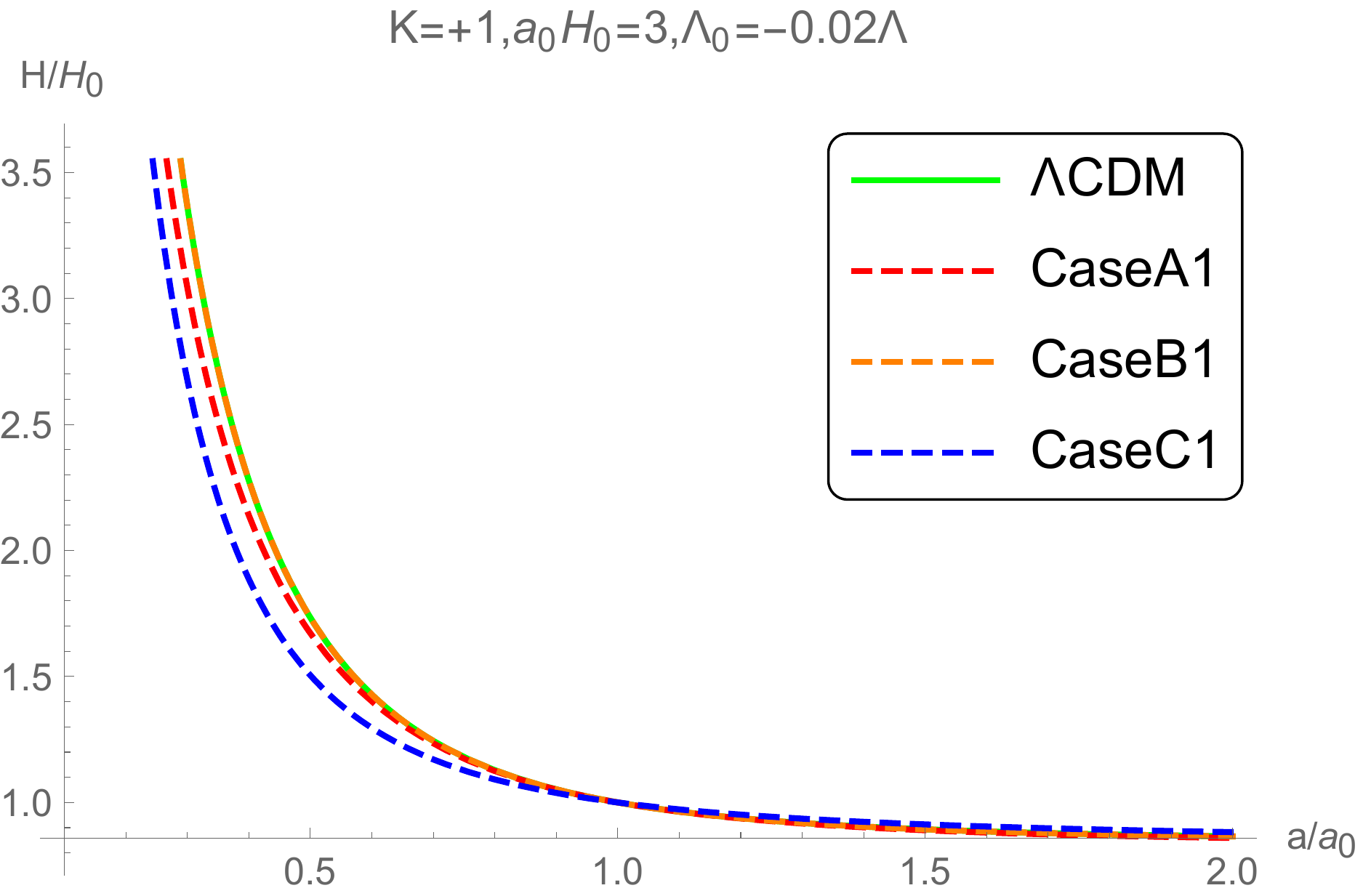}}
	\subfigure[]
	{\label{D1}
		\includegraphics[width=2.5in]{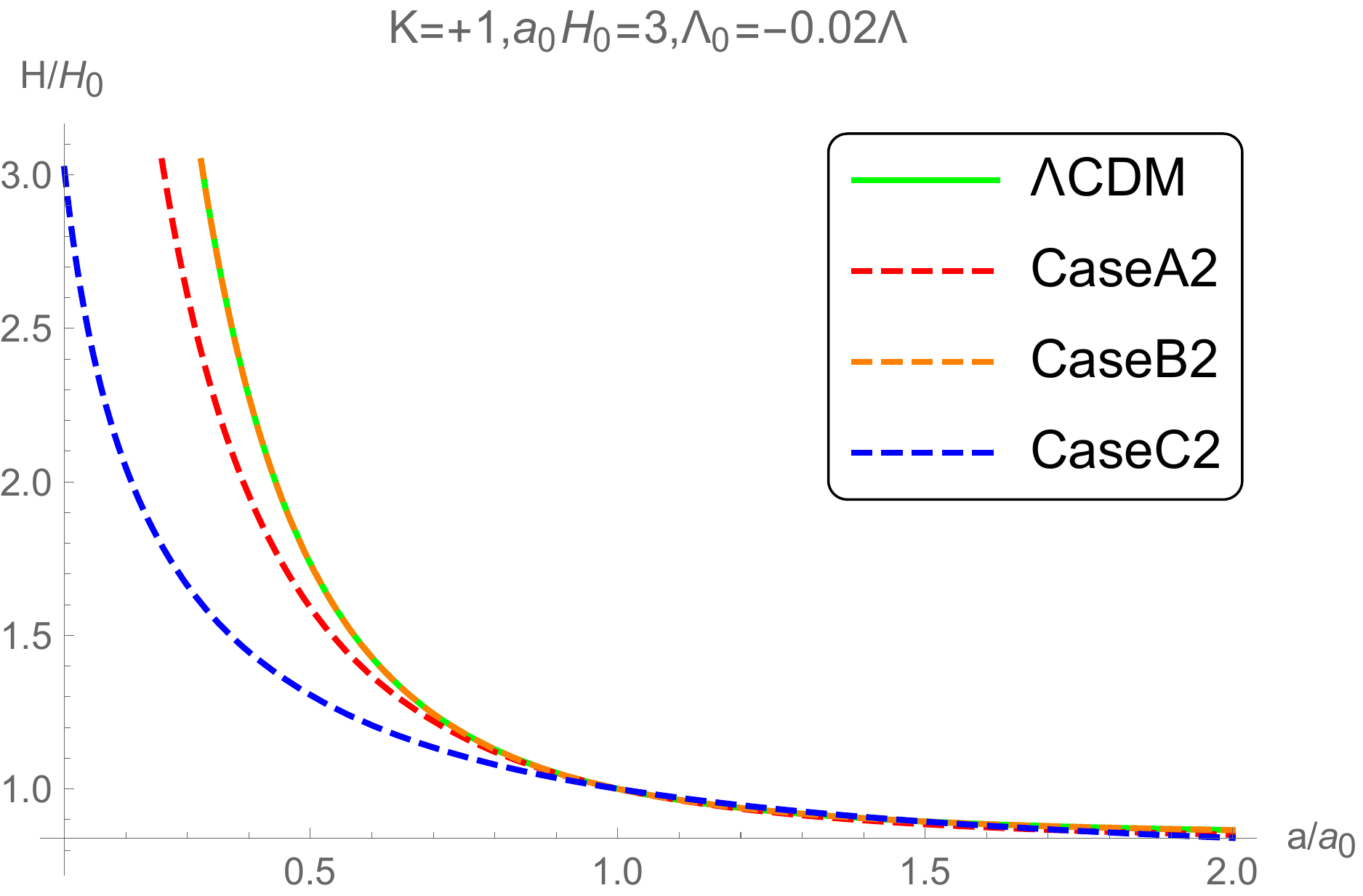}}
	\subfigure[]
	{\label{E1}
		\includegraphics[width=2.5in]{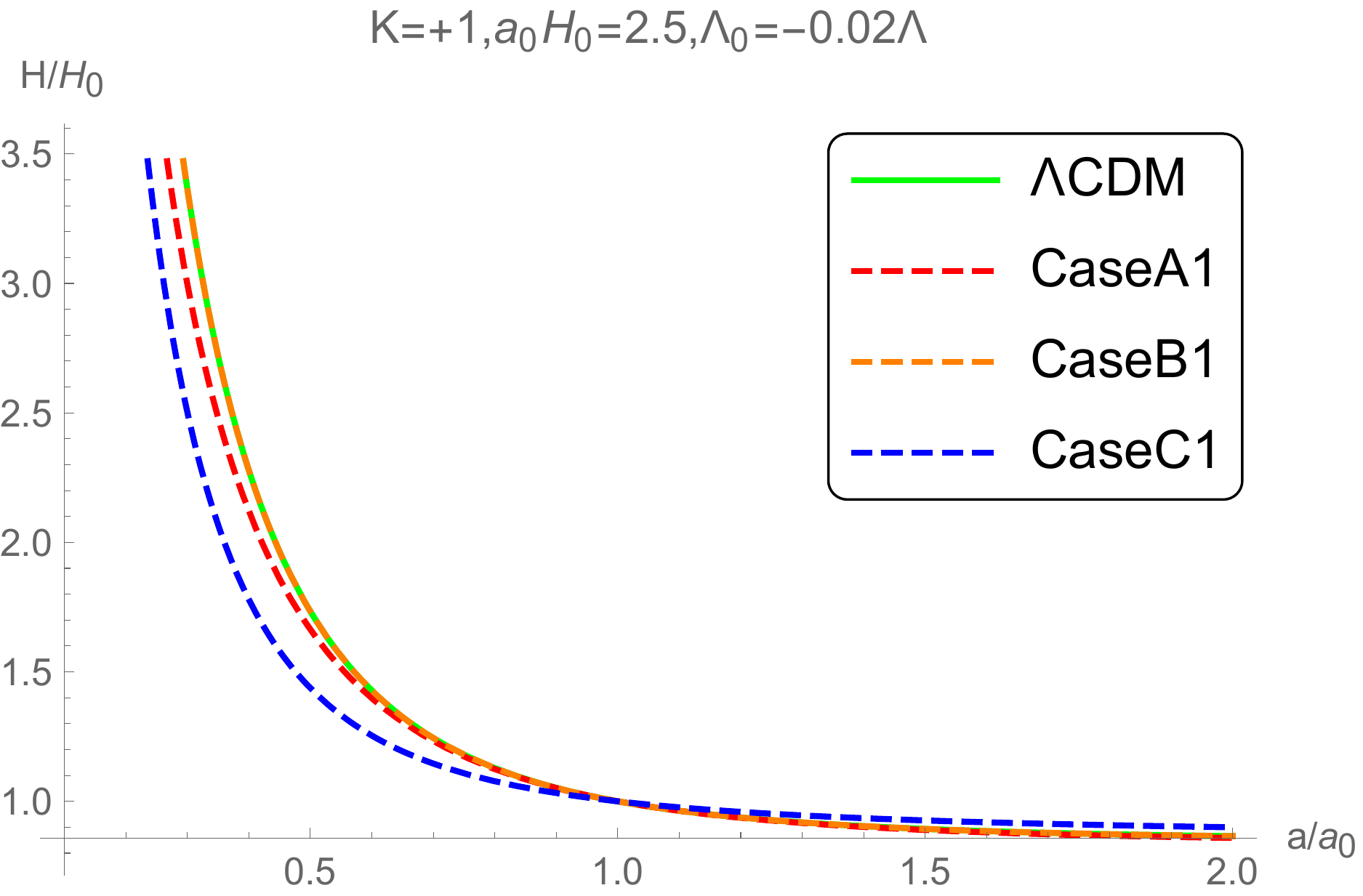}}
	\subfigure[]
	{\label{F1}
		\includegraphics[width=2.5in]{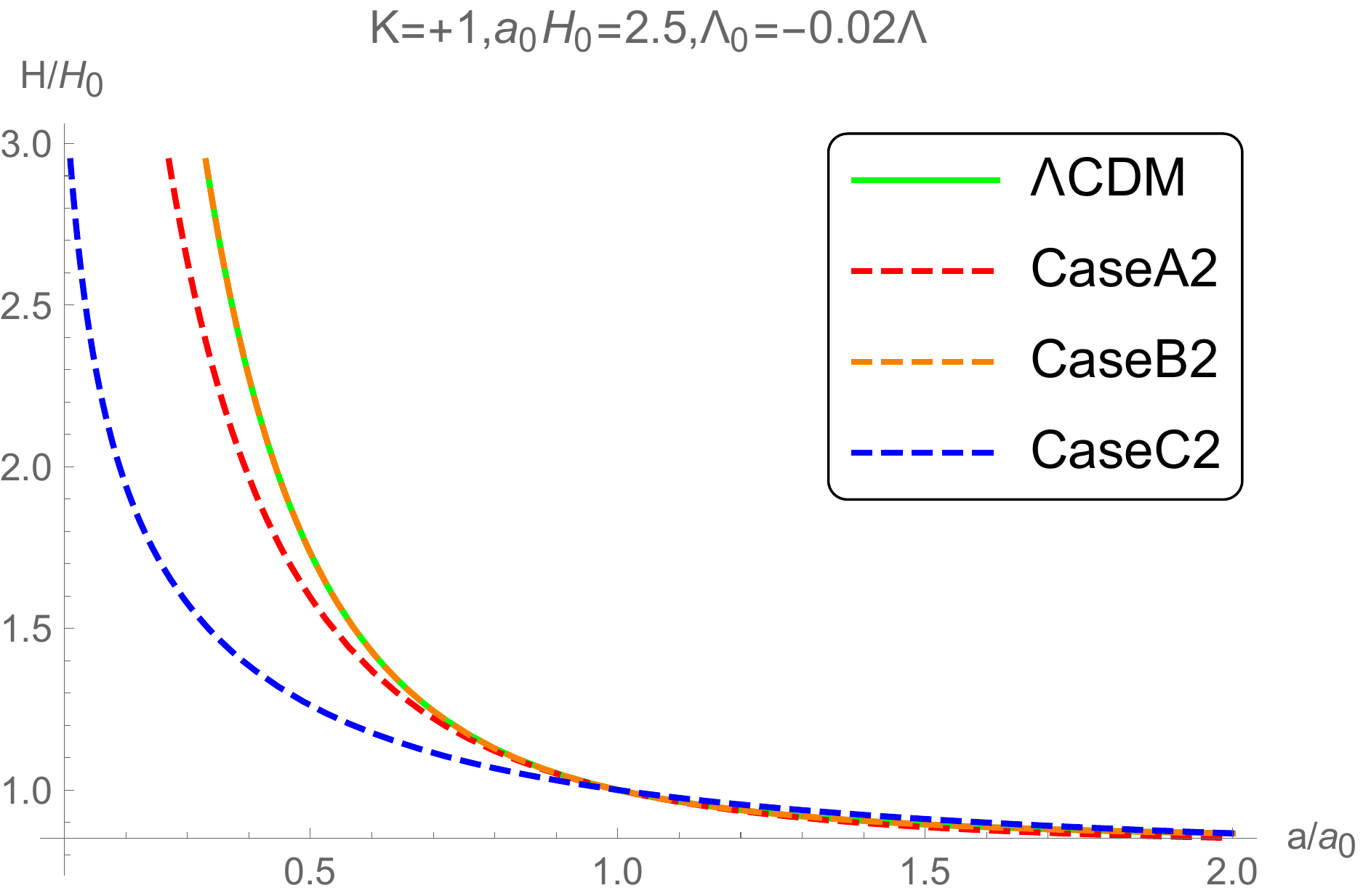}}
	\caption{The Hubble constant evolves with the scale factor when $K=+1,{{\Lambda }_{0}}=-0.02\Lambda $ .}\label{K+1x-0.02H0va}
\end{figure}
\begin{figure}[h]
	\centering
	\subfigure[]
	{\label{A2}
		\includegraphics[width=2.5in]{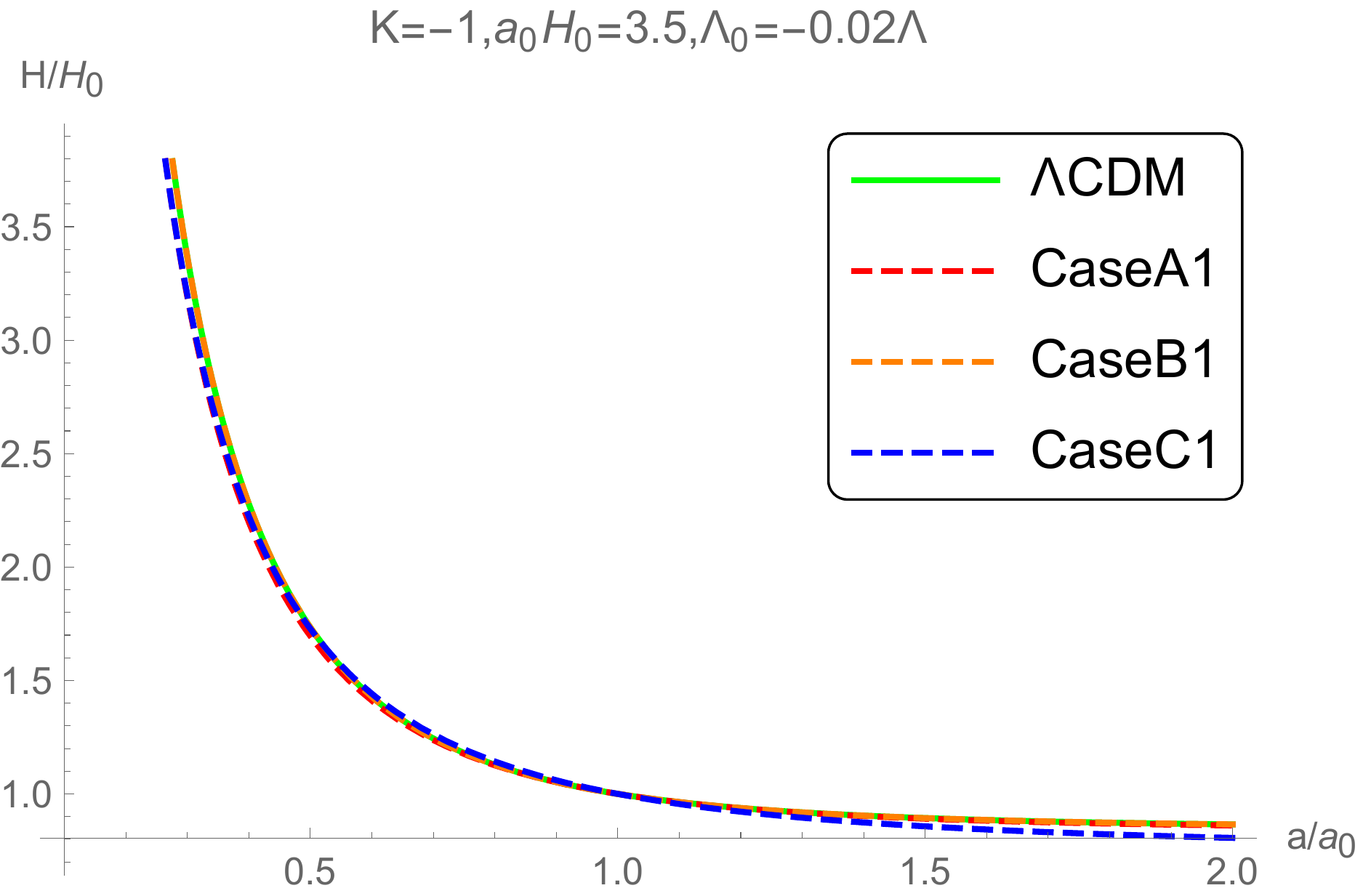}}
	\subfigure[]
	{\label{B2}
		\includegraphics[width=2.5in]{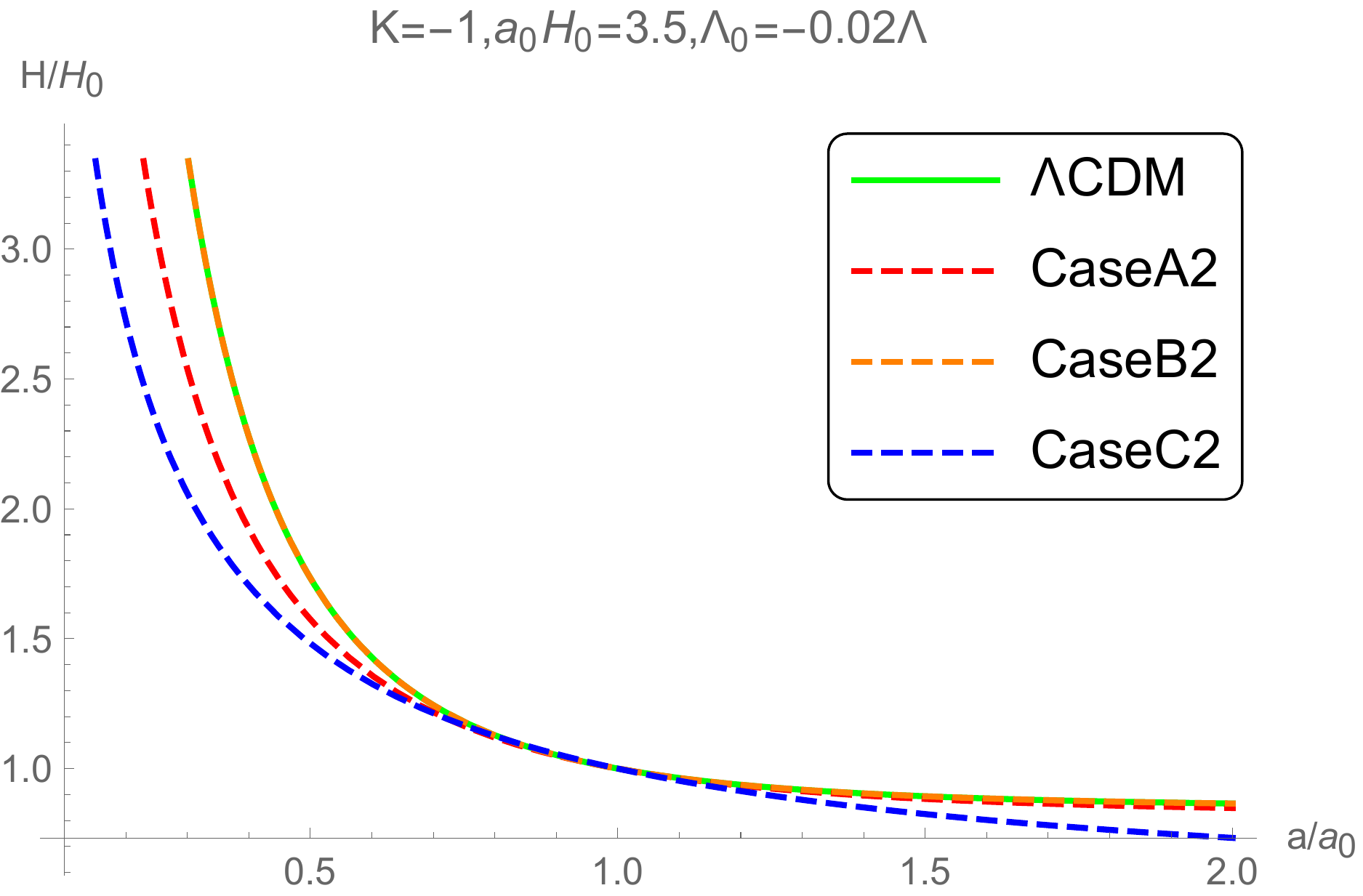}}
	\caption{The Hubble constant evolves with the scale factor when $K=-1,{{\Lambda }_{0}}=-0.02\Lambda ,{{a}_{0}}{{H}_{0}}=3.5$.}\label{K-1x-0.02H0a03.5H0va}
\end{figure}
\begin{figure}[h]
	\centering
	\subfigure[]
	{\label{A3}
		\includegraphics[width=2.5in]{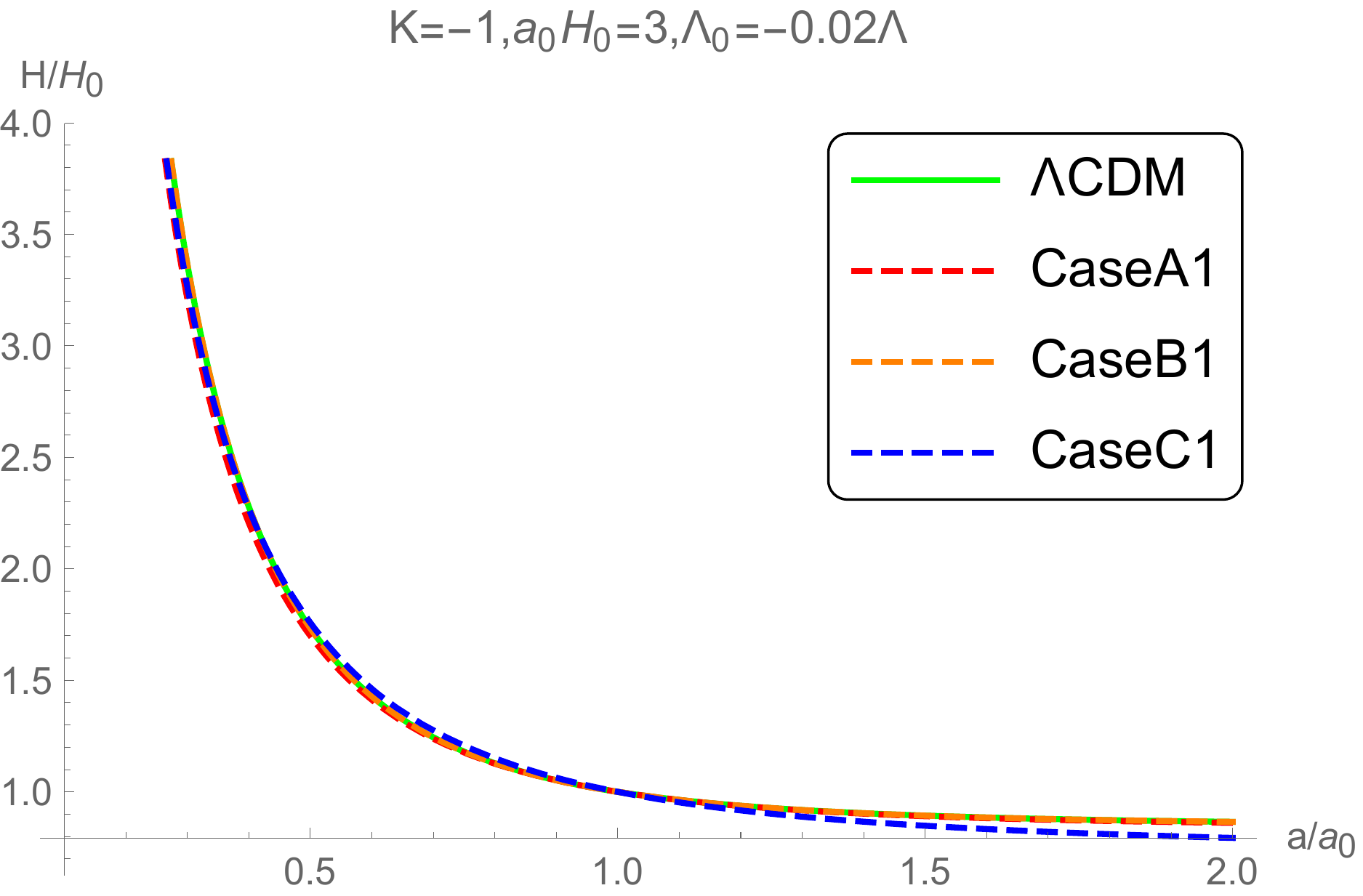}}
	\subfigure[]
	{\label{B3}
		\includegraphics[width=2.5in]{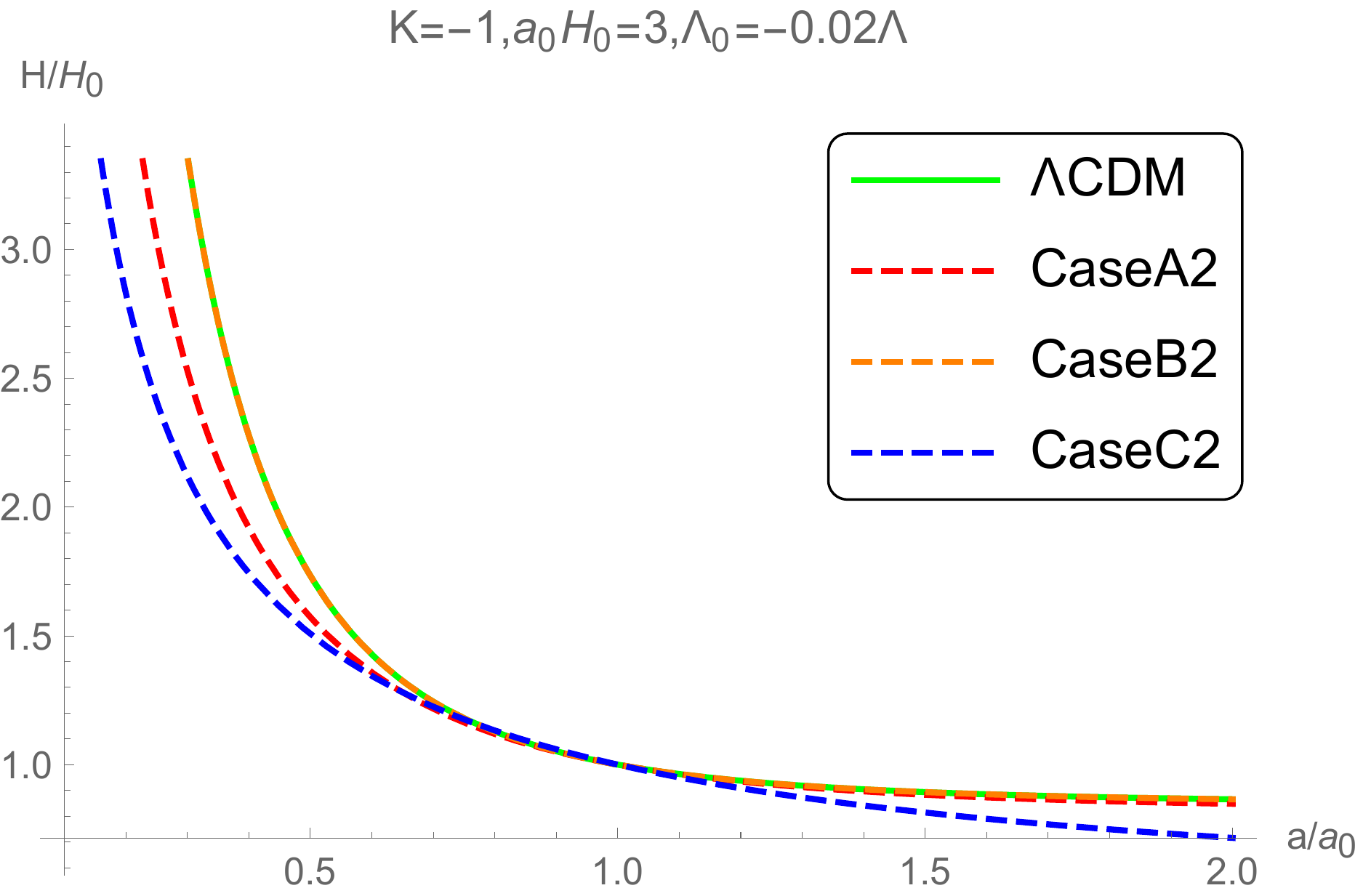}}
	\subfigure[]
	{\label{C3}
		\includegraphics[width=2.5in]{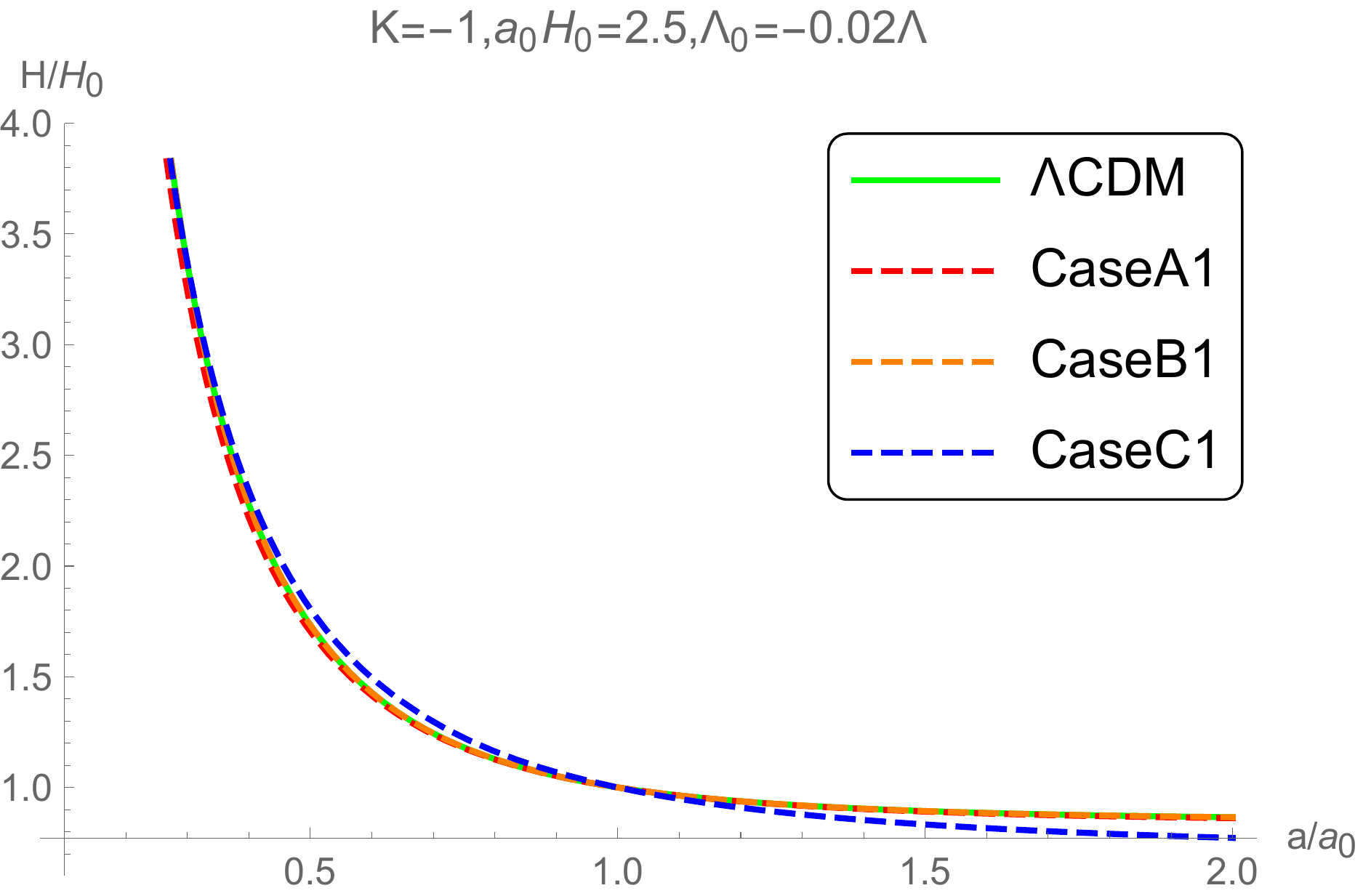}}
	\subfigure[]
	{\label{D3}
		\includegraphics[width=2.5in]{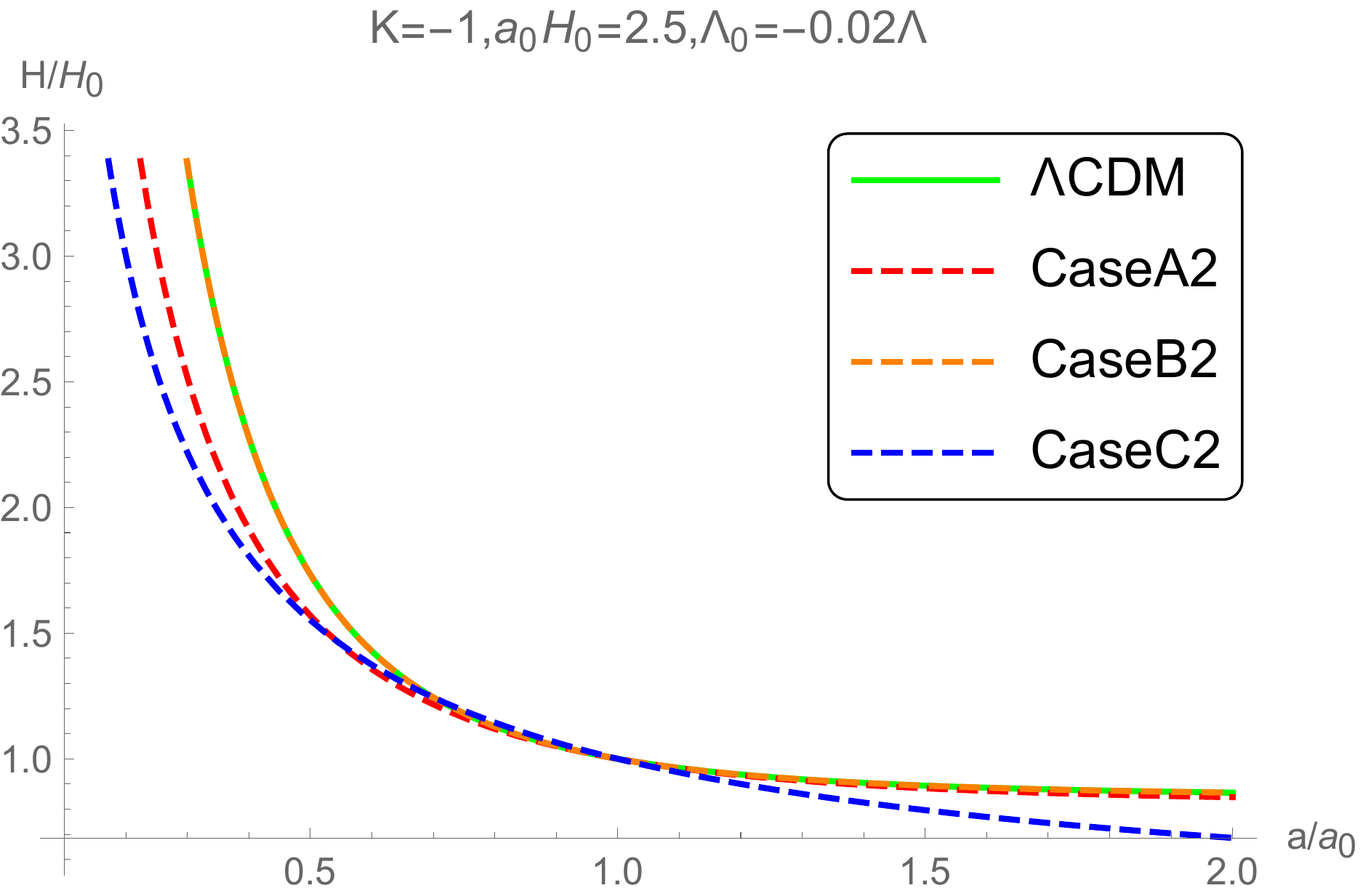}}
	\caption{The Hubble constant evolves with the scale factor when $K=-1$, ${{\Lambda }_{0}}=-0.02\Lambda$ , ${{a}_{0}}{{H}_{0}}=3$ and $2.5$}\label{K-1x-0.02H0a03H0va}
	\end{figure}

Since the accelerated expansion occurs in the late stage of the cosmic evolution, we mainly concentrate to consider the matter-dominated era. In this case of $w=0$, the evolution of the Hubble parameters and contortion can be obtained by the modified Friedmann equations \eqref{mofridmeq1}, \eqref{mofridmeq2} and one of the three kind of approximations, \eqref{appA}, \eqref{appB} and \eqref{appC}, with both the initial values of ${\mathcal K}\left( t \right)$ in \eqref{inivlu1} and \eqref{inivlu2}. The evolution of $H$ versus time in different cases are presented in Figures \ref{K+1x-0.02H0va}, \ref{K-1x-0.02H0a03.5H0va} and \ref{K-1x-0.02H0a03H0va}.
Figures \ref{K+1x-0.02H0va} show that the behavior of the evolution of Hubble parameter of Case C2 deviates relatively large from other cases and the $\Lambda$CDM model. However, deviation among different models become small after the scale factor of the universe, $a$, get greater than 1. The evolution of Case B and $\Lambda$CDM models tends to be identically the same globally.

\begin{figure}[]
	\centering
	\subfigure[]
	{\label{A4}
		\includegraphics[width=2.5in]{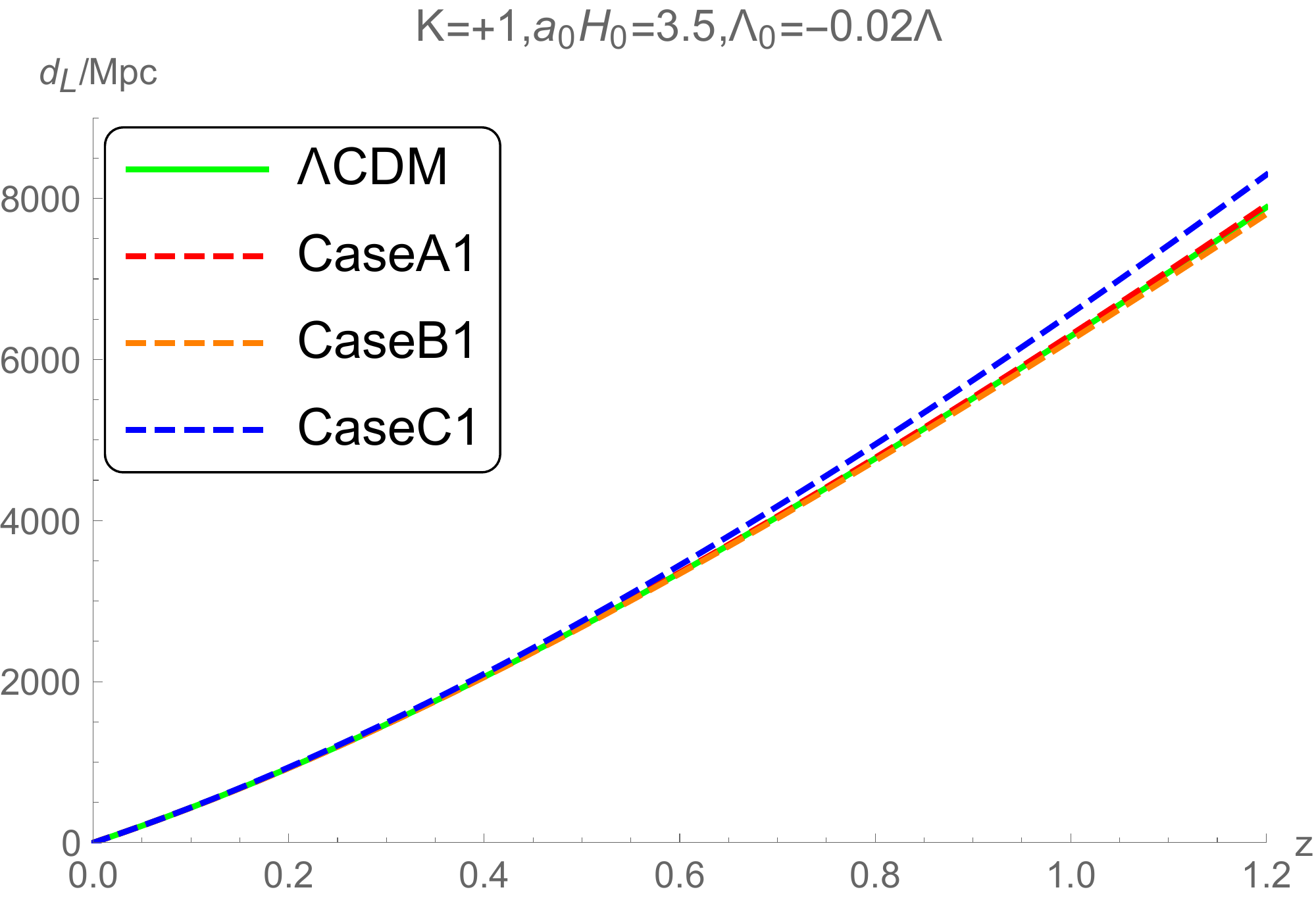}}
	\subfigure[]
	{\label{B4}
		\includegraphics[width=2.5in]{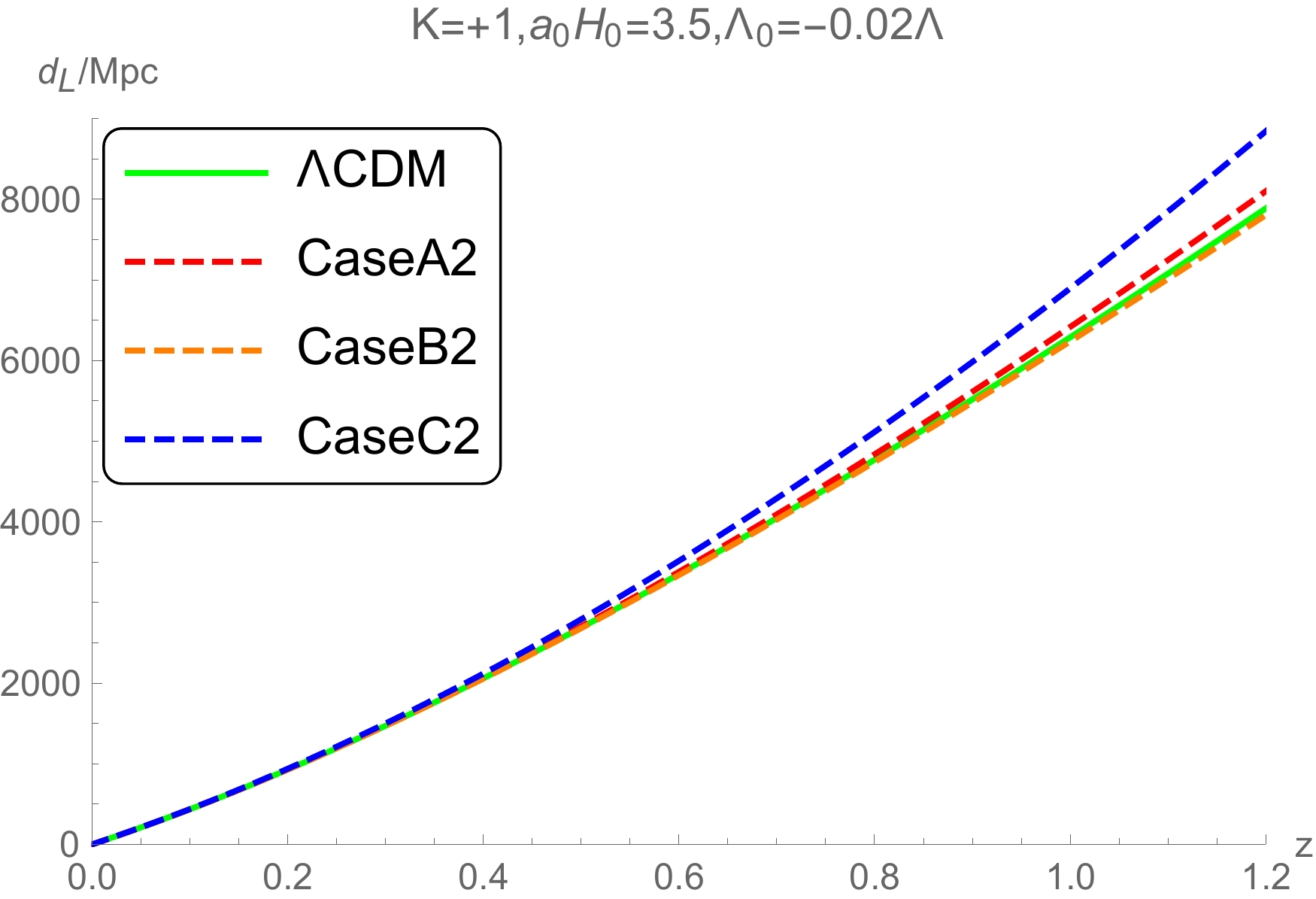}}
	\subfigure[]
	{\label{C4}
		\includegraphics[width=2.5in]{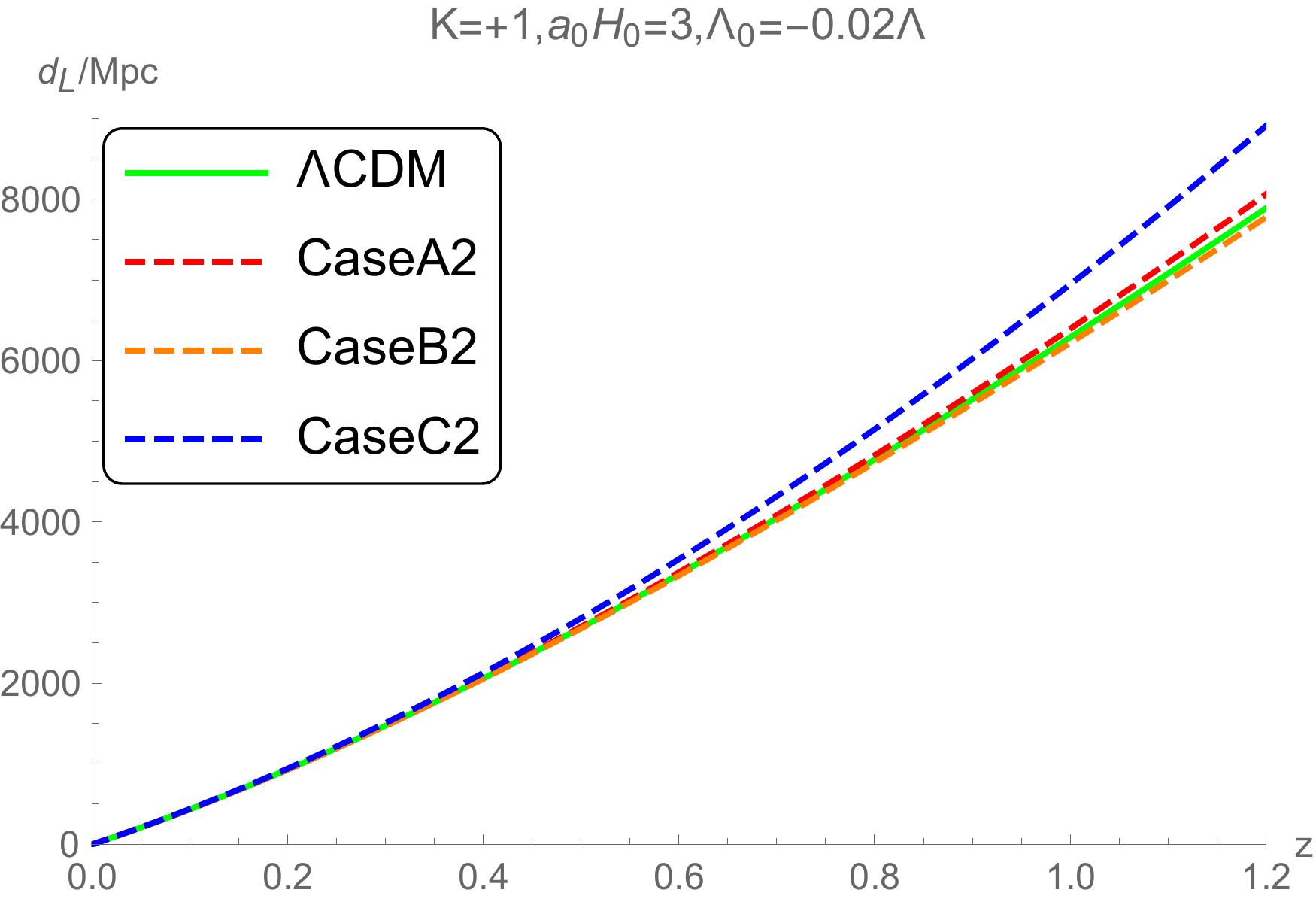}}
	\subfigure[]
	{\label{D4}
		\includegraphics[width=2.5in]{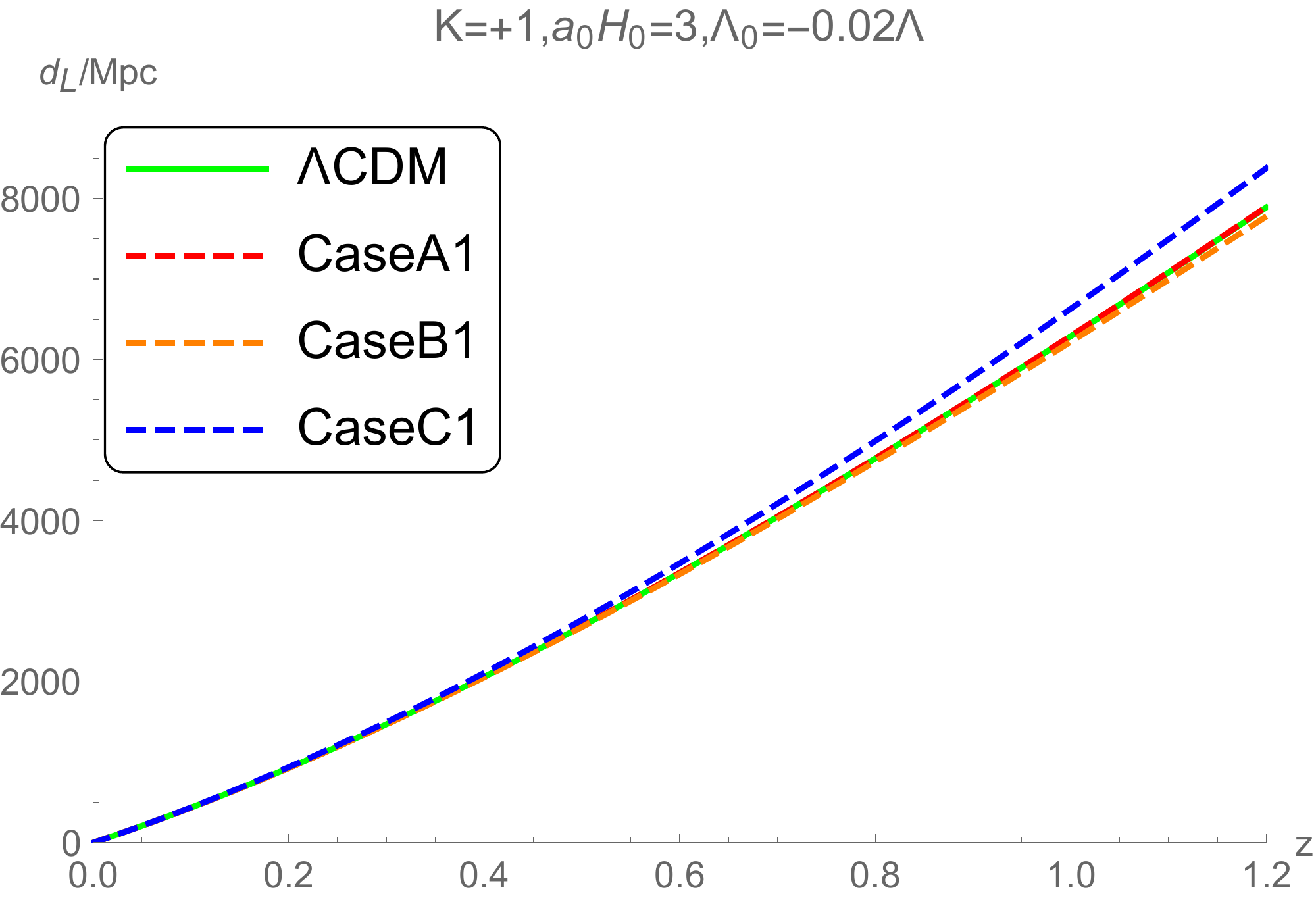}}
	\subfigure[]
	{\label{E4}
		\includegraphics[width=2.5in]{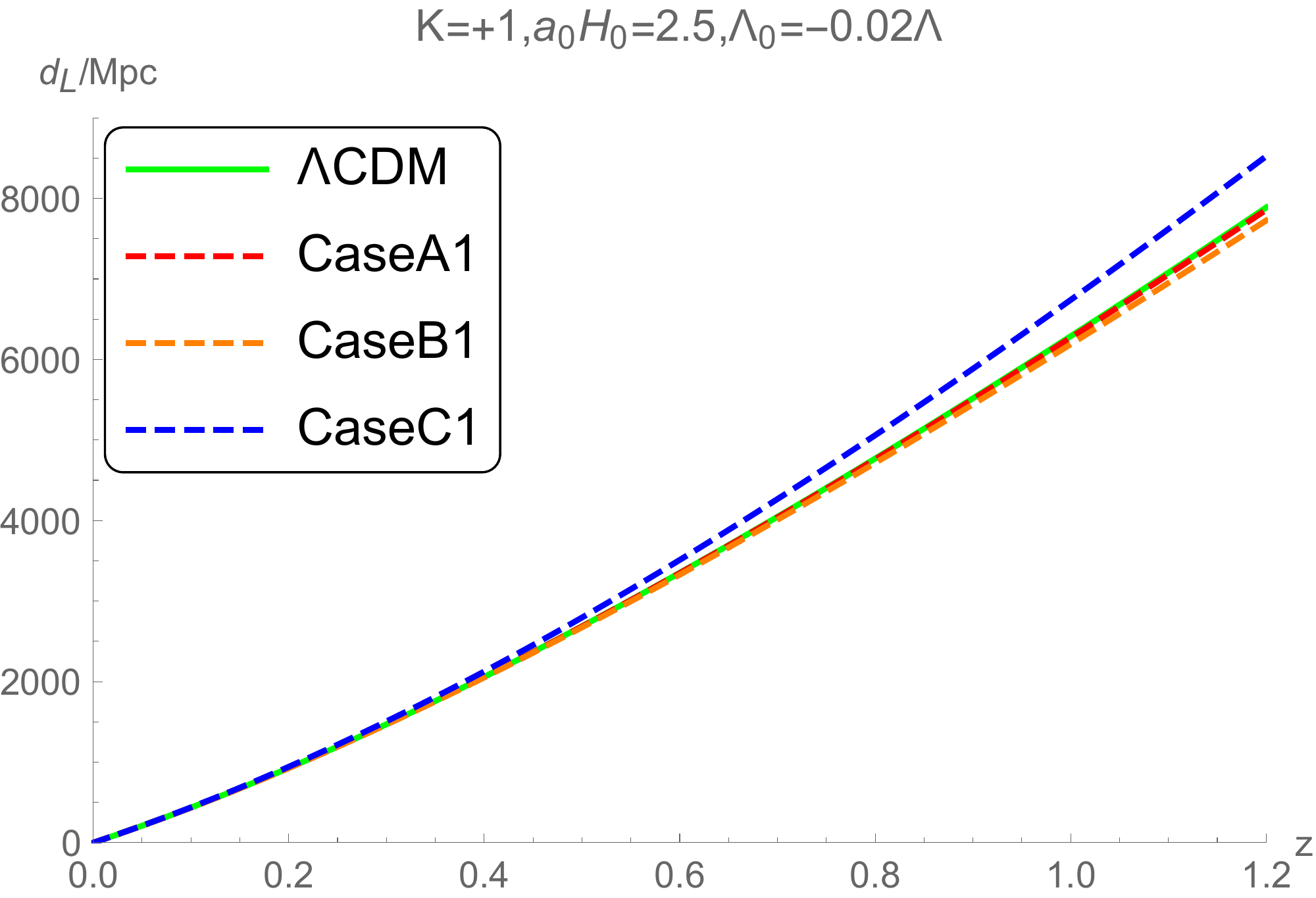}}
	\subfigure[]
	{\label{F4}
		\includegraphics[width=2.5in]{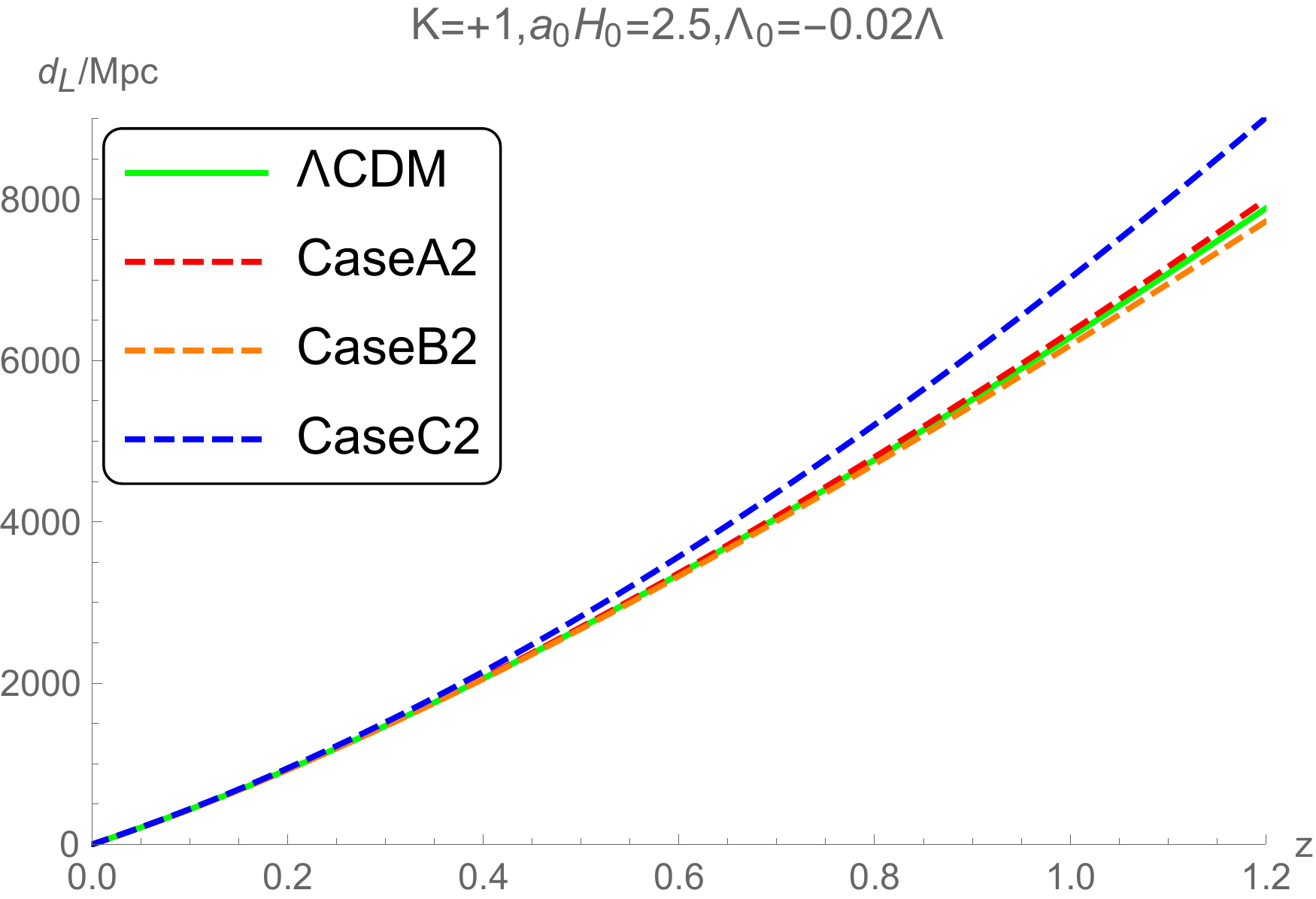}}
	\caption{the luminosity disyance evolves with the redshift when $K=+1, {{w}_{0}}=-1, {{\Lambda }_{0}}\text{=}-0.02\Lambda $.}\label{dLvszK+1w0-1x-0.02}
\end{figure}

Figures \ref{K-1x-0.02H0a03.5H0va} and \ref{K-1x-0.02H0a03H0va} show that the evolution of Case A2 and Case C2 in the case of $K=-1$ deviates from the behavior of the $\Lambda$CDM model, while the deviations among Case A1, Case B1, Case C1, Case B2 and $\Lambda$CDM models are small. The Figures \ref{K+1x-0.02H0va}, \ref{K-1x-0.02H0a03.5H0va} and \ref{K-1x-0.02H0a03H0va} show that the evolution in three kind of approximation respectively is very close to the evolution in the $\Lambda$CDM model when $K=-1$ and ${\mathcal K}\left( t \right)$ takes the first initial value.
Since the light ray travels along the geodesic line even in the case of non-zero torsion, the redshift formula for the large scale Lorentz violation model is the same as for the $\Lambda$CDM model,
\begin{equation}
1\text{+}z=\frac{{{a}_{0}}}{a}.
\end{equation}
From definition of luminosity distance, one can get 
\begin{equation}\label{dLvsz}
{{d}_{L}}\left( z \right)=\left( 1+z \right)\frac{{{d}_{H}}}{\sqrt{{{\Omega }_{K}}}}\sinh \left[ \sqrt{{{\Omega }_{K}}}\int_{0}^{z}{\frac{{{H}_{0}}}{H\left( {{z}'} \right)}d{z}'} \right],
\end{equation}
where ${{d}_{L}}$ is the luminosity distance, 
\begin{equation}\label{dH}
{{d}_{H}}\text{=}\frac{c}{{{H}_{0}}}	
\end{equation}
the Hubble distance and 
\begin{equation}\label{cuvtden}
{{\Omega }_{K}}=-\frac{K}{{{H}_{0}}^{2}{{a}_{0}}^{2}}
\end{equation}
is the curvature density at present. One can  derive the set of equations for ${\mathcal K(z)}$ and $d_L(z)$ with formula \eqref{dLvsz},

\begin{eqnarray}
&&\frac{{{d}_{L}}^{\prime \prime}\left( z \right){{\left( 1+z \right)}^{6}}{{\left( 1-\frac{K{{d}_{L}}^{2}\left( z \right)}{2{{a}_{0}}^{2}{{\left( 1+z \right)}^{2}}} \right)}^{2}}}{{{\left[ {{d}_{L}}^{\prime }\left( z \right)\left( 1+z \right)-{{d}_{L}}\left( z \right) \right]}^{3}}}-\frac{1}{2}\frac{{{\left( 1+z \right)}^{4}}{{\left( 1-\frac{K{{d}_{L}}^{2}\left( z \right)}{2{{a}_{0}}^{2}{{\left( 1+z \right)}^{2}}} \right)}^{2}}}{{{\left[ {{d}_{L}}^{\prime }\left( z \right)\left( 1+z \right)-{{d}_{L}}\left( z \right) \right]}^{2}}}\text{+}\frac{K}{{{a}_{0}}^{2}}\frac{{{d}_{L}}\left( z \right){{\left( 1+z \right)}^{2}}\left( 1-\frac{K{{d}_{L}}^{2}\left( z \right)}{2{{a}_{0}}^{2}{{\left( 1+z \right)}^{2}}} \right)}{{{d}_{L}}^{\prime }\left( z \right)\left( 1+z \right)-{{d}_{L}}\left( z \right)} \nonumber\\ 
&&+2\frac{{{\left( 1+z \right)}^{2}}\left( 1-\frac{K{{d}_{L}}^{2}\left( z \right)}{2{{a}_{0}}^{2}{{\left( 1+z \right)}^{2}}} \right)}{{{d}_{L}}^{\prime }\left( z \right)\left( 1+z \right)-{{d}_{L}}\left( z \right)}{\mathcal K}\left( z \right)-\frac{{{\left( 1+z \right)}^{3}}\left( 1-\frac{K{{d}_{L}}^{2}\left( z \right)}{2{{a}_{0}}^{2}{{\left( 1+z \right)}^{2}}} \right)}{{{d}_{L}}^{\prime }\left( z \right)\left( 1+z \right)-{{d}_{L}}\left( z \right)}{\mathcal K}'\left( z \right)\text{+}\frac{1}{2}{{\mathcal K}^{2}}\left( z \right)\text{+}\frac{K}{2{{a}^{2}}}-\frac{1}{2}{{\Lambda }_{0}}\text{=}0,                                   
\end{eqnarray}
together with the Case A, 
\begin{equation}
\frac{{{\left( 1+z \right)}^{2}}\left( 1-\frac{K{{d}_{L}}^{2}\left( z \right)}{2{{a}_{0}}^{2}{{\left( 1+z \right)}^{2}}} \right)}{{{d}_{L}}^{\prime }\left( z \right)\left( 1+z \right)-{{d}_{L}}\left( z \right)}\left[ {\mathcal K}\left( z \right)-\left( 1+z \right){\mathcal K}'\left( z \right) \right]=\frac{1}{3}\left( {{\Lambda }_{0}}-\Lambda  \right),
\end{equation}
the Case B,
\begin{equation}
\frac{{{\left( 1+z \right)}^{2}}\left( 1-\frac{K{{d}_{L}}^{2}\left( z \right)}{2{{a}_{0}}^{2}{{\left( 1+z \right)}^{2}}} \right)}{{{d}_{L}}^{\prime }\left( z \right)\left( 1+z \right)-{{d}_{L}}\left( z \right)}\left[ 2{\mathcal K}\left( z \right)-\left( 1+z \right){\mathcal K}'\left( z \right) \right]\text{+}\frac{1}{2}{{\mathcal K}^{2}}\left( z \right)\text{+}\frac{K}{2{{a}^{2}}}\text{+}\frac{1}{2}\left( \Lambda -{{\Lambda }_{0}} \right)\text{=0}
\end{equation}
and the Case C,
\begin{equation}
\frac{{{\left( 1+z \right)}^{2}}\left( 1-\frac{K{{d}_{L}}^{2}\left( z \right)}{2{{a}_{0}}^{2}{{\left( 1+z \right)}^{2}}} \right)}{{{d}_{L}}^{\prime }\left( z \right)\left( 1+z \right)-{{d}_{L}}\left( z \right)}\left[ {\mathcal K}\left( z \right)\text{+}\left( 1+z \right){\mathcal K}'\left( z \right) \right]=-{{\mathcal K}^{2}}\left( z \right),
\end{equation}
respectively. The distance modulus defined by
\begin{equation}\label{dismod}
\mu=25+5\text{log}\left(\frac{d_{\text L}}{M\text{pc}}\right)
\end{equation}
is often used in astronomical observation.
Figures \ref{dLvszK+1w0-1x-0.02} show the evolution of the luminosity distance versus redshift.

\section{The Numerical Predictions}
\begin{figure}[h]
	\centering
	\subfigure[]
	{\label{A5}
		\includegraphics[width=2.5in]{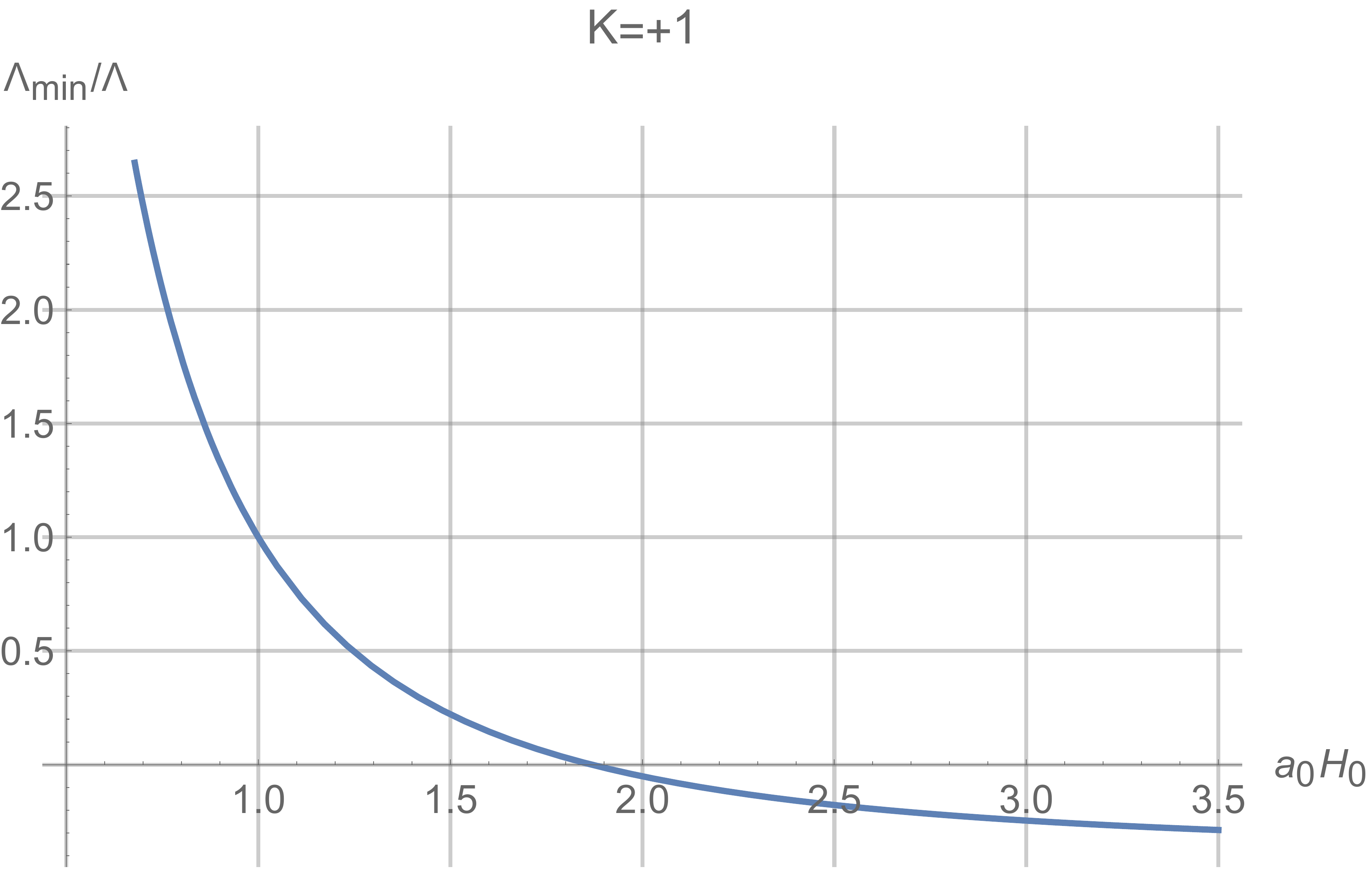}}
	\subfigure[]
	{\label{B5}
		\includegraphics[width=2.5in]{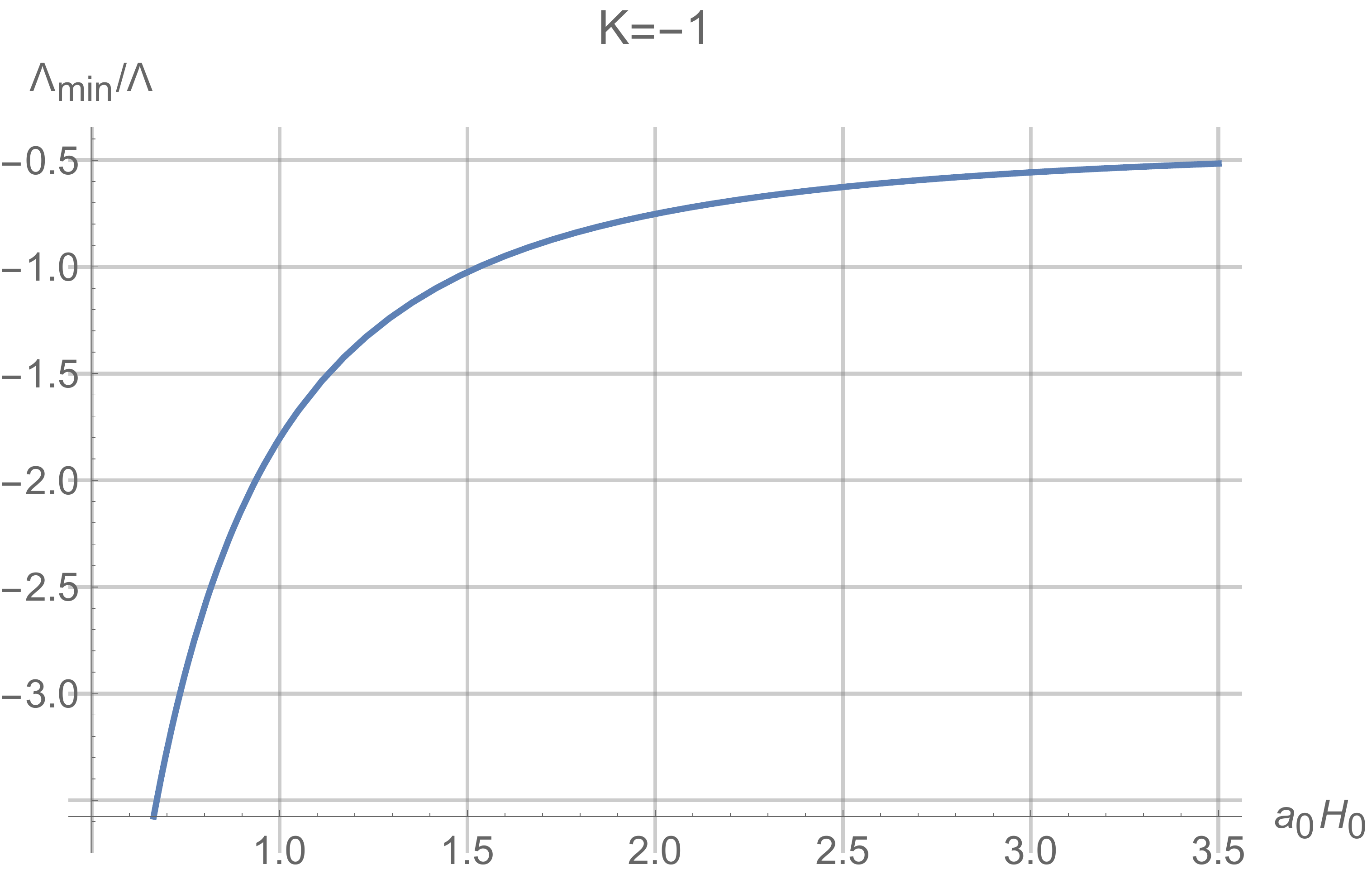}}
	\caption{ The evolution of the ${{\Lambda }_{\min }}$ versus ${{a}_{0}}{{H}_{0}}$}\label{Lminvsa0H0}
\end{figure}

\begin{figure}[h]
	\centering
	\subfigure[]
	{\label{A6}
		\includegraphics[width=2.5in]{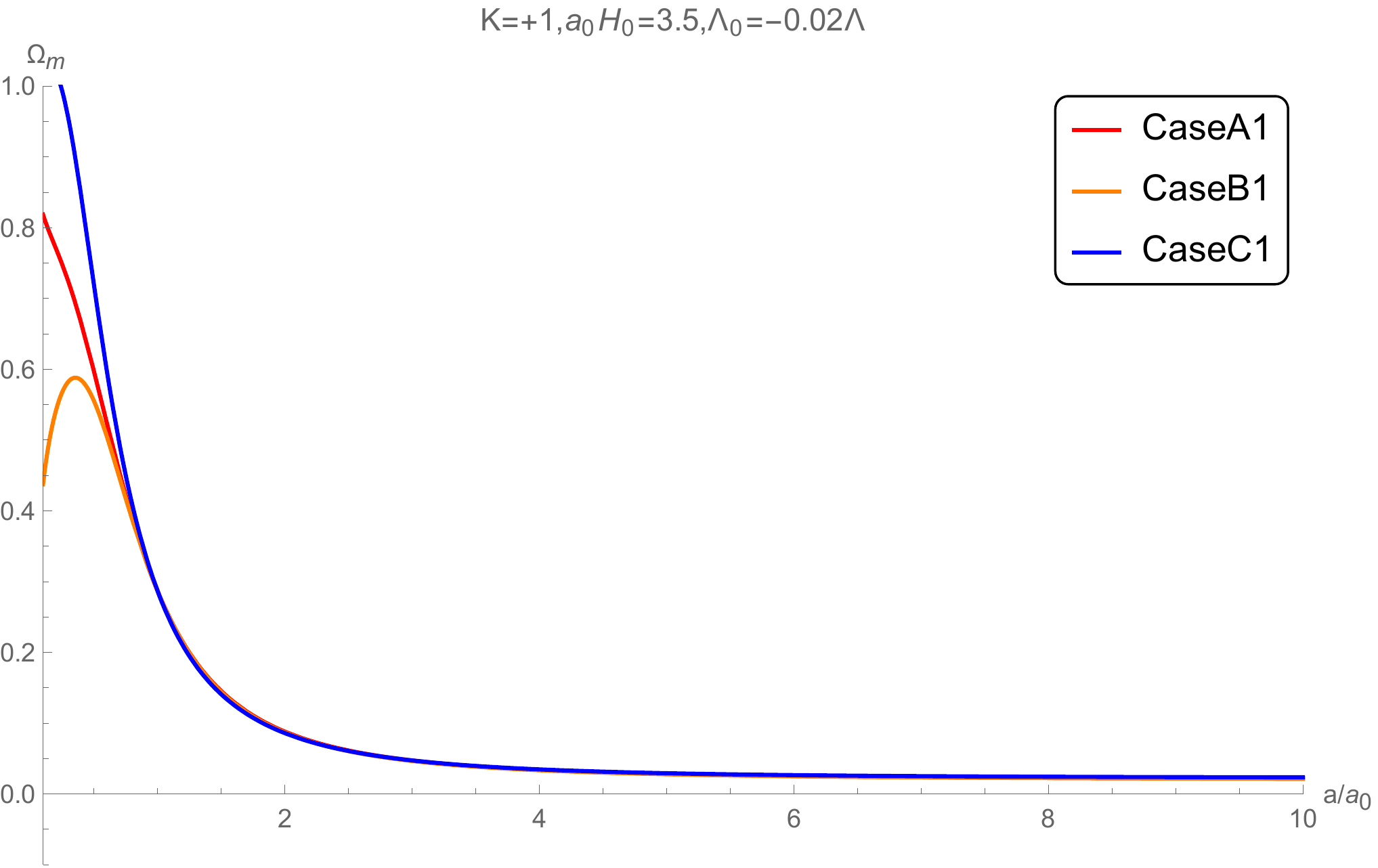}}
	\subfigure[]
	{\label{B6}
		\includegraphics[width=2.5in]{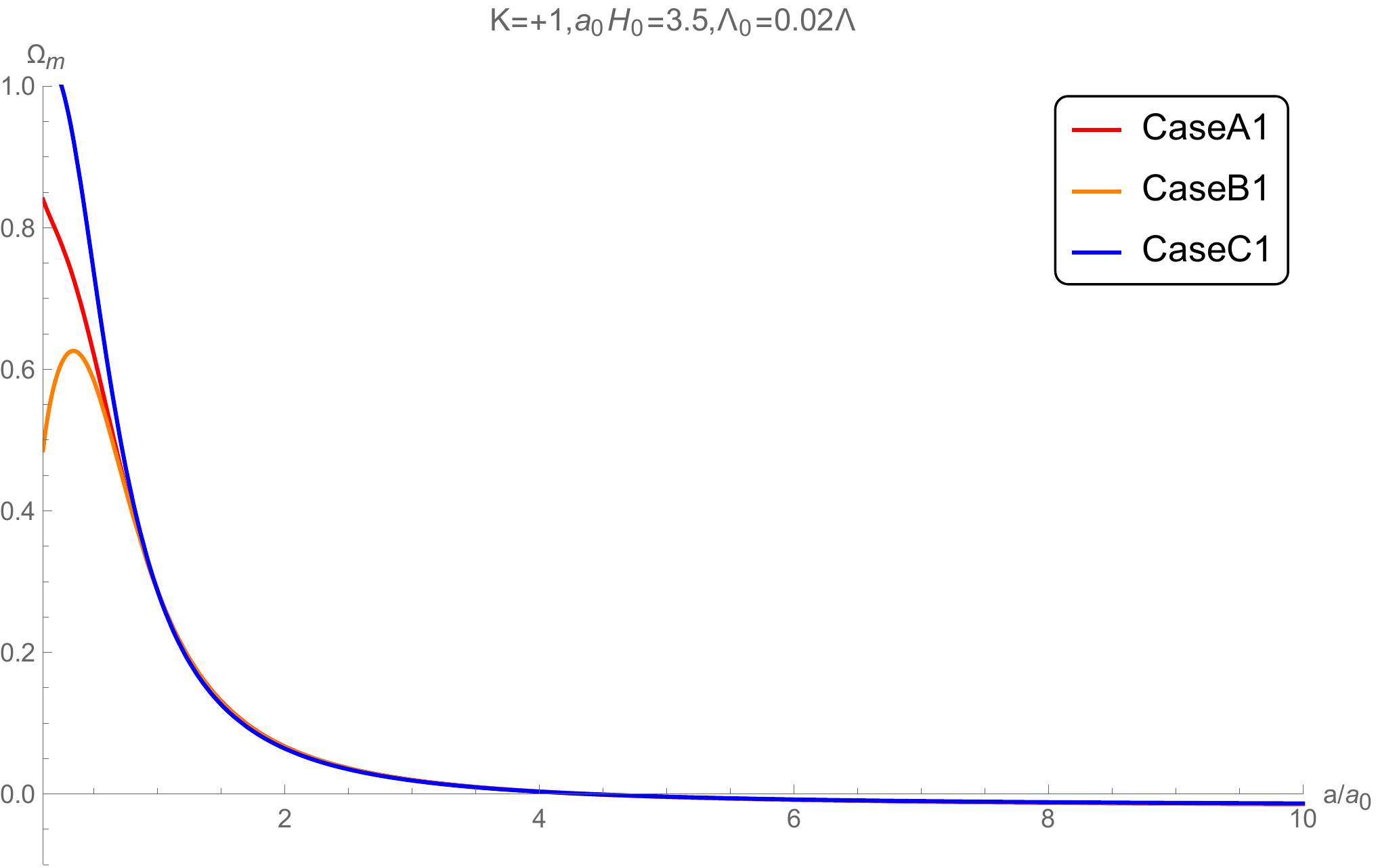}}
	\caption{ The evolution of the matter density $\Omega_{M}$ versus the scale factor}\label{omgmvsa}
\end{figure}

The modified Friedmann equation \eqref{mdffridmeq1} of the large scale Lorentz violation model can be written as
\begin{equation}
\frac{\rho }{3{{H}^{2}}}\text{+}\frac{{{\Lambda }_{0}}}{3{{H}^{2}}}-\frac{2\mathcal K}{H}-{{\left( \frac{\mathcal K}{H} \right)}^{2}}-\frac{K}{{{a}^{2}}{{H}^{2}}}\text{=}1
\end{equation}
and
\begin{equation}
	{{\Omega }_{M}}+{{\Omega }_{eff}}\text{=}1,
\end{equation}
where 
\begin{equation}\label{mattden}
{{\Omega }_{M}}\text{=}\frac{\rho }{3{{H}^{2}}}	
\end{equation}
is the matter density, ${{\Omega }_{eff}}=\frac{{{\Lambda }_{0}}}{3{{H}^{2}}}-\frac{2\mathcal K}{H}-{{\left( \frac{\mathcal K}{H} \right)}^{2}}-\frac{K}{{{a}^{2}}{{H}^{2}}}$ is the contribution of the bare cosmological constant, contortion and spatial curvature.  The minimum value for $\Lambda_0$ in the case of $K = +1, -1$ is shown in Figure \ref{Lminvsa0H0}.
Utilizing the modified Friedmann equation, for every kind of approximation models, one can obtain the evolution of the matter energy density with time. The evolution of ${{\Omega }_{M}}$ in the three kind of approximation models at different values of ${{\Lambda }_{0}}$ is shown in Figure \ref{omgmvsa}.
The ${{\Omega }_{M}}$ will decrease to be negative along with the increase of the ${{\Lambda }_{0}}$ beyond a critical value, $\Lambda _{max}$. It leads to a constrain condition on the $\Lambda_{0}$ value,
\begin{equation}\label{cntronLmax}
1-\frac{{{\Lambda }_{0}}}{3{{H}^{2}}}+\frac{2\mathcal K}{H}+{{\left( \frac{\mathcal K}{H} \right)}^{2}}+\frac{K}{{{a}^{2}}{{H}^{2}}}\ge 0,
\end{equation}
to ensure that ${{\Omega }_{M}}\ge 0$. Constrain \eqref{cntronLmax} sets a maximum $\Lambda_{\max}$, 
\begin{equation}\label{Lmax}
{{\Lambda }_{\max }}=3{{H}^{2}}+6H\mathcal{K}+3{{\mathcal{K}}^{2}}+\frac{3K}{{{a}^{2}}},
\end{equation}
for the $\Lambda_{0}$ value. When ${{\Lambda }_{0}} > {\Lambda }_{\max }$, the value of ${{\Omega }_{M}}$ will be less than 0. Table \ref{LmaxK+1} and \ref{LmaxK-1} show the ${\Lambda }_{\max }$ values for all the cases in consideration.

\begin{figure}[h]
	\centering
	{\label{A7}
		\includegraphics[width=2.5in]{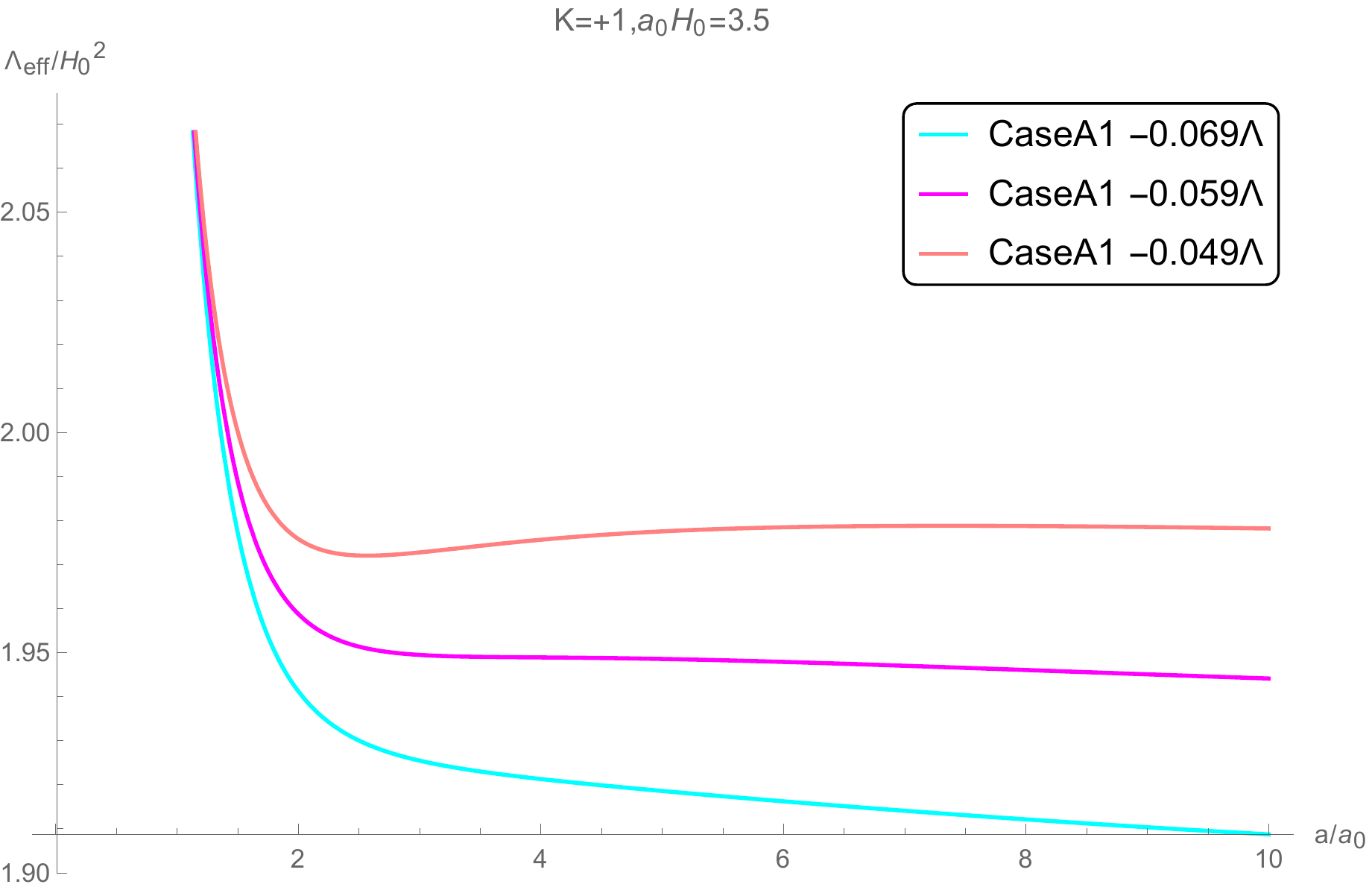}}
	\caption{The evolution pattern of ${{\Lambda }_{\text{eff}}}$ versus scale factor shifts from a monotonically decreasing quintessence type to a local minimal type with the dependence on $\Lambda_0$}\label{Lefftrs}
\end{figure}

\begin{table}[tp]
	
	\centering
	
	\fontsize{6.5}{8}\selectfont
	
	\caption{The value of $\Lambda_{max}$ when $K=+1$}
	
	\label{LmaxK+1}
	
	\begin{tabular}{|p{1.5cm}|p{1.8cm}|p{1.8cm}|p{1.8cm}|p{1.8cm}|}
		
		\hline
		
		\multicolumn{5}{|c|}{$K=+1$}\cr\cline{1-5}
		
		$\Lambda_{max}$&$a_0 H_0 = 2$\ $(\Omega_k=-0.25)$&$a_0 H_0 = 2.5$\ $(\Omega_k=-0.16)$&$a_0 H_0 = 3$\ $(\Omega_k=-0.11)$&$a_0 H_0 = 3.5$\ $(\Omega_k=-0.08)$\cr
		
		\hline
		
		CaseA1&0.0008&0.0007&0.0007&0.0006\cr\hline
		
		CaseA2&0.0012&0.0014&0.0015&0.0015\cr\hline
				
		CaseB1&0.0008&0.0007&0.0007&0.0006\cr\hline
				
		CaseB2&0.0012&0.0014&0.0015&0.0016\cr\hline
				
		CaseC1\ ($w_0=-1$)&0.0008&0.0007&0.0007&0.0006\cr\hline
				
		CaseC2\ ($w_0=-1$)&0.0012&0.0015&0.0015&0.0016\cr\hline
						
		CaseC1\ ($w_0=-8/9$)&0.0008&0.0007&0.0007&0.0006\cr\hline
						
		CaseC2\ ($w_0=-8/9$)&0.0013&0.0016&0.0017&0.0018\cr\hline
						
		CaseC1\ ($w_0=-7/9$)&0.0008&0.0007&0.0007&0.0006\cr\hline
						
		CaseC2\ ($w_0=-7/9$)&0.0013&0.0016&0.0018&0.0019\cr\hline
						
		CaseC1\ ($w_0=-2/3$)&0.0008&0.0007&0.0006&0.0006\cr\hline
						
		CaseC2\ ($w_0=-2/3$)&0.0014&0.0018&0.002&0.0022\cr\hline
						
		CaseC1\ ($w_0=-1/3$)&0.0008&0.0006&0.0006&0.0006\cr\hline
						
		CaseC2\ ($w_0=-1/3$)&0.0025&0.0046&0.0056&0.0062\cr\hline
		
	\end{tabular}
	
\end{table}
\begin{table}[tp]
	
	\centering
	
	\fontsize{6.5}{8}\selectfont
	
	\caption{The value of $\Lambda_{max}$ when $K=-1$}
	
	\label{LmaxK-1}
	
	\begin{tabular}{|p{1.5cm}|p{1.8cm}|p{1.8cm}|p{1.8cm}|p{1.8cm}|p{1.8cm}|}
		
		\hline
		
		\multicolumn{6}{|c|}{$K=-1$}\cr\cline{1-6}
		
		$\Lambda_{max}$&$a_0 H_0 = 1.5$\ $(\Omega_k=0.44)$&$a_0 H_0 = 2$\ $(\Omega_k=0.25)$&$a_0 H_0 = 2.5$\ $(\Omega_k=0.16)$&$a_0 H_0 = 3$\ $(\Omega_k=0.11)$&$a_0 H_0 = 3.5$\ $(\Omega_k=0.08)$\cr
		
		\hline
		
		CaseA1&0.0004&0.0005&0.0005&0.0005&0.0005\cr\hline
		
		CaseA2&0.0023&0.002&0.0019&0.0018&0.0018\cr\hline
		
		CaseB1&0.0004&0.0004&0.0005&0.0005&0.0005\cr\hline
		
		CaseB2&0.0023&0.002&0.0019&0.0019&0.0018\cr\hline
		
		CaseC1\ ($w_0=-1$)&0.0003&0.0004&0.0005&0.0005&0.0005\cr\hline
		
		CaseC2\ ($w_0=-1$)&0.015&0.0027&0.0023&0.0021&0.002\cr\hline
		
		CaseC1\ ($w_0=-8/9$)&0.0003&0.0004&0.0005&0.0005&0.0005\cr\hline
		
		CaseC2\ ($w_0=-8/9$)&0.0033&0.0033&0.0026&0.0024&0.0023\cr\hline
		
		CaseC1\ ($w_0=-7/9$)&0.0003&0.0004&0.0005&0.0005&0.0005\cr\hline
		
		CaseC2\ ($w_0=-7/9$)&0.0032&0.0032&0.0032&0.0028&0.0026\cr\hline
		
		CaseC1\ ($w_0=-2/3$)&0.0003&0.0004&0.0004&0.0004&0.0005\cr\hline
		
		CaseC2\ ($w_0=-2/3$)&0.0185&0.0091&0.0054&0.0037&0.0031\cr\hline
		
		CaseC1\ ($w_0=-1/3$)&0.001&0.0005&0.0005&0.0005&0.0005\cr\hline
		
		CaseC2\ ($w_0=-1/3$)&0.0147&0.0119&0.0105&0.0097&0.0092\cr\hline
		
	\end{tabular}
	
\end{table}
It can be observed that ${{\Omega }_{M}}$ will decrease to be negative whenever the value of ${{\Lambda }_{0}}$ is positive. So the reasonable values of ${{\Lambda }_{0}}$ should be non-positive, i.e. ${{\Lambda }_{max}}\approx 0$.
In \cite{Han-Yu:2019tmf, Zhai:2019std}, the authors point out that there is a critical value for ${{\Lambda }_{0}}$, named as ${{\Lambda }_{0-crit}}$,  which stands for the separation of two phases in the evolution pattern of ${{\Lambda }_{eff}}$ versus $t$, i.e. ${{\Lambda }_{eff}}$ decreases monotonically along with the increase of $t$ when ${{\Lambda }_{0}}\le {{\Lambda }_{0-crit}}$ while its evolution has a local minimum when ${{\Lambda }_{0}}>{{\Lambda }_{0-crit}}$, and the ${{\Lambda }_{0-crit}}$ is approximately zero. In other words, ${{\Lambda }_{eff}}$ is monotonically decreasing when ${{\Lambda }_{0}}$ is from the string landscape, while ${{\Lambda }_{eff}}$ has a local minimum when ${{\Lambda }_{0}}$ is from the swampland.
Table \ref{LcrtK+1} and \ref{LcrtK-1} summarize the values of ${{\Lambda }_{0-crit}}$ for all the cases of approximation in consideration.
In \cite{Li:2020tqx}, it is shown that $V\left( \phi  \right)$ is a monotonically decreasing quintessence potential if ${{\Lambda }_{0}}$ is negative or from the string landscape while $V\left( \phi  \right)$ is a meta-stable de Sitter one if ${{\Lambda }_{0}}$ is positive or from the swampland when ${{\Lambda }_{eff}}$ is simulated by a scalar quintessence field potential $V\left(\phi\right)$. The dependence of ${{\Lambda }_{0\text{-}crit}}$ on different values of $a_0H_0$ in both the cases of $K=+1$ and $K=-1$ is investigated in this paper and the numerical results are shown in Table \ref{LcrtK+1} and \ref{LcrtK-1}. A specific example solution of the ${{\Lambda }_{\text{eff}}}$ evolution pattern transition with the dependence on $\Lambda_{0}$ is shown in Figure \ref{Lefftrs}. 
Figure \ref{Lminvsa0H0} shows that ${{\Lambda }_{\min }}$ is positive if ${{a}_{0}}{{H}_{0}}$ takes a value less than 2 in the case of $K=+1$, i.e. ${{\Lambda }_{\min }}>{{\Lambda }_{\max }}$ so that $\Lambda_{0\text{-}crit}$ does not make any sense in the case of $K=+1$ and ${{a}_{0}}{{H}_{0}}\ge 2$.

\begin{figure}[h]
	\centering
	\subfigure[]
	{\label{A8}
		\includegraphics[width=2.5in]{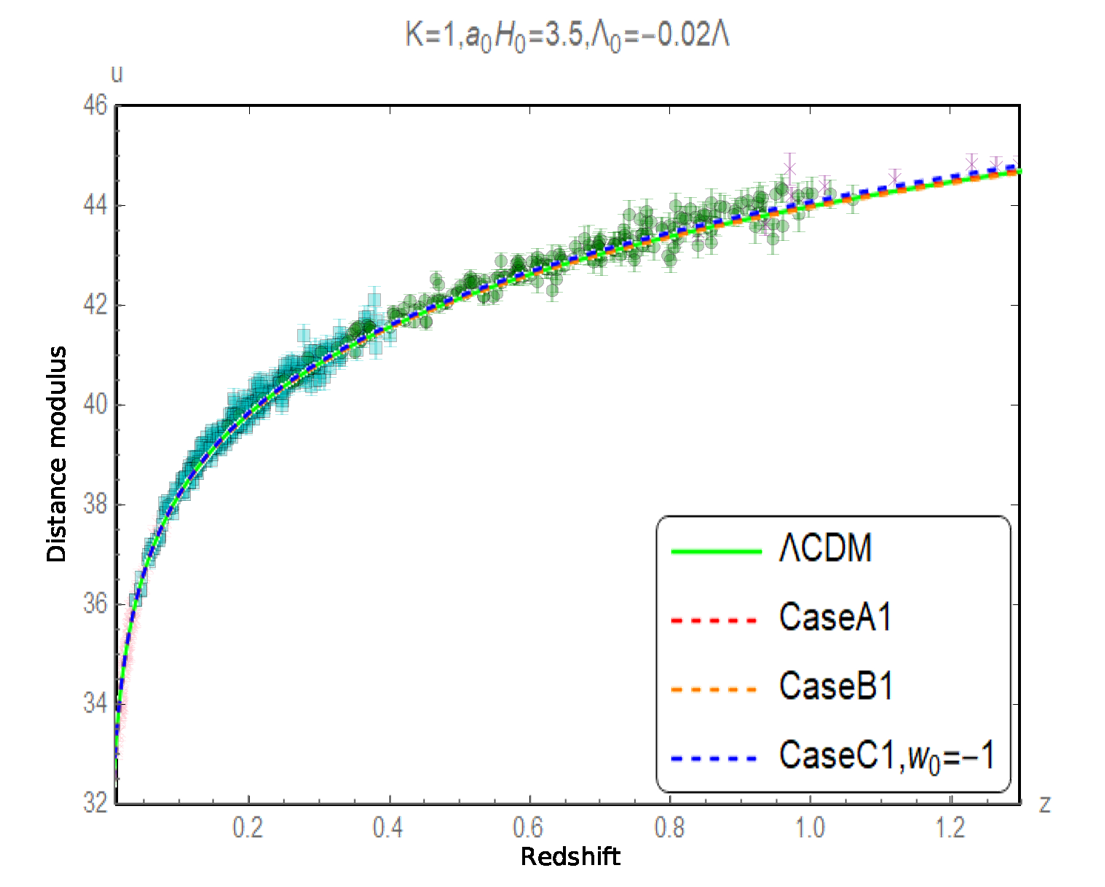}}
	\subfigure[]
	{\label{B8}
		\includegraphics[width=2.5in]{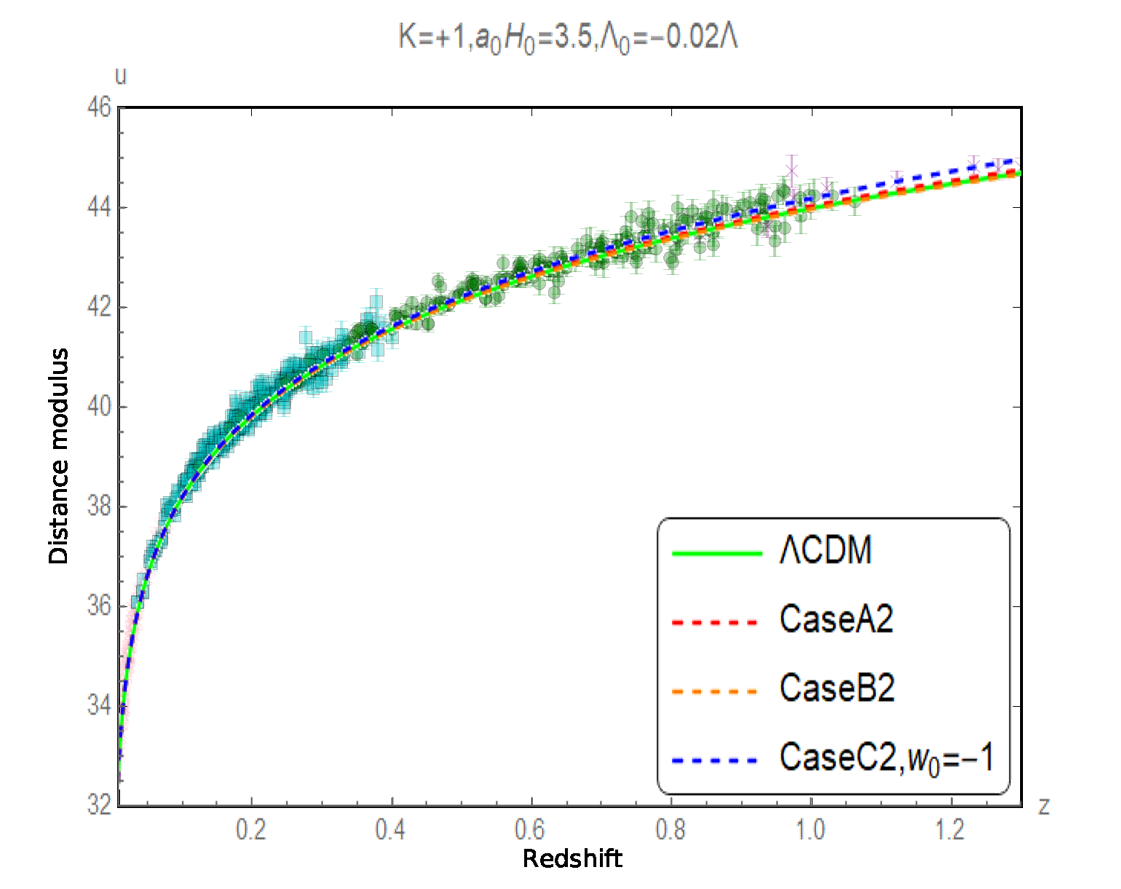}}
	\caption{Comparison of the measured distance modulus with the corresponding predicted values when $K=+1$, ${{\Lambda }_{0}}=-0.02\Lambda$ and ${{a}_{0}}{{H}_{0}}=3.5$}\label{dismodK+1L-0.02a0H03.5}
\end{figure}

\begin{figure}[h]
	\centering
	\subfigure[]
	{\label{A9}
		\includegraphics[width=2.5in]{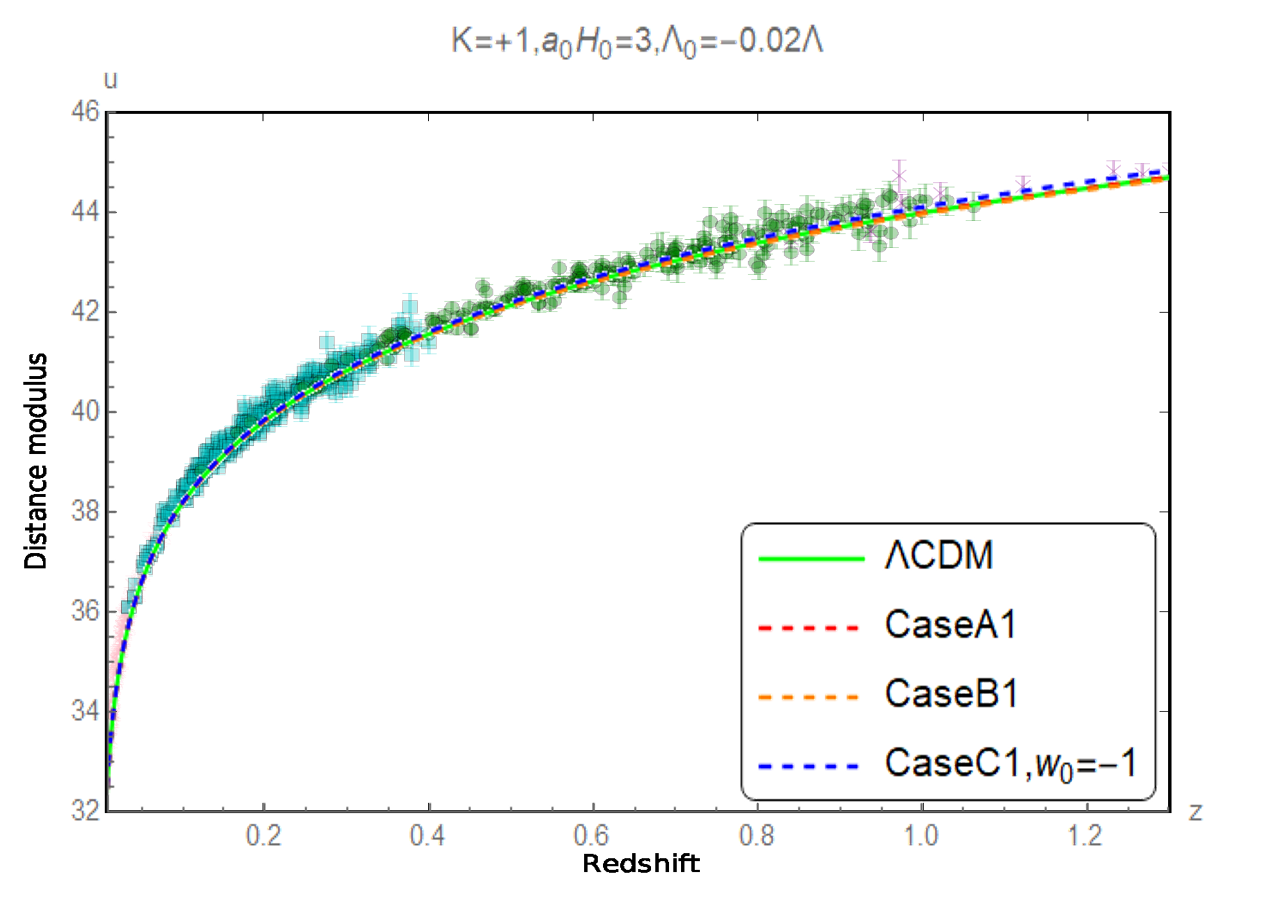}}
	\subfigure[]
	{\label{B9}
		\includegraphics[width=2.5in]{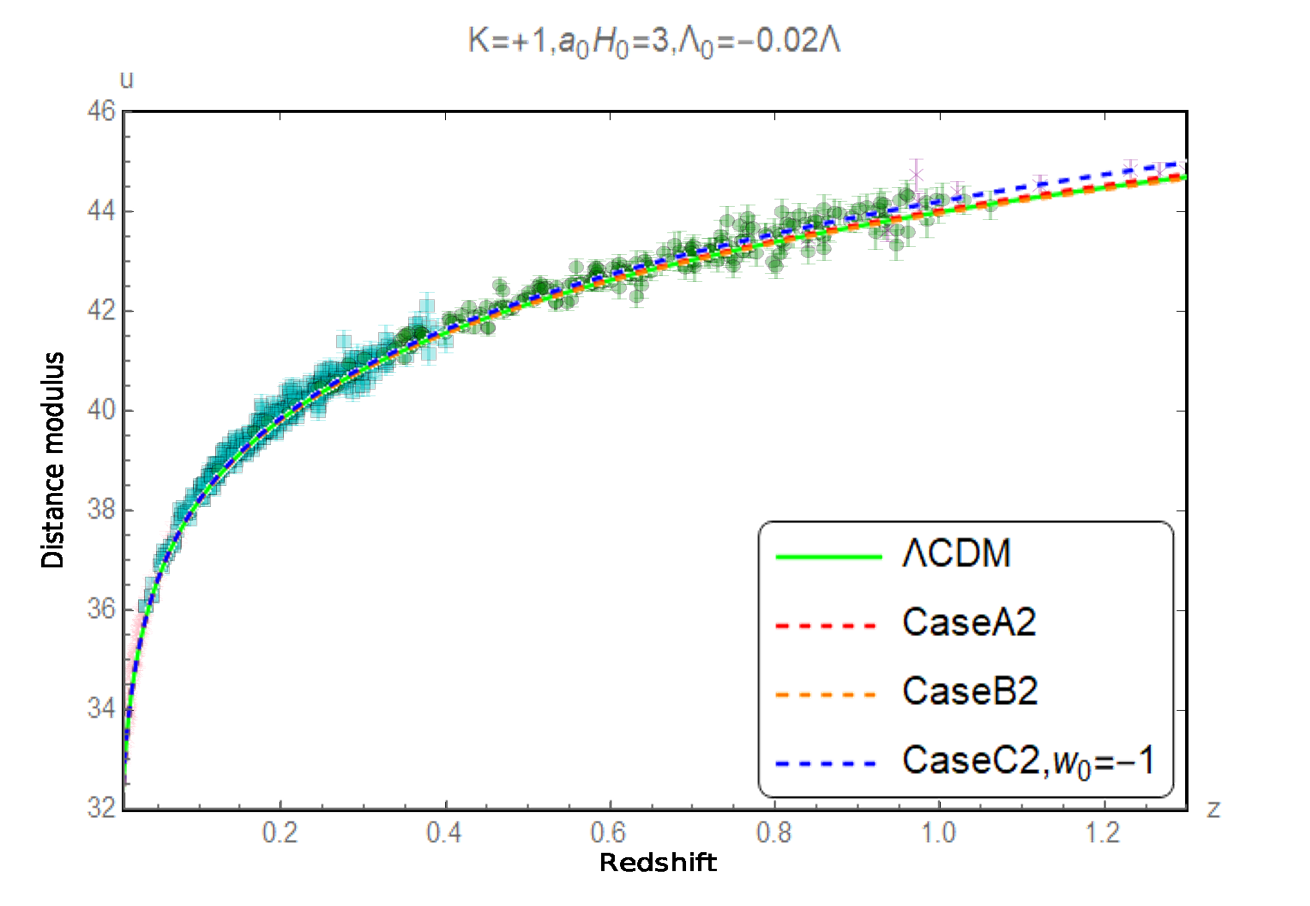}}
	\caption{Comparison of the measured distance modulus with the corresponding predicted values when $K=+1$, ${{\Lambda }_{0}}=-0.02\Lambda$ and ${{a}_{0}}{{H}_{0}}=3$}\label{dismodK+1L-0.02a0H03}
\end{figure}

\begin{figure}[h]
	\centering
	\subfigure[]
	{\label{A10}
		\includegraphics[width=2.5in]{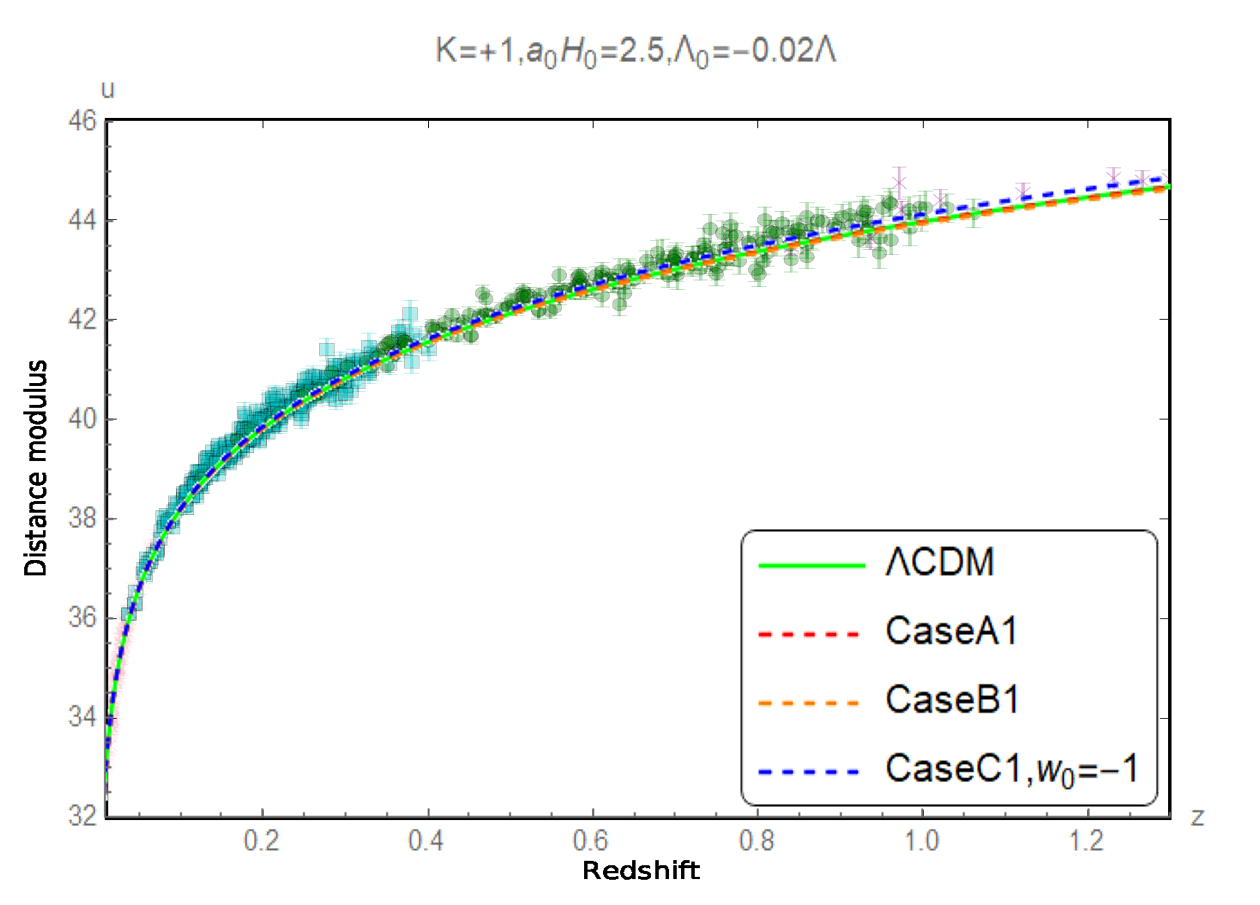}}
	\subfigure[]
	{\label{B10}
		\includegraphics[width=2.5in]{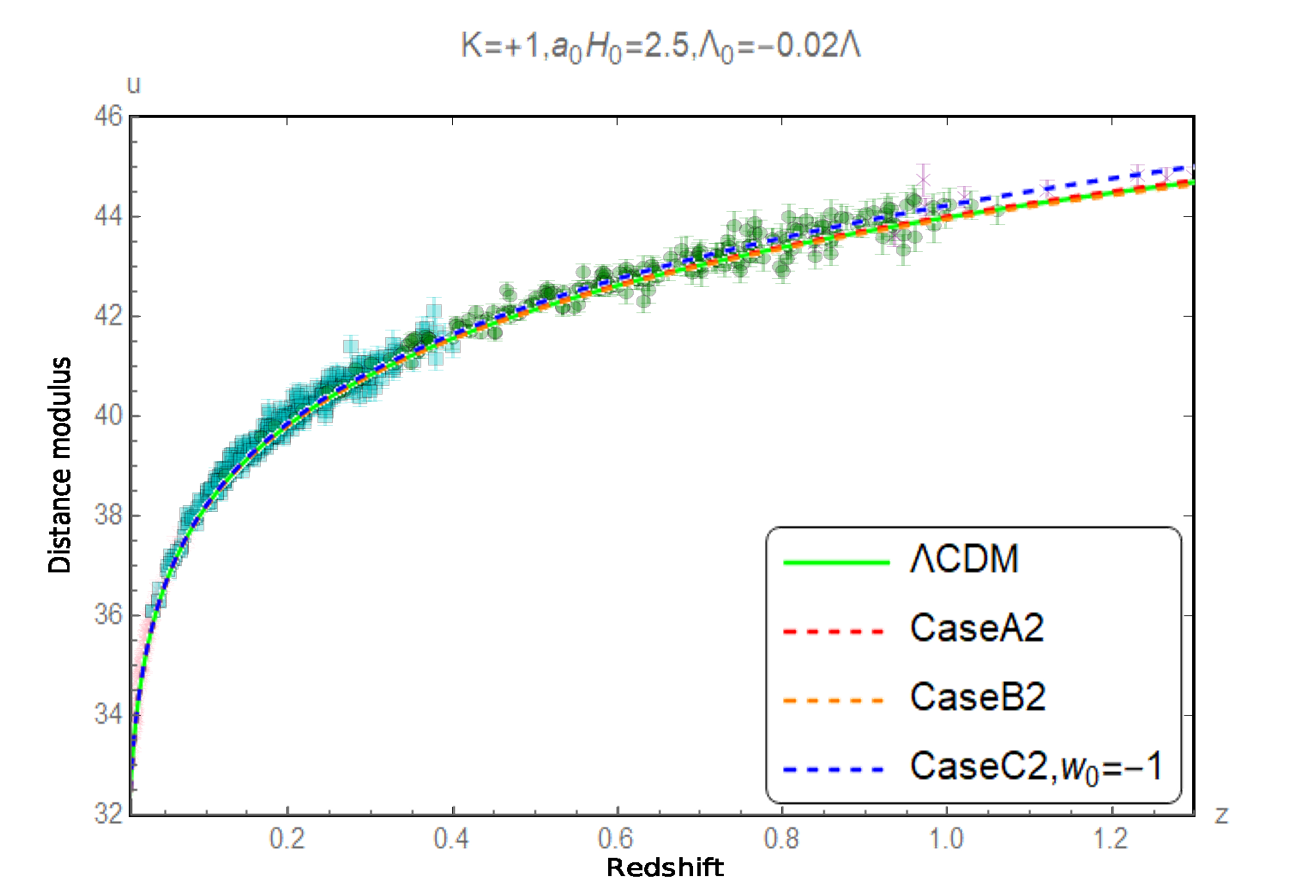}}
	\caption{Comparison of the measured distance modulus with the corresponding predicted values when $K=+1$, ${{\Lambda }_{0}}=-0.02\Lambda$ and ${{a}_{0}}{{H}_{0}}=2.5$}\label{dismodK+1L-0.02a0H02.5}
\end{figure}

\begin{figure}[h]
	\centering
	\subfigure[]
	{\label{A11}
		\includegraphics[width=2.5in]{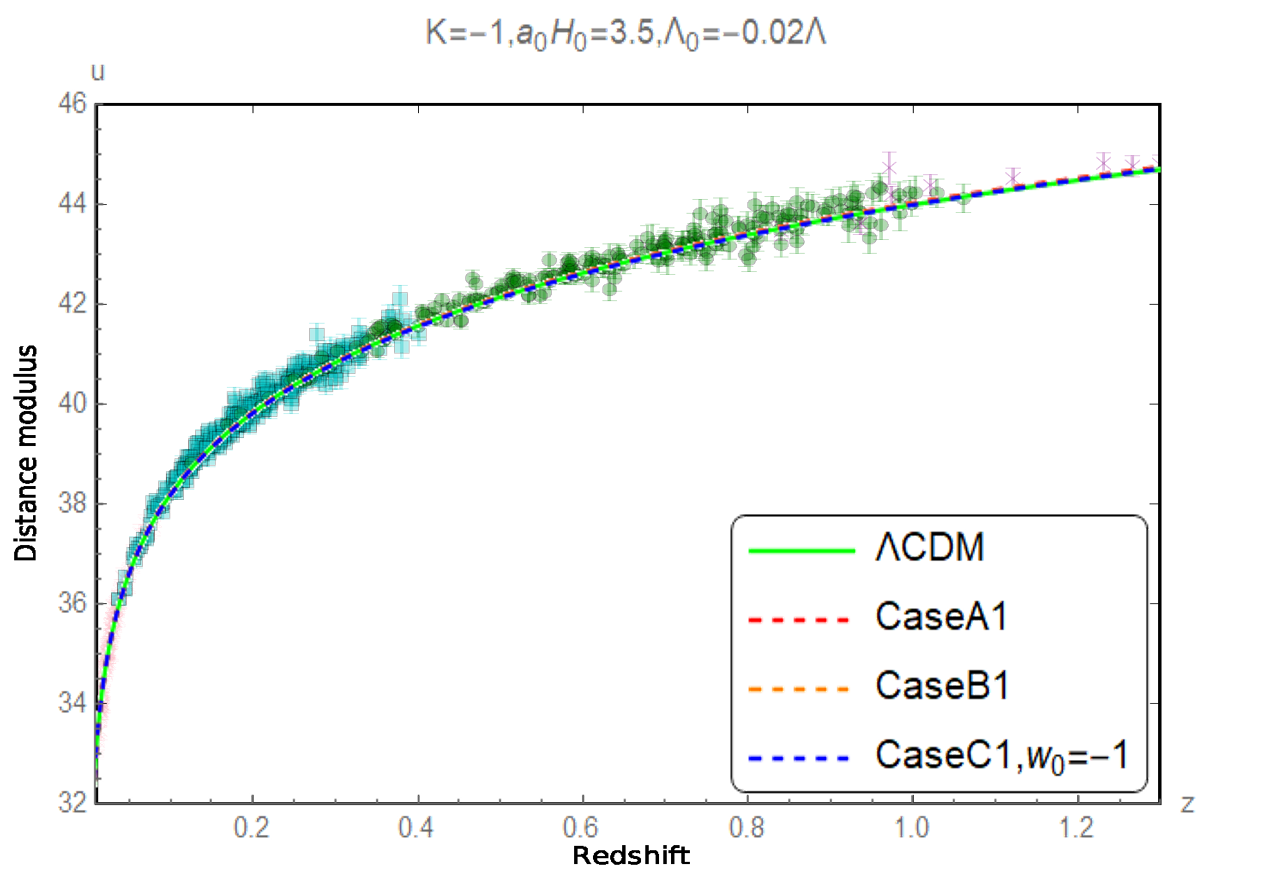}}
	\subfigure[]
	{\label{B11}
		\includegraphics[width=2.5in]{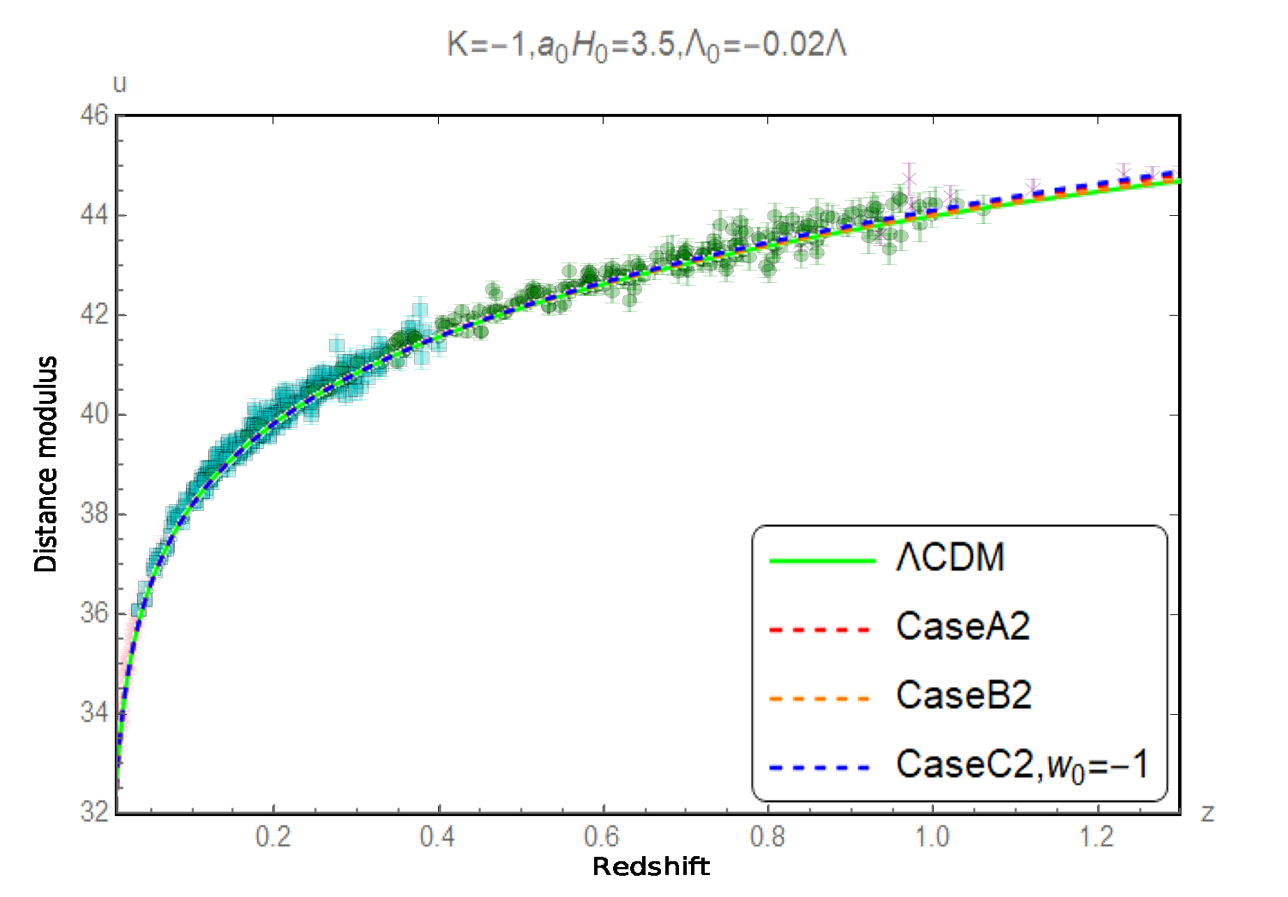}}
\caption{Comparison of the measured distance modulus with the corresponding predicted values when $K=-1$, ${{\Lambda }_{0}}=-0.02\Lambda$ and ${{a}_{0}}{{H}_{0}}=3.5$}\label{dismodK-1L-0.02a0H03.5}
\end{figure}

\begin{figure}[h]
	\centering
	\subfigure[]
	{\label{A12}
		\includegraphics[width=2.5in]{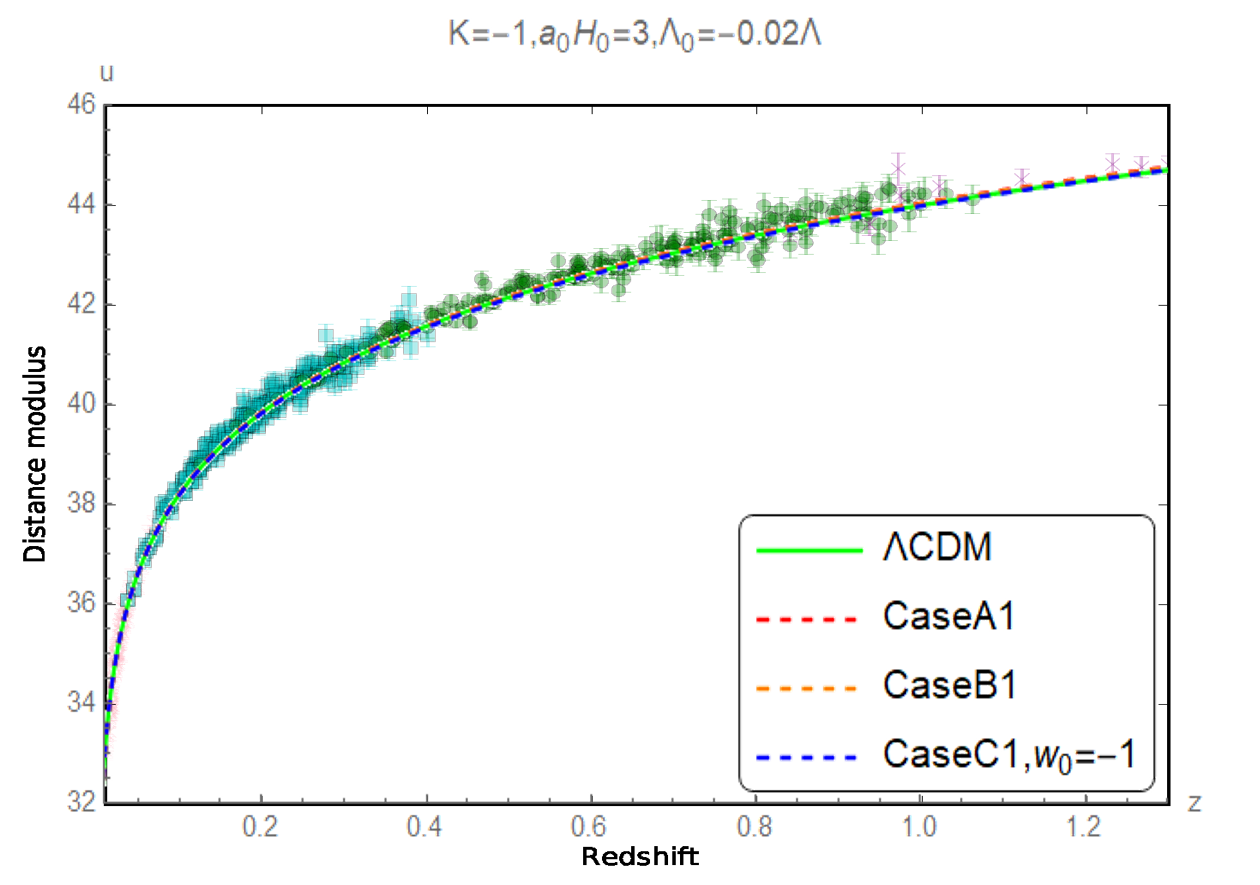}}
	\subfigure[]
	{\label{B12}
		\includegraphics[width=2.5in]{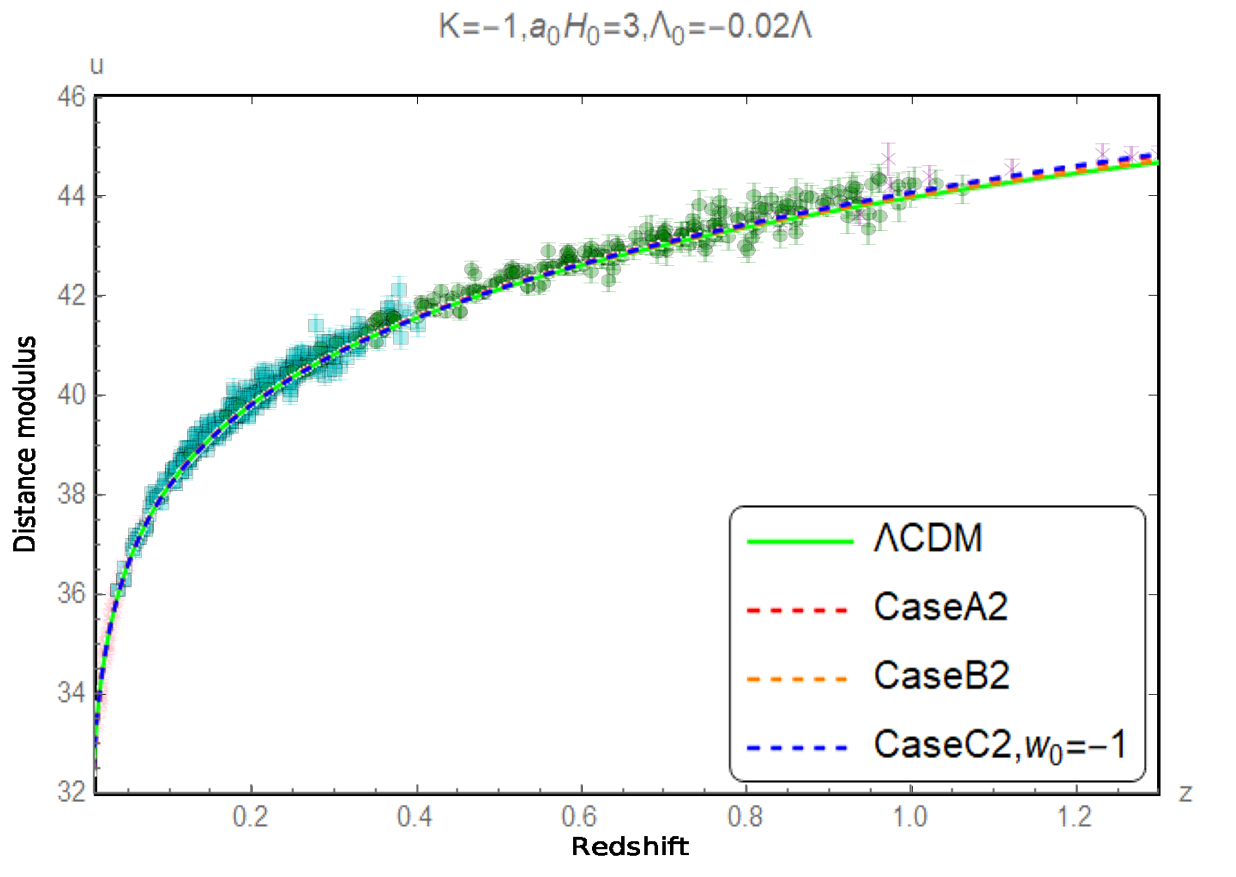}}
	\caption{Comparison of the measured distance modulus with the corresponding predicted values when $K=-1$, ${{\Lambda }_{0}}=-0.02\Lambda$ and ${{a}_{0}}{{H}_{0}}=3$}\label{dismodK-1L-0.02a0H03}
\end{figure}

\begin{figure}[h]
	\centering
	\subfigure[]
	{\label{A13}
		\includegraphics[width=2.5in]{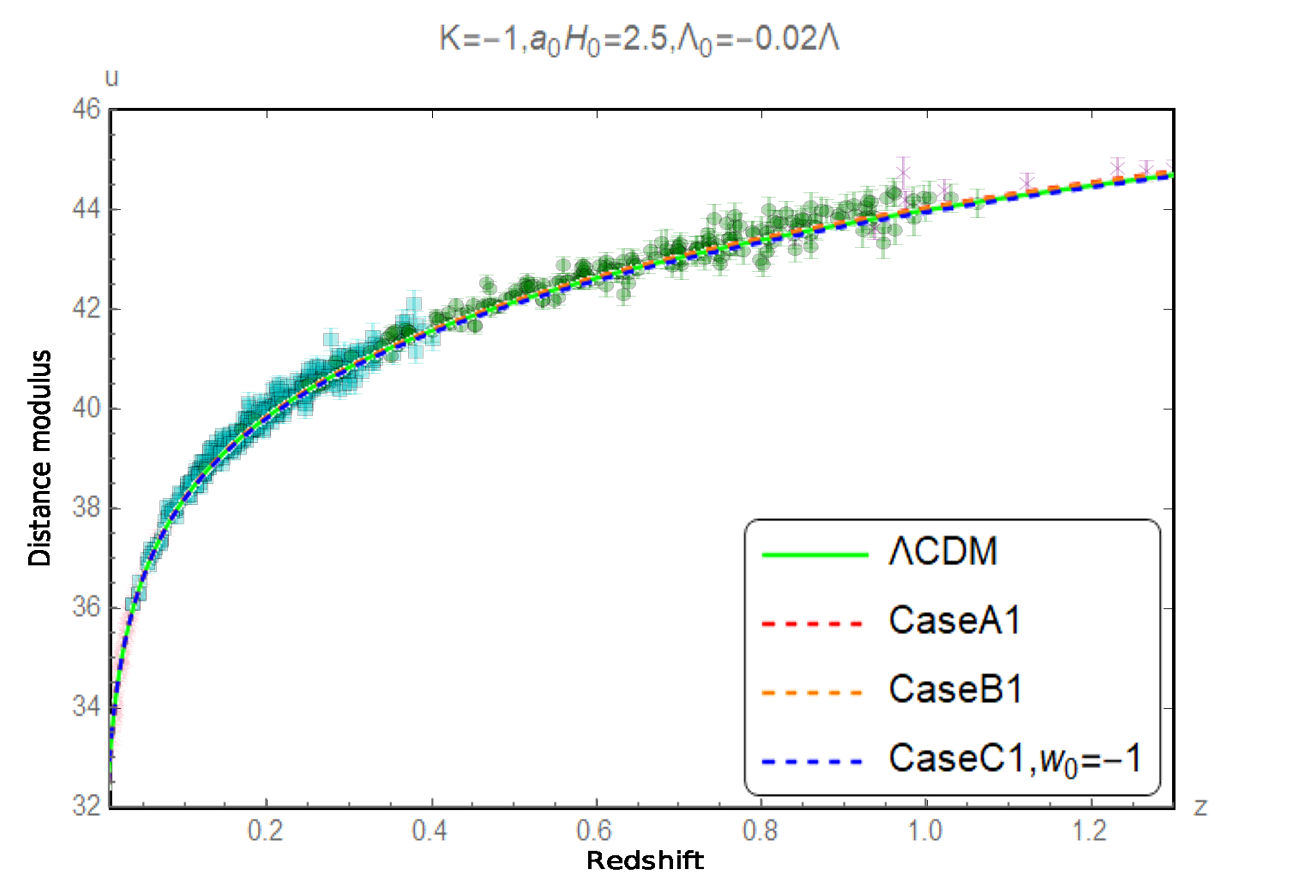}}
	\subfigure[]
	{\label{B13}
		\includegraphics[width=2.5in]{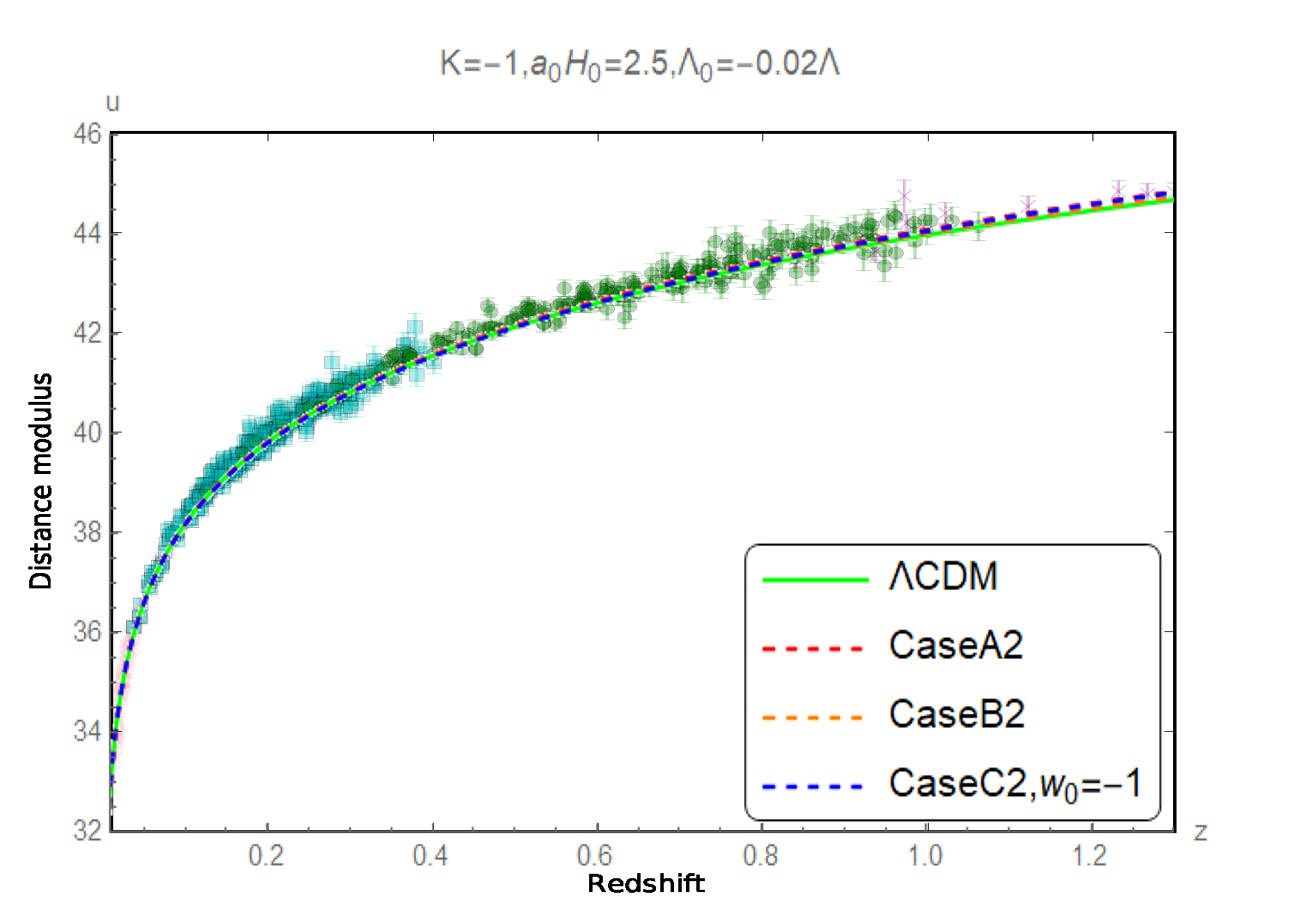}}
\caption{Comparison of the measured distance modulus with the corresponding predicted values when $K=-1$, ${{\Lambda }_{0}}=-0.02\Lambda$ and ${{a}_{0}}{{H}_{0}}=2.5$}\label{dismodK-1L-0.02a0H02.5}
\end{figure}

\begin{table}[tp]
	
	\centering
	
	\fontsize{6.5}{8}\selectfont
	
	\caption{The value of $\Lambda_{0-crit}$ in case of $K=+1$}
	
	\label{LcrtK+1}
	
	\begin{tabular}{|p{1.5cm}|p{1.8cm}|p{1.8cm}|p{1.8cm}|p{1.8cm}|}
		
		\hline
		
		\multicolumn{4}{|c|}{$K=+1$}&\multicolumn{1}{c|}{$K=0$}\cr\cline{1-5}
		
		$\Lambda_{0-crit}$&$a_0 H_0 = 2.5$&$a_0 H_0 = 3$&$a_0 H_0 = 3.5$&\ \cr
		
		\hline
		
		CaseA1&-0.072&-0.064&-0.059&0.05\cr\hline
		
		CaseA2&-0.14&-0.0159&-0.167&-0.18\cr\hline
		
		CaseB1&-0.094&-0.083&-0.078&-0.066\cr\hline
		
		CaseB2&-0.154&-0.176&-0.187&-0.2144\cr\hline
		
		CaseC1\ ($w_0=-1$)&0&0&0&0\cr\hline
		
		CaseC2\ ($w_0=-1$)&0&0&0&0\cr\hline
		
		CaseC1\ ($w_0=-8/9$)&0.162&0.152&0.146&0.119\cr\hline
		
		CaseC2\ ($w_0=-8/9$)&0.09&0.086&0.083&0.075\cr\hline
		
		CaseC1\ ($w_0=-7/9$)&none&none&none&none\cr\hline
		
		CaseC2\ ($w_0=-7/9$)&0.173&0.164&0.164&0.152\cr\hline
		
		CaseC1\ ($w_0=-2/3$)&none&none&none&none\cr\hline
		
		CaseC2\ ($w_0=-2/3$)&0.246&0.235&0.228&0.225\cr\hline
		
		CaseC1\ ($w_0=-1/3$)&none&none&none&none\cr\hline
		
		CaseC2\ ($w_0=-1/3$)&0.354&0.368&0.375&0.397\cr\hline
		
	\end{tabular}
	
\end{table}
When the $w_0$ value of Case C2 is greater than $-\frac{8}{9}$, the $\Lambda_{eff}$ shows a quintessence like potential that decreases monotonically over time without a local minimum, i.e. there is no solution for ${{\Lambda }_{0-crit}}$. It is possible that the behavior is caused by the fixed value of $w_0$  other than an evolved one in the equation of state for dark partner. A similar conclusion is reached based on the large-scale Lorentz violation model without spatial curvature in \cite{Han-Yu:2019tmf, Zhai:2019std}. When ${{w}_{0}}>-\frac{8}{9}$ , the model prediction on luminosity distance modulus versus redshift curve of Case C2 does not match the observation one in case of $K=0$ while the prediction on evolution luminosity distance modulus is compatible with observation one in case of $K\ne 0$. The spatial non-flat Case C2 with ${{w}_{0}}>-\frac{8}{9}$ can not be excluded by observation just like spatial flat case as in \cite{Han-Yu:2019tmf, Zhai:2019std}. However, we find that the ${{\Lambda }_{0-crit}}$ of Case C is always greater than ${{\Lambda }_{\max }}$ value when ${{w}_{0}}>-1$ from Tables \ref{LmaxK+1}, \ref{LmaxK-1}, \ref{LcrtK+1} and \ref{LcrtK-1}, i.e. the solution of ${{\Lambda }_{0-crit}}$ for Case C does not exist within the range of values that can be taken for $\Lambda_{0}$ when ${{w}_{0}}>-1$.  Comparisons of the luminosity distance ${{d}_{L}}$ curve versus redshift $z$  among three models of approximation and $\Lambda$CDM model are presented in Figure \ref{dismodK+1L-0.02a0H03.5} to Figure \ref{dismodK-1L-0.02a0H02.5}.
 
\begin{table}[tp]
	
	\centering
	
	\fontsize{6.5}{8}\selectfont
	
	\caption{The value of $\Lambda_{0-crit}$ in case of $K=-1$}
	
	\label{LcrtK-1}
	
	\begin{tabular}{|p{1.5cm}|p{1.8cm}|p{1.8cm}|p{1.8cm}|p{1.8cm}|p{1.8cm}|p{1.8cm}|}
		
		\hline
		
		\multicolumn{6}{|c|}{$K=-1$}&\multicolumn{1}{c|}{$K=0$}\cr\cline{1-7}
		
		$\Lambda_{0-crit}$&$a_0 H_0 = -1.5$&$a_0 H_0 = -2$&$a_0 H_0 = -2.5$&$a_0 H_0 = -3$&$a_0 H_0 = -3.5$&\ \cr
		
		\hline
		
		CaseA1&-0.023&-0.033&-0.038&-0.041&-0.044&-0.05\cr\hline
		
		CaseA2&-0.214&-0.208&-0.203&-0.198&-0.195&-0.187\cr\hline
		
		CaseB1&-0.03&-0.042&-0.049&-0.053&-0.056&-0.066\cr\hline
		
		CaseB2&-0.284&-0.262&-0.248&-0.239&-0.233&-0.2144\cr\hline
		
		CaseC1\ ($w_0=-1$)&0&0&0&0&0&0\cr\hline
		
		CaseC2\ ($w_0=-1$)&0&0&0&0&0&0\cr\hline
		
		CaseC1\ ($w_0=-8/9$)&0.05&0.086&0.102&0.107&0.115&0.119\cr\hline
		
		CaseC2\ ($w_0=-8/9$)&none&0.03&0.052&0.059&0.064&0.075\cr\hline
		
		CaseC1\ ($w_0=-7/9$)&none&none&none&none&none&none\cr\hline
		
		CaseC2\ ($w_0=-7/9$)&none&0.079&0.104&0.118&0.131&0.152\cr\hline
		
		CaseC1\ ($w_0=-2/3$)&none&none&none&none&none&none\cr\hline
		
		CaseC2\ ($w_0=-2/3$)&0.03&0.15&0.173&0.189&0.198&0.225\cr\hline
		
		CaseC1\ ($w_0=-1/3$)&none&none&none&none&none&none\cr\hline
		
		CaseC2\ ($w_0=-1/3$)&0.203&0.25&0.278&0.309&0.343&0.397\cr\hline
		
	\end{tabular}
	
\end{table}

The Figures \ref{dismodK+1L-0.02a0H03.5} to \ref{dismodK-1L-0.02a0H02.5} show that the predictions of the luminosity distance modulus of the three cases of approximation models for different spatial curvature values with the restricted range of $\Lambda-0$ values are all compatible with observational data within the error range. That is, the luminosity distance observations are not efficiency in choosing the models.

\begin{figure}[h]
	\centering
	\subfigure[]
	{\label{A14}
		\includegraphics[width=2.5in]{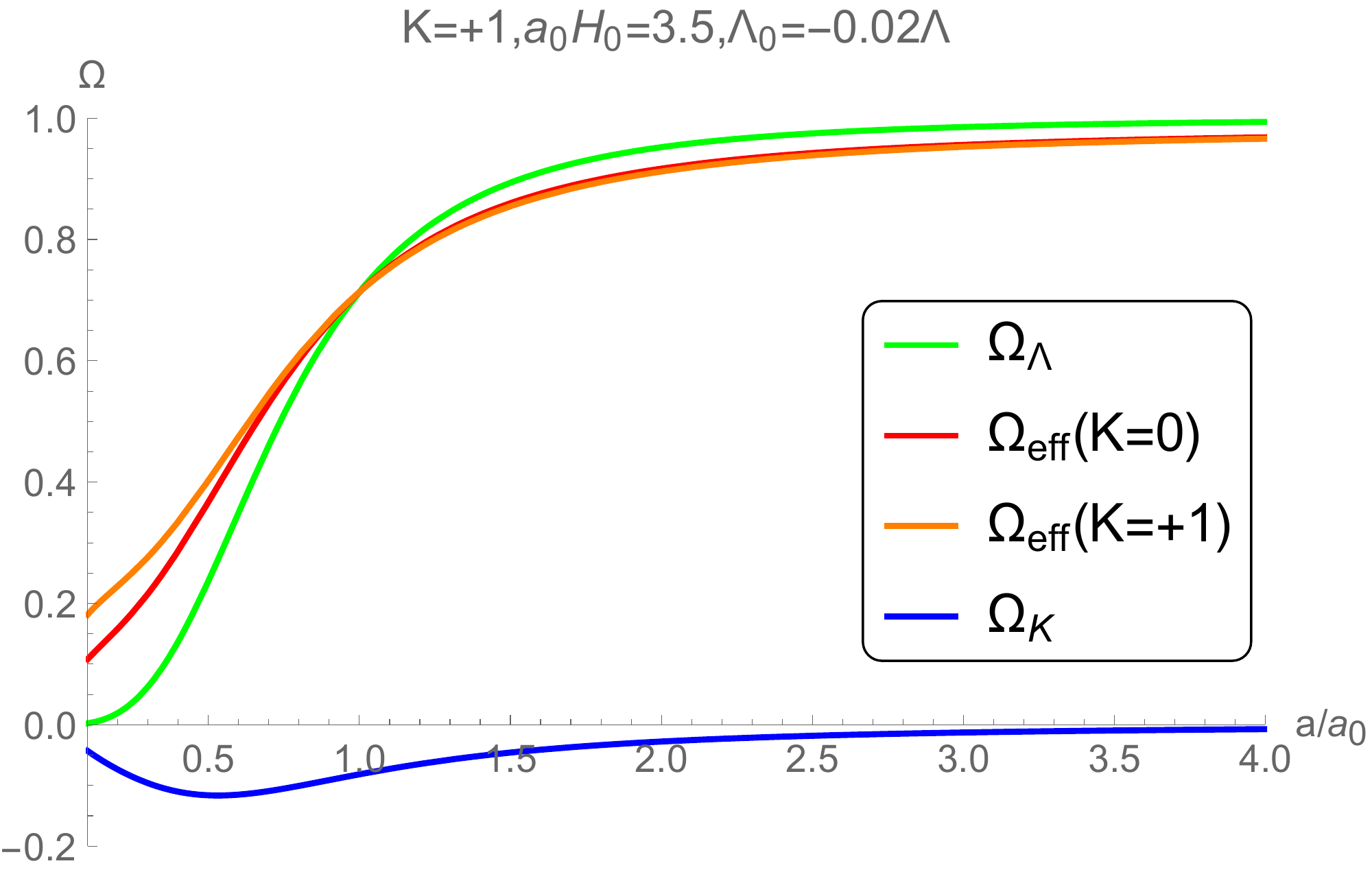}}
	\subfigure[]
	{\label{B14}
		\includegraphics[width=2.5in]{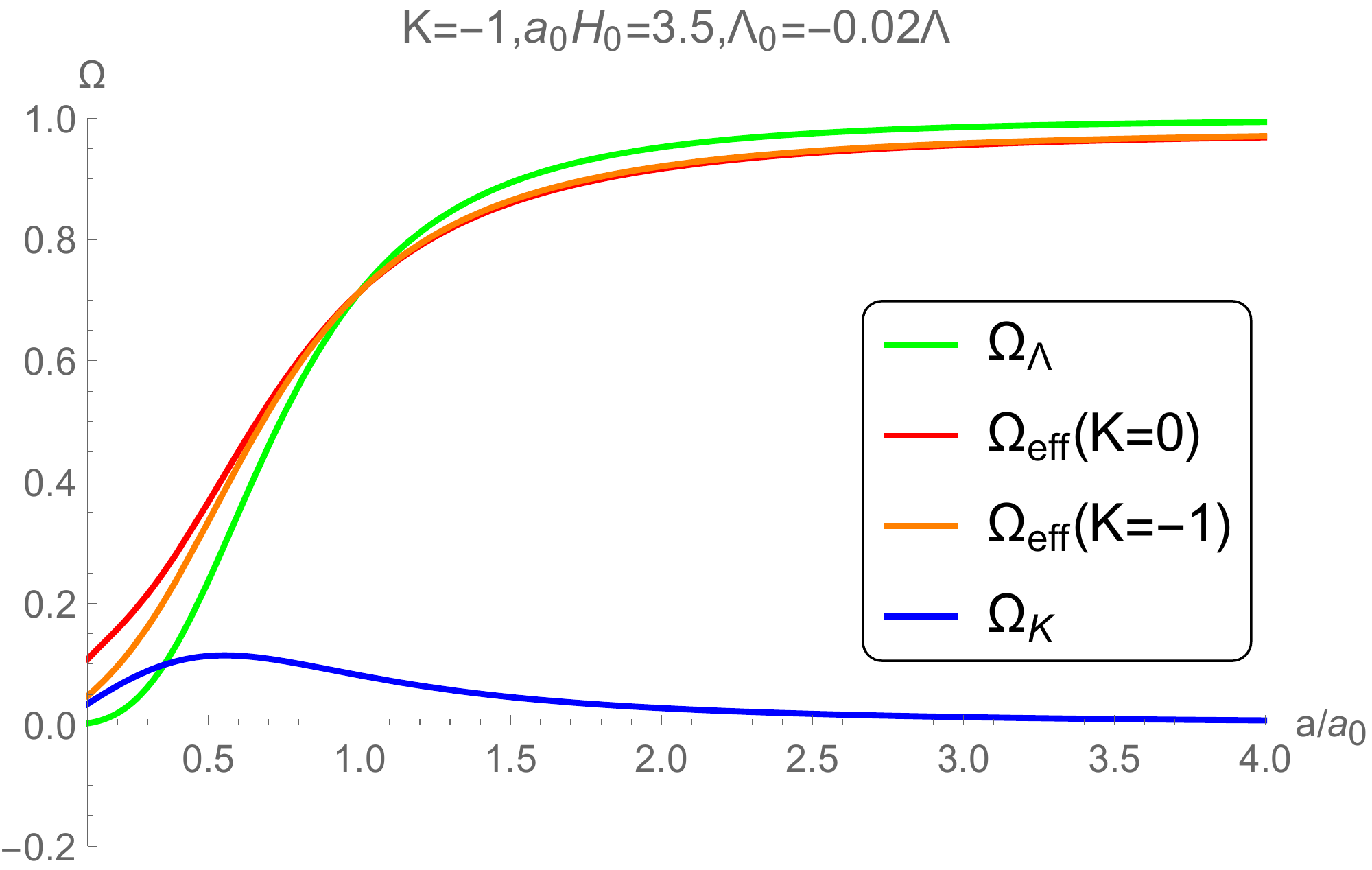}}
	\caption{The evolution of ${{\Omega }_{\Lambda }}$, ${{\Omega }_{eff}}\left( K=0 \right)$, ${{\Omega }_{eff}}\left( K\text{=}\pm 1 \right)$ and $\Omega_k$}\label{evlOmega}
\end{figure}

The evolution of ${{\Omega }_{\Lambda }}=\frac{{{\Lambda }}}{3{{H}^{2}}}$ in $\Lambda$CDM model, ${{\Omega }_{eff}}\left( K=0 \right)$ in the large-scale Lorentz violation model without spatial curvature and ${{\Omega }_{eff}}\left( K\text{=}\pm 1 \right)$ in the large-scale Lorentz violation model with spatial curvature are presented in Figure \ref{evlOmega}, which shows that the evolution behaviors of ${{\Omega }_{\Lambda }}$, ${{\Omega }_{eff}}\left( K=0 \right)$ and ${{\Omega }_{eff}}\left( K\text{=}\pm 1 \right)$ are similar, so their contributions are degenerate in result.

\begin{figure}[h]
	\centering
	\subfigure[]
	{\label{A15}
		\includegraphics[width=2.5in]{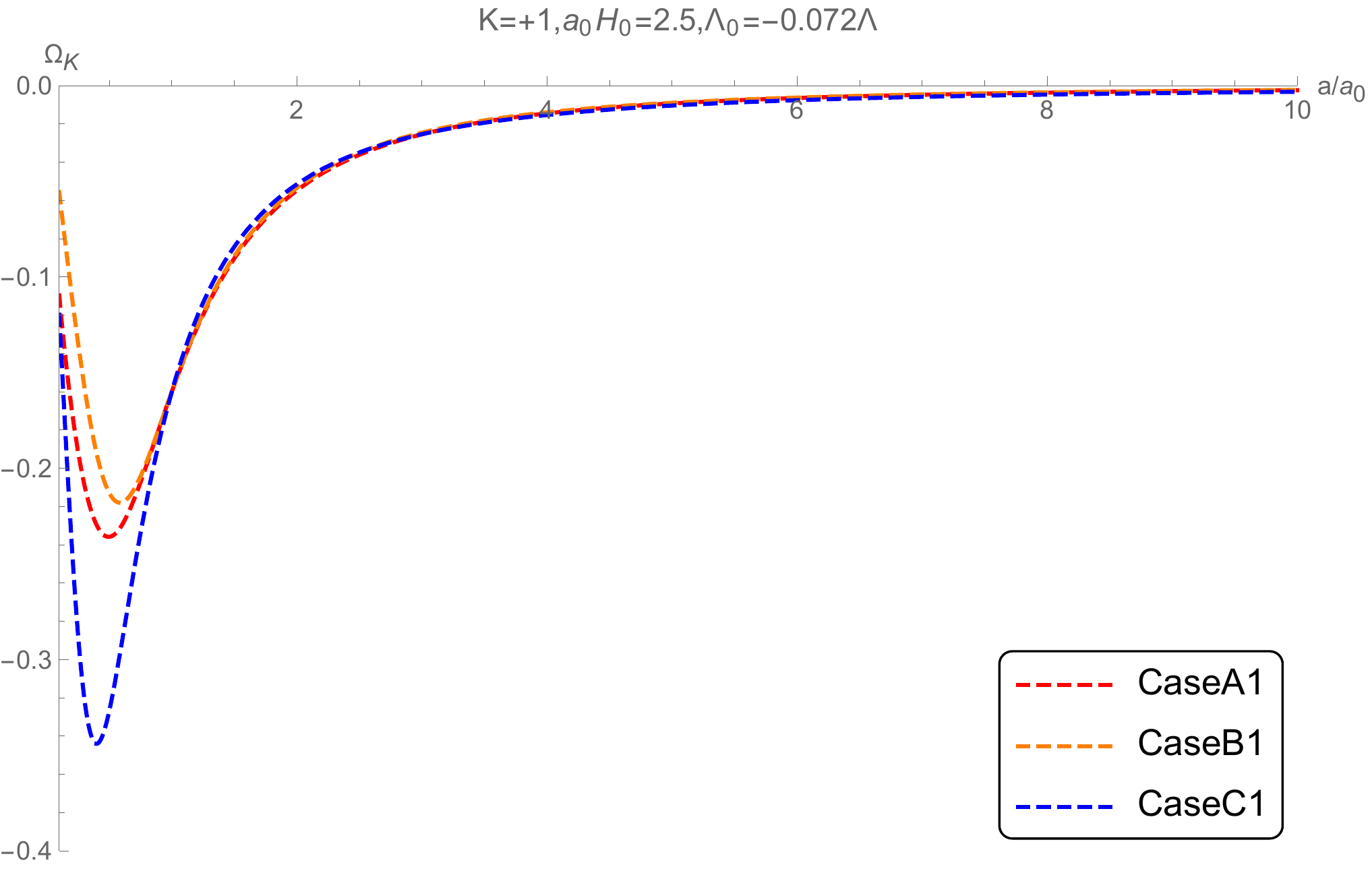}}
	\subfigure[]
	{\label{B15}
		\includegraphics[width=2.5in]{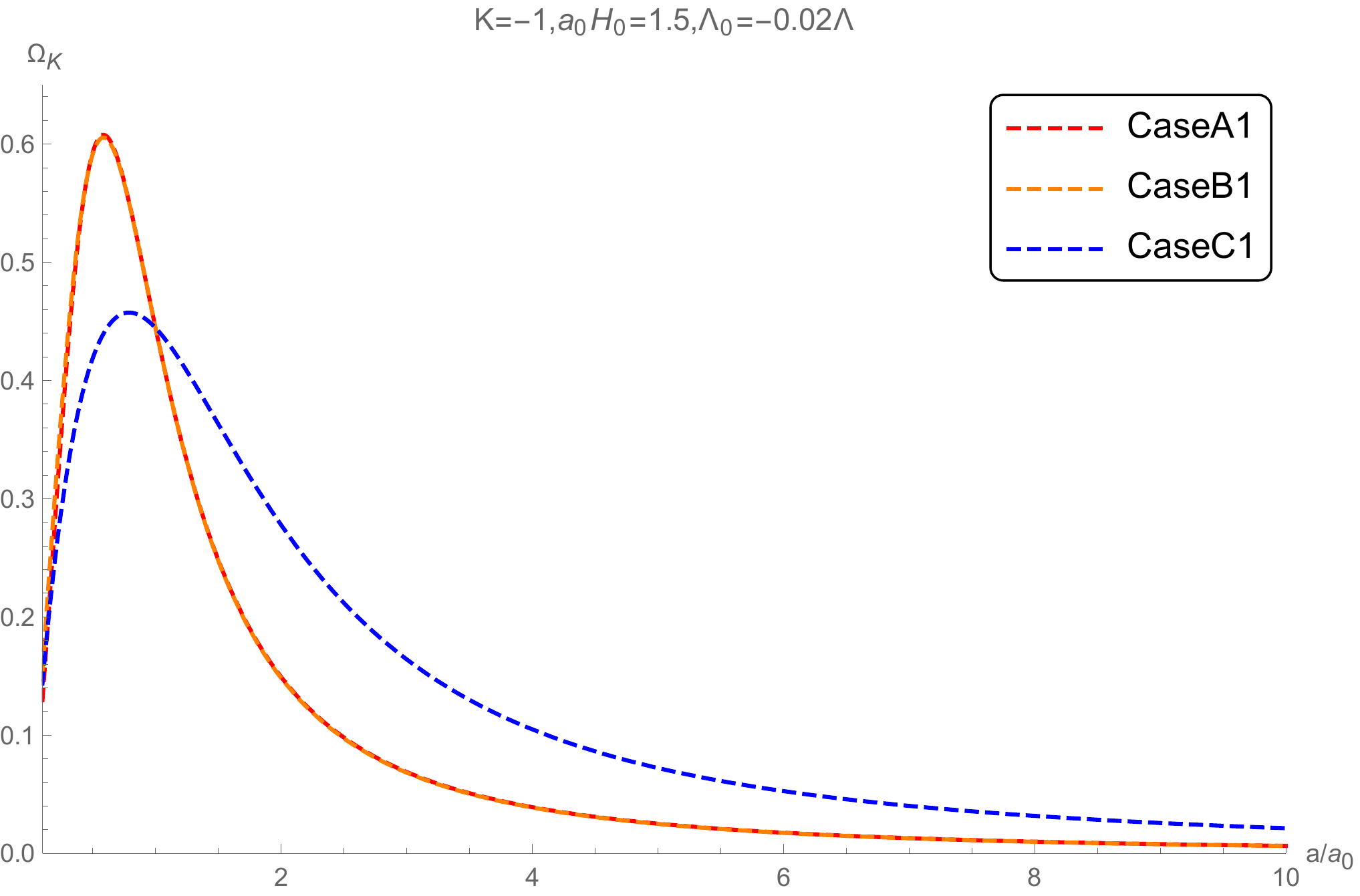}}
	\caption{The ${{\Omega }_{K}}$ evolves with the scale factor.}\label{OmgKvsa}
\end{figure} 

It is shown in Figure \ref{OmgKvsa} that the evolution of the energy density of the spatial curvature $\Omega_K=-\frac{K}{{{a}^{2}}{{H}^{2}}}$ versus scale factor $a$ based on the modified Friedmann equations for both the cases of $K=+1$ and $K=-1$. The maximum and minimum of ${{\Omega }_{K}}$ can also be obtained from the evolution of $\Omega_K$. When $K=+1$, the Case C1 has the minimum curvature energy density ${{\Omega }_{K}}=-0.34$, while the Case A1 has the maximum one with ${{\Omega }_{K}}=0.6$ when $K=-1$. The evolution of the absolute values of the curvature energy densities always increase at first stage then decrease to zero after reaching the maximum. Figures \ref{K+1x-0.02H0va} to \ref{K-1x-0.02H0a03H0va} can also supply the evolution tendency of the absolute values of the curvature densities, where the evolution of $H$ is decreasing and ${{a}^{2}}{{H}^{2}}$ has a minimum value during evolution. 
\begin{figure}[h]
	\centering
	\subfigure[]
	{\label{A16}
		\includegraphics[width=2.5in]{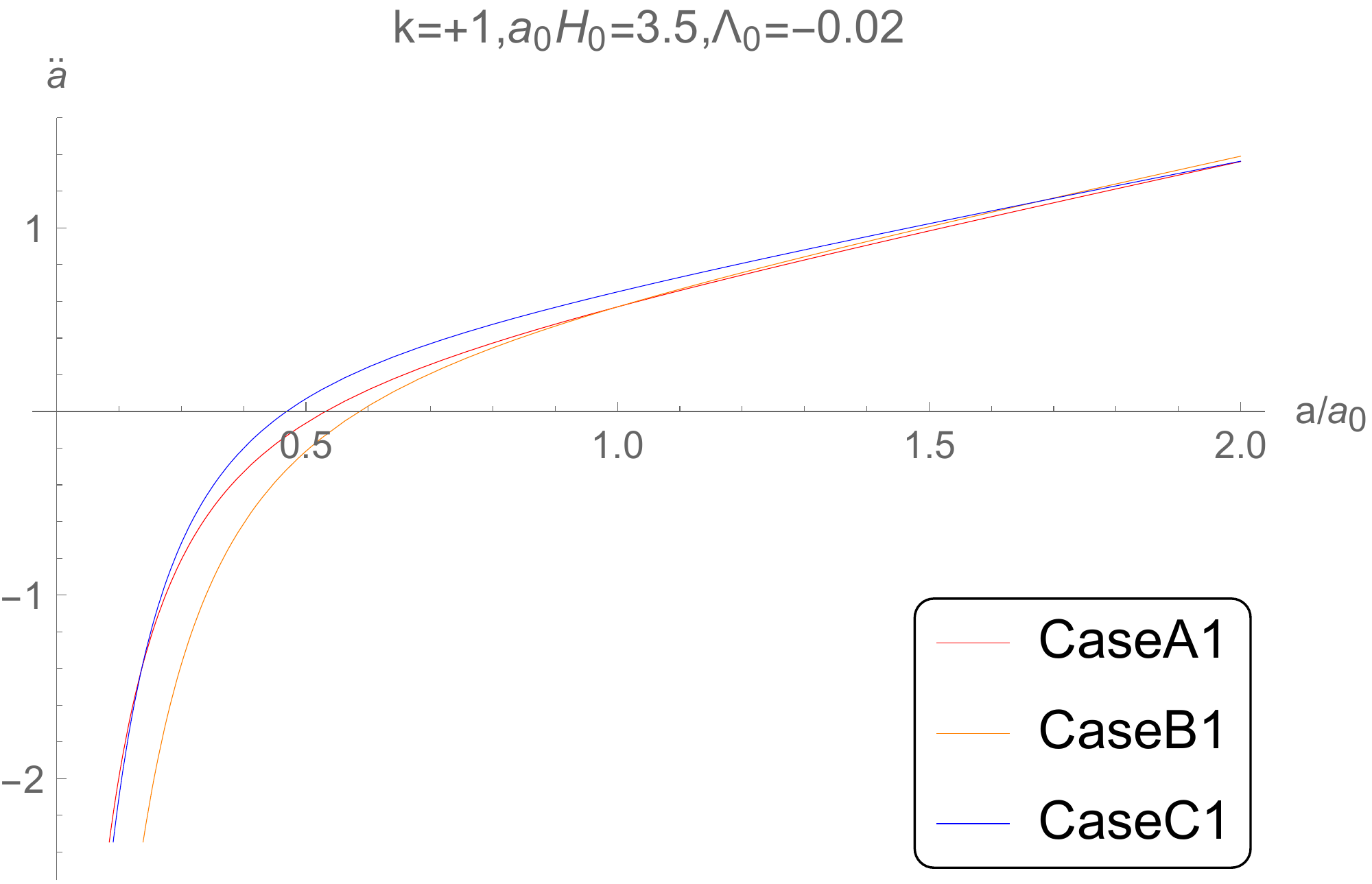}}
	\subfigure[]
	{\label{B16}
		\includegraphics[width=2.5in]{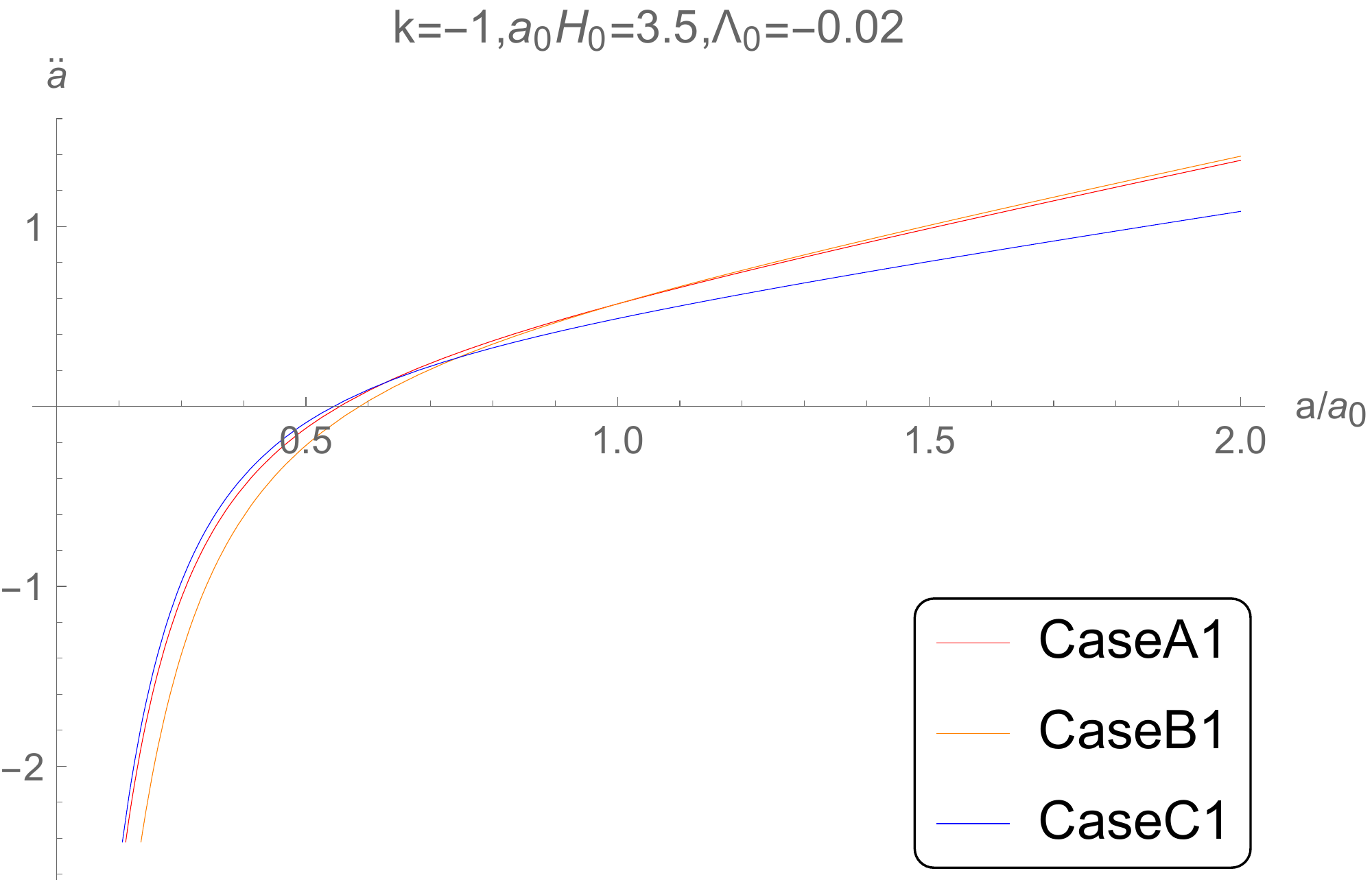}}
	\caption{The evolution of $\ddot{a}$ versus the scale factor}\label{evlacl}
\end{figure} 

The evolution of the acceleration of the universe expansion $\ddot a$ can also be obtained by the modified Friedmann equations and the approximations considered. The evolution of $\ddot{a}$ versus scale factor $a$ can be found in Figure \ref{evlacl}, from which it can be roughly obtained that the expansion of the universe is decelerating before the cosmic scale factor reaches $0.5a_0$ and the universe becomes accelerating afterwards. The conclusion is consistent with the observed universe experiencing the transition from decelerating expansion to accelerating one, thus supplying a further side proof for the confidence level of the large scale Lorentz violation model with non-vanishing spatial curvature.

\section{Summary and Outlook}

The cause of the $H_0$ tension between high and low redshift measurements is so far unknown. Many solutions to $H_0$ tension have been proposed. It was suggested that this was due to systematic error, however these claims were quickly disproved\cite{Shanks:2018rka, Riess:2018kzi, vonMarttens:2018bvz, Bengaly:2018xko}.
The gravitational waves from the binary neutron star (BNS) merger may be able to make independent measurements of $H_0$\cite{LIGOScientific:2017vwq}, although future observations should be able to reduce the error\cite{LIGOScientific:2018gmd, Mortlock:2018azx, Feeney:2018mkj, Hotokezaka:2018dfi, Chen:2017rfc, Vitale:2018wlg}. However, this current measurement method has a large error (due to being a single event) and therefore does not resolve the tension between $H_0$ measurements. Furthermore, as one of the simplest alternatives to the $\Lambda$CDM model: quintessence fields with smooth potential energy, rolling slowly or at moderate velocity in the non-flat FRW universe, are expected to resolve the tension between Hubble constant measurements. But after combining the latest CMB, SNe and BAO data, it was found that this model does not reduce the Hubble constant tension\cite{Miao:2018zpw}.

Literature \cite{Bolejko:2017fos} presents a new solution to the Hubble constant problem. The solution is based on the Simsilun simulation (relativistic simulation of the large scale structure of the Universe) with the ray-tracing algorithm implemented, within the Simsilun simulation relativistic and nonlinear evolution of cosmic structures leads to the phenomenon of emerging spatial curvature, where the mean spatial curvature evolves from spatial flatness of the early universe towards slightly curved present-day universe. The deduced Hubble constant, $H_0=(68.1\pm 2.0)\text{km}/(\text{s}\cdot \text{Mpc})$, alleviates the tension between the CMB and distance ladder measurements of the Hubble constant and it is argued that the $H_0$ tension is a manifestation of rigidity of the FLRW geometry.

At the current stage, the existence of spatial curvature seems to be a viable solution to the Hubble constant problem. In fact, we can also counter that the tension is an indirect evidence of the emerging spatial curvature. From the point of view of astronomical observations, no direct measurement of spatial curvature has yet been made at low redshifts (the constraint obtained is also only the result of fitting FLRW to the data, which is not equivalent to a direct measurement).

In conclusion, although the large scale Lorentz violation model with non-vanishing spatial curvature has one more cosmological scale contortion parameter than the $\Lambda$CDM model with non-vanishing spatial curvature, the inputs are well constrained in a variety of ways, and the theoretical predictions of distance modulus, cosmic expansion acceleration, etc. are in good agreement with observations.

The temperature fluctuation in CMB power spectra can fix the cosmological parameters of $\Lambda$CDM model strictly. The Hubble constant predicted by the large scale Lorentz violation with non-vanishing spatial curvature is also constrained by the CMB data and be compared with one predicted by CMB data based on $\Lambda$CDM model to evaluate the improvement on the $H_0$ tension problem by the large scale Lorentz violation with non-vanishing spatial curvature model. The cosmological parameters of the specific cosmological model such as $\Lambda$CDM model can be obtained from CMB data by the numerical software like Code for Anisotropies in the Microwave Background(CAMB). It is pointed out that the spatial flatness assumption may be responsible for the $H_0$ tension, the non-vanishing spatial curvature is an alternative choice. This paper provides a viable solution to $H_0$ tension by investigating the constrain on the spatial curvature provided by the large scale Lorentz violation cosmological model with spatial curvature. We find that the non-vanishing of the aptial curvature is compatible with the present observation. The model provides an viable premise to solve the $H_0$ tension problem by the spatial non-flat cosmological model. The numerical software like CAMB needs to be extended to the model with non-trivial contortion distribution to present the $H_0$ value in this kind of models.

\section*{Acknowledgment}
This work is supported by the National Natural Science Foundation of China, under Grant No. 11775080 and Grant No. 11865016.


\bibliographystyle{utcaps}
\bibliography{References}

\providecommand{\href}[2]{#2}\begingroup\raggedright\begin{thebibliography}{10}

\bibitem{Planck:2018vyg}
{\bfseries Planck} Collaboration, N.~Aghanim {\em et~al.}, ``{Planck 2018
  results. VI. Cosmological parameters},''
  \href{http://dx.doi.org/10.1051/0004-6361/201833910}{{\em Astron. Astrophys.}
  {\bfseries 641} (2020) A6}, \href{http://arxiv.org/abs/1807.06209}{{\ttfamily
  arXiv:1807.06209 [astro-ph.CO]}}. [Erratum: Astron.Astrophys. 652, C4
  (2021)].

\bibitem{ade2014planckI}
P.~A. Ade {\em et~al.}, ``Planck 2013 results. I. Overview of products and
  scientific results,''
  \href{http://dx.doi.org/10.1051/0004-6361/201321529}{{\em Astronomy \&
  Astrophysics} {\bfseries 571} (2014) A1}.

\bibitem{Riess:2018byc}
A.~G. Riess {\em et~al.}, ``{Milky Way Cepheid Standards for Measuring Cosmic
  Distances and Application to Gaia DR2: Implications for the Hubble
  Constant},'' \href{http://dx.doi.org/10.3847/1538-4357/aac82e}{{\em
  Astrophys. J.} {\bfseries 861} no.~2, (2018) 126},
  \href{http://arxiv.org/abs/1804.10655}{{\ttfamily arXiv:1804.10655
  [astro-ph.CO]}}.

\bibitem{Virey:2008nu}
J.~M. Virey, D.~Talon-Esmieu, A.~Ealet, P.~Taxil, and A.~Tilquin, ``{On the
  determination of curvature and dynamical Dark Energy},''
  \href{http://dx.doi.org/10.1088/1475-7516/2008/12/008}{{\em JCAP} {\bfseries
  12} (2008) 008}, \href{http://arxiv.org/abs/0802.4407}{{\ttfamily
  arXiv:0802.4407 [astro-ph]}}.

\bibitem{Wang:2007mza}
Y.~Wang and P.~Mukherjee, ``{Observational Constraints on Dark Energy and
  Cosmic Curvature},'' \href{http://dx.doi.org/10.1103/PhysRevD.76.103533}{{\em
  Phys. Rev. D} {\bfseries 76} (2007) 103533},
  \href{http://arxiv.org/abs/astro-ph/0703780}{{\ttfamily
  arXiv:astro-ph/0703780}}.

\bibitem{Clarkson:2007bc}
C.~Clarkson, M.~Cortes, and B.~A. Bassett, ``{Dynamical Dark Energy or Simply
  Cosmic Curvature?},''
  \href{http://dx.doi.org/10.1088/1475-7516/2007/08/011}{{\em JCAP} {\bfseries
  08} (2007) 011}, \href{http://arxiv.org/abs/astro-ph/0702670}{{\ttfamily
  arXiv:astro-ph/0702670}}.

\bibitem{Rest:2013mwz}
A.~Rest {\em et~al.}, ``{Cosmological Constraints from Measurements of Type Ia
  Supernovae discovered during the first 1.5 yr of the Pan-STARRS1 Survey},''
  \href{http://dx.doi.org/10.1088/0004-637X/795/1/44}{{\em Astrophys. J.}
  {\bfseries 795} no.~1, (2014) 44},
  \href{http://arxiv.org/abs/1310.3828}{{\ttfamily arXiv:1310.3828
  [astro-ph.CO]}}.

\bibitem{Kumar:2015saa}
S.~Kumar, ``{Consistency of nonflat $\Lambda$CDM model with the new result from
  BOSS},'' \href{http://dx.doi.org/10.1103/PhysRevD.92.103512}{{\em Phys. Rev.
  D} {\bfseries 92} no.~10, (2015) 103512},
  \href{http://arxiv.org/abs/1507.04684}{{\ttfamily arXiv:1507.04684 [gr-qc]}}.

\bibitem{DiValentino:2019qzk}
E.~Di~Valentino, A.~Melchiorri, and J.~Silk, ``{Planck evidence for a closed
  Universe and a possible crisis for cosmology},''
  \href{http://dx.doi.org/10.1038/s41550-019-0906-9}{{\em Nature Astron.}
  {\bfseries 4} no.~2, (2019) 196--203},
  \href{http://arxiv.org/abs/1911.02087}{{\ttfamily arXiv:1911.02087
  [astro-ph.CO]}}.

\bibitem{DiValentino:2020hov}
E.~Di~Valentino, A.~Melchiorri, and J.~Silk, ``{Investigating Cosmic
  Discordance},'' \href{http://dx.doi.org/10.3847/2041-8213/abe1c4}{{\em
  Astrophys. J. Lett.} {\bfseries 908} no.~1, (2021) L9},
  \href{http://arxiv.org/abs/2003.04935}{{\ttfamily arXiv:2003.04935
  [astro-ph.CO]}}.

\bibitem{Huang:2004ai}
Q.-G. Huang and M.~Li, ``{The Holographic dark energy in a non-flat
  universe},'' \href{http://dx.doi.org/10.1088/1475-7516/2004/08/013}{{\em
  JCAP} {\bfseries 08} (2004) 013},
  \href{http://arxiv.org/abs/astro-ph/0404229}{{\ttfamily
  arXiv:astro-ph/0404229}}.

\bibitem{Shen:2018elj}
J.~Shen and X.~Xue,
  \href{http://dx.doi.org/10.1142/9789811207402_0038}{``{Large Scale Lorentz
  Violation Gravity and Dark Energy},''} in {\em {Proceedings, 28th
  International Symposium on Lepton Photon Interactions at High Energies
  (LP17)}: {Guangzhou (Guangdong), China, August 7-12, 2017}}, W.~Wang and
  Z.-z. Xing, eds., pp.~459--475.
\newblock 2020.
\newblock \href{http://arxiv.org/abs/1802.03502}{{\ttfamily arXiv:1802.03502
  [gr-qc]}}.

\bibitem{Han-Yu:2019tmf}
H.~Zhai, J.~Shen, and X.~Xue, ``{Effective quintessence from string
  landscape},'' \href{http://dx.doi.org/10.7498/aps.68.20190282}{{\em Acta
  Phys. Sin.} {\bfseries 68} no.~13, (2019) 139501}.

\bibitem{Zhai:2019std}
H.~Zhai, J.~Shen, and X.~Xue, ``{Uplifting of AdS type to quintessence-like
  potential induced by frozen large-scale Lorentz violation},''
  \href{http://dx.doi.org/10.1088/1674-1137/44/8/085101}{{\em Chin. Phys. C}
  {\bfseries 44} no.~8, (2020) 085101},
  \href{http://arxiv.org/abs/1906.11860}{{\ttfamily arXiv:1906.11860
  [hep-th]}}.

\bibitem{Li:2020tqx}
Q.~Li, J.~Li, Y.~Zhou, and X.~Xue, ``{The Effective Potential Originating from
  Swampland and the Non-trivial Brans-Dicke Coupling},''
  \href{http://dx.doi.org/10.1088/1674-1137/abab8e}{{\em Chin. Phys. C}
  {\bfseries 44} no.~10, (2020) 105108},
  \href{http://arxiv.org/abs/2003.09121}{{\ttfamily arXiv:2003.09121 [gr-qc]}}.

\bibitem{Shanks:2018rka}
T.~Shanks, L.~Hogarth, and N.~Metcalfe, ``{Gaia Cepheid parallaxes and 'Local
  Hole' relieve $H_0$ tension},''
  \href{http://dx.doi.org/10.1093/mnrasl/sly239}{{\em Mon. Not. Roy. Astron.
  Soc.} {\bfseries 484} no.~1, (2019) L64--L68},
  \href{http://arxiv.org/abs/1810.02595}{{\ttfamily arXiv:1810.02595
  [astro-ph.CO]}}.

\bibitem{Riess:2018kzi}
A.~G. Riess, S.~Casertano, D.~Kenworthy, D.~Scolnic, and L.~Macri, ``{Seven
  Problems with the Claims Related to the Hubble Tension in
  arXiv:1810.02595},'' \href{http://arxiv.org/abs/1810.03526}{{\ttfamily
  arXiv:1810.03526 [astro-ph.CO]}}.

\bibitem{vonMarttens:2018bvz}
R.~von Marttens, V.~Marra, L.~Casarini, J.~E. Gonzalez, and J.~Alcaniz, ``{Null
  test for interactions in the dark sector},''
  \href{http://dx.doi.org/10.1103/PhysRevD.99.043521}{{\em Phys. Rev. D}
  {\bfseries 99} no.~4, (2019) 043521},
  \href{http://arxiv.org/abs/1812.02333}{{\ttfamily arXiv:1812.02333
  [astro-ph.CO]}}.

\bibitem{Bengaly:2018xko}
C.~A.~P. Bengaly, U.~Andrade, and J.~S. Alcaniz, ``{How does an incomplete sky
  coverage affect the Hubble Constant variance?},''
  \href{http://dx.doi.org/10.1140/epjc/s10052-019-7284-4}{{\em Eur. Phys. J. C}
  {\bfseries 79} no.~9, (2019) 768},
  \href{http://arxiv.org/abs/1810.04966}{{\ttfamily arXiv:1810.04966
  [astro-ph.CO]}}.

\bibitem{LIGOScientific:2017vwq}
{\bfseries LIGO Scientific, Virgo} Collaboration, B.~P. Abbott {\em et~al.},
  ``{GW170817: Observation of Gravitational Waves from a Binary Neutron Star
  Inspiral},'' \href{http://dx.doi.org/10.1103/PhysRevLett.119.161101}{{\em
  Phys. Rev. Lett.} {\bfseries 119} no.~16, (2017) 161101},
  \href{http://arxiv.org/abs/1710.05832}{{\ttfamily arXiv:1710.05832 [gr-qc]}}.

\bibitem{LIGOScientific:2018gmd}
{\bfseries LIGO Scientific, Virgo} Collaboration, M.~Fishbach {\em et~al.},
  ``{A Standard Siren Measurement of the Hubble Constant from GW170817 without
  the Electromagnetic Counterpart},''
  \href{http://dx.doi.org/10.3847/2041-8213/aaf96e}{{\em Astrophys. J. Lett.}
  {\bfseries 871} no.~1, (2019) L13},
  \href{http://arxiv.org/abs/1807.05667}{{\ttfamily arXiv:1807.05667
  [astro-ph.CO]}}.

\bibitem{Mortlock:2018azx}
D.~J. Mortlock, S.~M. Feeney, H.~V. Peiris, A.~R. Williamson, and S.~M.
  Nissanke, ``{Unbiased Hubble constant estimation from binary neutron star
  mergers},'' \href{http://dx.doi.org/10.1103/PhysRevD.100.103523}{{\em Phys.
  Rev. D} {\bfseries 100} no.~10, (2019) 103523},
  \href{http://arxiv.org/abs/1811.11723}{{\ttfamily arXiv:1811.11723
  [astro-ph.CO]}}.

\bibitem{Feeney:2018mkj}
S.~M. Feeney, H.~V. Peiris, A.~R. Williamson, S.~M. Nissanke, D.~J. Mortlock,
  J.~Alsing, and D.~Scolnic, ``{Prospects for resolving the Hubble constant
  tension with standard sirens},''
  \href{http://dx.doi.org/10.1103/PhysRevLett.122.061105}{{\em Phys. Rev.
  Lett.} {\bfseries 122} no.~6, (2019) 061105},
  \href{http://arxiv.org/abs/1802.03404}{{\ttfamily arXiv:1802.03404
  [astro-ph.CO]}}.

\bibitem{Hotokezaka:2018dfi}
K.~Hotokezaka, E.~Nakar, O.~Gottlieb, S.~Nissanke, K.~Masuda, G.~Hallinan,
  K.~P. Mooley, and A.~T. Deller, ``{A Hubble constant measurement from
  superluminal motion of the jet in GW170817},''
  \href{http://dx.doi.org/10.1038/s41550-019-0820-1}{{\em Nature Astron.}
  {\bfseries 3} no.~10, (2019) 940--944},
  \href{http://arxiv.org/abs/1806.10596}{{\ttfamily arXiv:1806.10596
  [astro-ph.CO]}}.

\bibitem{Chen:2017rfc}
H.-Y. Chen, M.~Fishbach, and D.~E. Holz, ``{A two per cent Hubble constant
  measurement from standard sirens within five years},''
  \href{http://dx.doi.org/10.1038/s41586-018-0606-0}{{\em Nature} {\bfseries
  562} no.~7728, (2018) 545--547},
  \href{http://arxiv.org/abs/1712.06531}{{\ttfamily arXiv:1712.06531
  [astro-ph.CO]}}.

\bibitem{Vitale:2018wlg}
S.~Vitale and H.-Y. Chen, ``{Measuring the Hubble constant with neutron star
  black hole mergers},''
  \href{http://dx.doi.org/10.1103/PhysRevLett.121.021303}{{\em Phys. Rev.
  Lett.} {\bfseries 121} no.~2, (2018) 021303},
  \href{http://arxiv.org/abs/1804.07337}{{\ttfamily arXiv:1804.07337
  [astro-ph.CO]}}.

\bibitem{Miao:2018zpw}
H.~Miao and Z.~Huang, ``{The $H_0$ Tension in Non-flat QCDM Cosmology},''
  \href{http://dx.doi.org/10.3847/1538-4357/aae523}{{\em Astrophys. J.}
  {\bfseries 868} no.~1, (2018) 20},
  \href{http://arxiv.org/abs/1803.07320}{{\ttfamily arXiv:1803.07320
  [astro-ph.CO]}}.

\bibitem{Bolejko:2017fos}
K.~Bolejko, ``{Emerging spatial curvature can resolve the tension between
  high-redshift CMB and low-redshift distance ladder measurements of the Hubble
  constant},'' \href{http://dx.doi.org/10.1103/PhysRevD.97.103529}{{\em Phys.
  Rev. D} {\bfseries 97} no.~10, (2018) 103529},
  \href{http://arxiv.org/abs/1712.02967}{{\ttfamily arXiv:1712.02967
  [astro-ph.CO]}}.

\end{thebibliography}\endgroup

\end{document}